\documentclass[longauth]{aa} 
\usepackage{graphicx}
\usepackage{txfonts}
\usepackage{grffile}
\usepackage{longtable}
\def\targ{IRAS~21078$+$5211}

\def\nh3{NH$_{3}$}
\def\kms{km~s$^{-1}$}

\def\Vlsr{$V_{\rm LSR}$}
\def\Jyb{Jy~beam$^{-1}$}

\def\G24{G24.78$+$0.08}
\def\HII{H{\sc ii}}

\newcommand{\ms}{$M_{\odot}$}
\newcommand{\ls}{$L_{\odot}$}

\newcommand{\pss}{$\rlap{.}^{\rm s}$}

\begin{document} 

   \title{Multi-scale view of star formation in \targ: From clump fragmentation to disk wind}
    
  \subtitle{}

  \titlerunning{Multi-scale view of \targ}
   
   \author{L. Moscadelli\inst{1}
         \and
         H. Beuther \inst{2}
         \and
         A. Ahmadi \inst{3}
         \and
         C. Gieser \inst{2}
         \and
         F. Massi \inst{1}
         \and
         R. Cesaroni \inst{1}
         \and
        \'A S{\'a}nchez-Monge \inst{4}
         \and
         F. Bacciotti \inst{1}
         \and
         M.~T. Beltr{\'a}n \inst{1} 
         \and 
         T. Csengeri \inst{5}
         \and
         R. Galv\'an-Madrid \inst{6}
         \and
         Th. Henning \inst{2}
         \and
         P.~D. Klaassen \inst{7}
         \and
         R. Kuiper \inst{8}
         \and
         S. Leurini \inst{9}
         \and
         S.~N. Longmore \inst{10}
         \and 
         L.~T. Maud \inst{11}
         \and
         T. M\"oller \inst{4}
         \and 
         A. Palau \inst{6} 
         \and 
         T. Peters \inst{12}
         \and
         R.~E. Pudritz \inst{13}
          \and
         A. Sanna \inst{9}
          \and
         D. Semenov \inst{2,14}
          \and
         J.~S. Urquhart \inst{15}
         \and
         J.~M. Winters \inst{16} 
         \and
         H. Zinnecker \inst{17}
         }
  
   \institute{INAF-Osservatorio Astrofisico di Arcetri, Largo E. Fermi 5, 50125 Firenze, Italy \\
              \email{mosca@arcetri.astro.it}
             \and
             Max Planck Institut for Astronomy, K\"{o}nigstuhl 17, 69117 Heidelberg, Germany 
             \and
             Leiden University, Niels Bohrweg 2, 2333 CA Leiden, Netherlands
             \and
              I.\ Physikalisches Institut, Universit\"at zu K\"oln, Z\"ulpicher Str.\ 77, D-50937 K\"oln, Germany 
             \and
             Max Planck Institut f\"ur Radioastronomie, Auf dem H\"ugel 69, 53121, Bonn, Germany 
             \and
              Instituto de Radioastronom\'ia y Astrofísica, Universidad Nacional Aut\'onoma de M\'exico, P.O. Box 3-72, 58090, Morelia, Michoac\'an, M\'exico
             \and
              UK Astronomy Technology Centre, Royal Observatory Edinburgh, Blackford Hill, Edinburgh EH9 3HJ, UK
             \and
             Institute of Astronomy and Astrophysics, University of T\"ubingen, Auf der Morgenstelle 10, D-72076 T\"ubingen, Germany 
             \and
             INAF - Osservatorio Astronomico di Cagliari, Via della Scienza 5, 09047 Selargius (CA), Italy
             \and
              Astrophysics Research Institute, Liverpool John Moores University, 146 Brownlow Hill, Liverpool L3 5RF, UK
             \and
             European Southern Observatory, Karl-Schwarzschild-Str. 2, 85748 Garching bei M\"unchen, Germany
             \and
             Max-Planck-Institut f\"ur Astrophysik, Karl-Schwarzschild-Str. 1,
85748 Garching bei M\"unchen, Germany
              \and
              Department of Physics and Astronomy, McMaster University, 1280 Main St. W, Hamilton, ON L8S 4K1, Canada
              \and
              Department of Chemistry, Ludwig Maximilian University, Butenandtstr. 5-13, 81377 Munich, Germany
              \and
              Centre for Astrophysics and Planetary Science, University of Kent, Canterbury CT2 7NH, UK
             \and
             IRAM, 300 rue de la Piscine, Domaine Universitaire de Grenoble, 38406, St.-Martin-d'Hères, France
             \and
             Universidad Autonoma de Chile, Avda Pedro de Valdivia 425, Santiago de Chile, Chile 
             }
   \date{}

 
  \abstract
 {Star formation (SF) is a multi-scale process in which the mode of fragmentation of the collapsing clump on scales of \ 0.1--1~pc determines the mass reservoir and affects the accretion process of the individual protostars on scales of \ 10--100~au.}
   {We want to investigate the nearby (located at 1.63$\pm$0.05~kpc) high-mass star-forming region \targ\ at linear scales from \ $\sim$1~pc \ down to \ $\sim$10~au.}
   {We combine the data of two recent programs: the NOrthern Extended Millimeter Array (NOEMA) large project CORE and the Protostellar Outflows at the EarliesT Stages (POETS) survey. The former provides images of the 1~mm dust continuum and molecular line emissions with a linear resolution of \ $\approx$~600~au \ covering a field of view up to \ $\approx$~0.5~pc. The latter targets the ionized gas and 22~GHz water masers, mapping linear scales from a few 10$^3$~au down to a few astronomical units.}
   {In \targ, a highly fragmented cluster (size $\sim$0.1~pc) of molecular cores is observed, located at the density peak of an elongated (size $\sim$1~pc) molecular cloud. A small ($\approx$~1~\kms\ per 0.1~pc) LSR velocity (\Vlsr) gradient is detected across the major axis of the molecular cloud. Assuming we are observing a mass flow from the harboring cloud to the cluster, we derive a mass infall rate of \ $\approx$~10$^{-4}$~\ms~yr$^{-1}$. The most massive cores (labeled 1,~2,~and~3) are found at the center of the cluster, and these are the only ones that present a signature of protostellar activity in terms of emission from high-excitation molecular lines or a molecular outflow. The masses of the young stellar objects (YSOs) inside these three cores are estimated in the range \ 1--6~\ms. We reveal an extended (size $\sim$0.1~pc), bipolar collimated molecular outflow emerging from core~1. We believe this is powered by the compact (size $\lesssim$~1000~au) radio jet discovered in the POETS survey, ejected by a YSO embedded in core~1 (named YSO-1), since the molecular outflow and the radio jet are almost parallel and have a comparable momentum rate. By means of high-excitation lines, we find a large ($\approx$~14~\kms\ over 500~au) \Vlsr\ gradient at the position of YSO-1, oriented approximately perpendicular to the radio jet. Assuming this is an edge-on, rotating disk and fitting a Keplerian rotation pattern, we determine the YSO-1 mass to be \ 5.6 $\pm$ 2.0~\ms. The water masers observed in the POETS survey emerge within 100--300~au from YSO-1 and are unique tracers of the jet kinematics. Their three-dimensional (3D) velocity pattern reveals that the gas flows along, and rotates about, the jet axis. We show that the 3D maser velocities are fully consistent with the magneto-centrifugal disk-wind models predicting a cylindrical rotating jet. Under this hypothesis, we determine the jet radius to be \ $\approx$~16~au and the corresponding launching radius and terminal velocity to be \ $\approx$~2.2~au \ and \ $\approx$~200~\kms, respectively.}
   {Complementing high-angular resolution, centimeter~and~millimeter interferometric observations in thermal tracers with Very Long Baseline Interferometry (VLBI) of molecular masers, is invaluable in studying high-mass SF. The combination of these two datasets allows us to connect the events that we see at large scales, as clump fragmentation and mass flows, with the physical processes identified at small scales, specifically, accretion and ejection in disk-jet systems.}


   \maketitle
\section{Introduction}
\label{intro}
High-mass ($\ga$ 8~\ms \ and \ $\ga$ 10$^4$~\ls) stars form within young stellar object (YSO) clusters deeply embedded inside thick dust cocoons,
whose study requires the superior angular resolution and sensitivity achieved only recently by the new or upgraded (sub)millimeter interferometers, such as the Atacama Large Millimeter/submillimeter Array (ALMA), the NOrthern Extended Millimeter Array (NOEMA), and the SubMillimeter Array (SMA). The processes of the formation of these clusters via the collapse and fragmentation of the parental molecular clump into multiple smaller cores, and mass accretion on individual cores, are not yet clear. Observing similar scales of 0.1--1~pc with comparable angular resolution of $\ga$1\arcsec, recent SMA and ALMA studies of infrared-quiet massive clumps have revealed different fragmentation properties: in some cases, limited fragmentation with a large fraction of cores with masses in the range \ 8--120~\ms\ \citep{Wan14,Cse17,Neu20}, and, in other cases, a large population of low-mass ($\le$1~\ms) cores with a maximum core mass of \ 11~\ms\ \citep{Sanh19}. There is a clear need for extending these kind of studies to identify the main agent(s) of molecular clump fragmentation. In the very early evolutionary phases, one would expect negligible support against gravitational collapse from thermal pressure and protostellar feedback (such as outflows or \HII~regions), and theoretical models predict that the main competitors of gravity should be the internal clump turbulence and magnetic pressure \citep[e.g.,][]{Fed15,Kle16,Hen19}.

The fragmentation modality determines the mass reservoir for the formation of individual stars and affects the accretion process as well, as discussed by the two main competing theories: the ``core-accretion'' model \citep{McK03}, which predicts the existence of quasi-equilibrium massive cores providing all the material to form high-mass stars, and the ``competitive accretion'' model \citep{Bon04}, in which the parent clump fragments into many low-mass cores that competitively accrete the surrounding gas. 
Within a core cluster, tidal interactions among nearby ($\sim$1000~au) YSOs can have a strong impact on accretion disks, causing them to precess \citep{Ces05} and to suffer perturbations \citep{Win18} or truncation \citep{God11c}, depending on the relative mass and separation of the interacting YSOs.   

Surveys at infrared and radio wavelengths have discovered that filaments are ubiquitous in massive molecular clouds \citep{Mol10a,Lu18} and have 
 a wide range in lengths (a few to 100~pc) and line
masses (a few hundreds to thousands of \ms~pc$^{-1}$). Young stellar object clusters and high-mass YSOs are often found at the junctions (named ``hubs'') of this filamentary structures in molecular clouds \citep[see, for instance,][]{Mye09}, and longitudinal mass flows (with typical rates of \ 10$^{-4}$--10$^{-3}$~\ms~yr$^{-1}$) along the filaments converging toward the hubs have been identified \citep{Chen19,Schw19,Tre19}. These recent findings strongly suggest that hub-filament systems can play a fundamental role in the formation of high-mass stars and have led to complementary or alternative theories to the  core-accretion and competitive-accretion models, such as the ``global hierarchical collapse'' \citep{Vaz19}, ``conveyor belt'' \citep{Long14}, and ``inertial-inflow'' \citep{Pad20} models. 
The main difference among these theories regards the origin of the hub-filament structures inside molecular clouds and the main driver (turbulence, density gradient, cloud-cloud collision, etc.) of the mass flows, but they all agree that the material to form the YSO clusters can be gathered from very large scales of \ 1--10~pc, channeled through the molecular cloud filaments.

Aiming at a statistical study of clump fragmentation and disk properties in high-mass YSOs, we are carrying out the large program "CORE" (P.I.:~Henrik Beuther; team web-page:~"http://www.mpia.de/core"), observing a sample of 20 high-mass star-forming regions with the IRAM interferometer NOEMA  in the 1.37~mm continuum and line emissions at high-angular resolution (0\farcs4). The analysis of the continuum emission from cold dust shows a large variety in the fragmentation properties of the sample, from regions where a single massive core dominates to regions where as many as 20 cores, with comparable masses, are detected \citep{Beu18}. In several CORE targets, employing dense-gas molecular tracers such as the \ CH$_3$CN and CH$_3$OH rotational transitions, LSR velocity (\Vlsr) gradients are often detected toward the most massive cores of the cluster \citep{Ahm18,Bos19,Ces19,Gie19}. Extending over typical linear scales of \ $\lesssim$1000~au, these \Vlsr\ gradients are interpreted in terms of the edge-on rotation of either a disk around a high-mass YSO or an unresolved binary (or multiple) system, or the combination of both motions. 

\targ, at a distance of \ 1.63$\pm$0.05~kpc \citep{Xu13}, is one of the closest and, with a bolometric luminosity  of 
\ 1.3$\, \times \,$10$^4$~\ls, least luminous CORE sources \citep[][see Table~1]{Beu18}. \targ\ (alias G092.69$+$3.08) also belongs to the target list of the Protostellar Outflows at the EarliesT Stages (POETS) survey \citep{Mos16,San18}, which has recently been carried out with the goal of imaging the inner (10--100~au) outflow scales in a sample of 38 high-mass YSOs. Employing Very Long Baseline Array (VLBA) observations of the 22~GHz water masers and Jansky Very Large Array (JVLA) multi-frequency continuum observations, we investigated both the molecular and ionized component of the outflows. In \targ, a jet is detected in the radio continuum emission,
consisting of two slightly resolved components separated by \ $\approx$~0\farcs5 \ along the SW-NE direction \citep[][see Fig.~8]{Mos16}. While the NE component has a positive spectral index and traces the ionized core of the jet close to the YSO, the SW lobe has a negative spectral index owing to synchrotron emission. The 22~GHz water masers are located to the NE from the YSO at a separation between \ 100~au and 300~au. The masers are clearly tracing the jet kinematics because their positions and proper motions are collimated along the radio jet. 

In this paper we combine the CORE and POETS data to study the process of star formation (SF) in \targ\ over linear scales from 10$^4$~au to 10~au. Recent Large Binocular Telescope (LBT) observations are also employed to reveal the structure of the molecular outflows at the largest scales. The CORE and LBT observational setups, data reduction and analysis are described in Sect.~\ref{obs}, including a brief summary of the POETS observations of \targ. In Sect.~\ref{res}, we report on the main observational results going from the large scale, that is the distribution and kinematics of the molecular cores in the cluster, to the small scale, that is the detailed view of the accretion-ejection structures around the most massive YSO(s). In Sect.~\ref{discu}, these results are interpreted within the frame of a coherent picture connecting the multi-scale physical processes. Finally, our conclusions are drawn in Sect.~\ref{conclu}. 

\section{Observations}
\label{obs}

\subsection{CORE}
The CORE program employs IRAM/NOEMA multi-configuration interferometric observations complemented with short-spacing IRAM~30~m single-dish data. A full description of the IRAM/NOEMA observational strategy and sample selection are given in \citet{Beu18} and the IRAM~30~m observations are detailed in \citet{Ahm18} and \citet{Mot20}.
Hereafter, we report on the observations, data reduction, and analysis specifically for \targ.

\subsubsection{NOEMA}

The IRAM/NOEMA interferometer observed \targ\ at 1.37~mm, in track sharing with another CORE target G100.3779$-$03.578, in three configurations (A, B, and D) between December 2014 and December 2016. The phase center for \targ\ was \ RA(J2000):~21$^{\rm h}$~09$^{\rm m}$~21\pss64 \ and \ DEC(J2000):~$+$52\degr~22$^{\prime}$~37\farcs50, and the systemic \Vlsr\ was assumed equal to \ $V_{\rm sys}$ = $-$6.1~\kms. In the course of the CORE observing campaign, the number of the NOEMA antennas increased from 6 to 8. The baseline lengths in the $uv$-plane range from 34~m to 765~m, corresponding to spatial frequencies from \ 0\farcs37 to 8\farcs3. The gain calibrators were the quasars \ J2201$+$508 \ and \ 2037$+$511, and \ 3C454.3 \ and \ MWC349 \ were the bandpass and flux calibrators.

By employing the WIDEX correlator, we recorded a broad band covering the frequency range \ 217.167--220.834~GHz at 1.95~MHz spectral resolution (corresponding to $\approx$~2.7~\kms), and eight narrow-band high-spectral resolution ($\approx$~0.43~\kms) units distributed over the broad band. The wide band is used to extract the line-free continuum, to get a chemical census of the region, and to study the distribution of more diffuse gas and the outflow kinematics (through low-excitation lines of the \ $^{13}$CO, H$_2$CO and SO molecules). The eight narrow-band spectral units are centered at specific spectral locations to observe typical dense-gas tracers (such as high-excitation lines of CH$_3$CN, CH$_3$OH, and CH$_2$CO) for investigating the gas kinematics and physical conditions at high-angular resolution. A detailed description of the spectral line coverage can be found in \citet[][see Table~8]{Beu18} and \citet{Ahm18}.
 Table~\ref{mol_lin} shows the parameters, taken from the CDMS \citep[Cologne Database for Molecular
Spectroscopy\footnote{http://www.astro.uni-koeln.de/cdms/},][]{Muel01,Muel05} and JPL \citep[Jet Propulsion Laboratory Catalog of Molecular Spectroscopy\footnote{https://spec.jpl.nasa.gov/},][]{Pick98} databases, of all the molecular lines analyzed in this article.

\begin{table}
\caption{Parameters of the molecular lines employed in the analysis}             
\label{mol_lin}      
\centering          
\begin{tabular}{c c c c}     
\hline\hline       
Frequency & Molecule & Transition & $E_{\rm u} / k_{\rm B}$ \\ 
 (GHz)    &          &            &   (K)          \\
\hline 
218.222  & H$_2$CO   &  3$_{0,3}$--2$_{0,2}$  &  21 \\        
218.325  & HC$_3$N   & 24--23  & 131   \\          
219.560  & C$^{18}$O & 2--1 & 16 \\
219.949  & SO &  6$_5$--5$_4$ & 35 \\
220.178  & CH$_2$CO & 11$_{1,11}$--10$_{1,10}$  &  76 \\
220.399  & $^{13}$CO & 2--1 & 16 \\
220.594  & CH$_3$CN & 12$_6$--11$_6$ & 326 \\
220.600  & CH$_3$$^{13}$CN & 12$_3$--11$_3$ & 133 \\
220.621  & CH$_3$$^{13}$CN & 12$_2$--11$_2$ & 97 \\
220.634  & CH$_3$$^{13}$CN & 12$_1$--11$_1$ & 76 \\
220.638  & CH$_3$$^{13}$CN & 12$_0$--11$_0$ & 69 \\
220.641  & CH$_3$CN & 12$_5$--11$_5$ & 247 \\
220.679  & CH$_3$CN & 12$_4$--11$_4$ & 183 \\
220.709  & CH$_3$CN & 12$_3$--11$_3$ & 133 \\
220.730  & CH$_3$CN & 12$_2$--11$_2$ & 97 \\
220.743  & CH$_3$CN & 12$_1$--11$_1$ & 76 \\
220.747  & CH$_3$CN & 12$_0$--11$_0$ & 69 \\
\hline                  
\end{tabular}
\tablefoot{\\
Column~1 report the rest frequency of the transition; Col.~2 the molecular species; Col.~3 the quantum numbers; Col.~4 the upper state energy.
The quantum numbers are given depending on the symmetry of
the molecule: $J^{\rm upper} - J^{\rm lower}$ for linear, 
$J^{\rm upper}_K - J^{\rm lower}_K$ for symmetric top, 
and $J^{\rm upper}_{K_a,K_c} - J^{\rm lower}_{K_a,K_c}$ for asymmetric top molecules.
}
\end{table}



The NOEMA data were calibrated with the CLIC and imaged with the MAPPING package in gildas\footnote{\url{http://www.iram.fr/IRAMFR/GILDAS/}}. The continuum data were extracted from line-free channels of the wide-band WideX spectra. Phase self-calibration was performed on the continuum data using the SELFCAL procedure. The gain table containing the self-calibration solutions was then applied to the narrow- and wide-band spectral line data using the UV\_CAL task. A detailed description of the phase self-calibration of the CORE continuum and spectral line data can be found in Gieser et al.~(submitted).

The NOEMA continuum data were imaged with uniform weighting using the Clark algorithm \citep{Cla80}. The synthesized beam of the continuum image of \targ\ has full width at half maximum (FWHM) major and minor sizes of \ 0\farcs48 and 0\farcs33, and \ PA~=~41\degr. For the extended $^{13}$CO and SO emissions, we used the low-resolution WideX spectral line data smoothed to a spectral resolution of 3.0~\kms. In order to avoid missing flux due to missing short-spacings, the data were combined with the IRAM~30~m observations and imaged with a robust parameter of 3 using the SDI algorithm \citep{Ste84}. 
 For \targ, the synthesized beam of the images produced by merging the NOEMA and IRAM~30~m spectral line data has FWHM major and minor sizes of \ 0\farcs70 and 0\farcs61, and \ PA~=~66\degr.
For the compact emissions of \ CH$_{3}$CN, HC$_3$N, \ and \ CH$_{2}$CO, we used the narrow-band NOEMA-only data smoothed to a spectral resolution of 0.5~\kms \ and imaged with uniform weighting using the Clark algorithm. The NOEMA images of \targ  \ have FWHM major and minor beam sizes of \ 0\farcs46 and 0\farcs31, with beam PA~=~40\degr.

\subsubsection{IRAM~30~m}
\targ\ was observed with the IRAM~30~m telescope in the 1.4~mm band employing the on-the-fly (OTF) mode and producing maps of sizes typically of \ $1^{\prime} \times 1^{\prime}$. Using double-sideband receivers, the IRAM~30~m data cover a broader range of frequencies than the NOEMA data, between $\approx$~213 and $\approx$~221~GHz in the lower sideband
and between $\approx$~229 and $\approx$~236~GHz in the upper sideband. At 1.37~mm, the angular and spectral resolution of the single-dish observations is  $\approx$~11\arcsec\ 
and \ 0.195~MHz (or 0.27~\kms), respectively. Merging the IRAM~30~m telescope data and the NOEMA visibilities yields coverage of the uv-plane in the inner 15~m. The IRAM~30~m data were calibrated using the CLASS package in gildas. To match the NOEMA observations, the IRAM~30~m data were smoothed to a spectral resolution of 3~\kms. A more detailed description of the calibration of the single-dish data is provided in \citet{Mot20}.


\begin{figure*}
\centering
\includegraphics[width=0.73\textwidth]{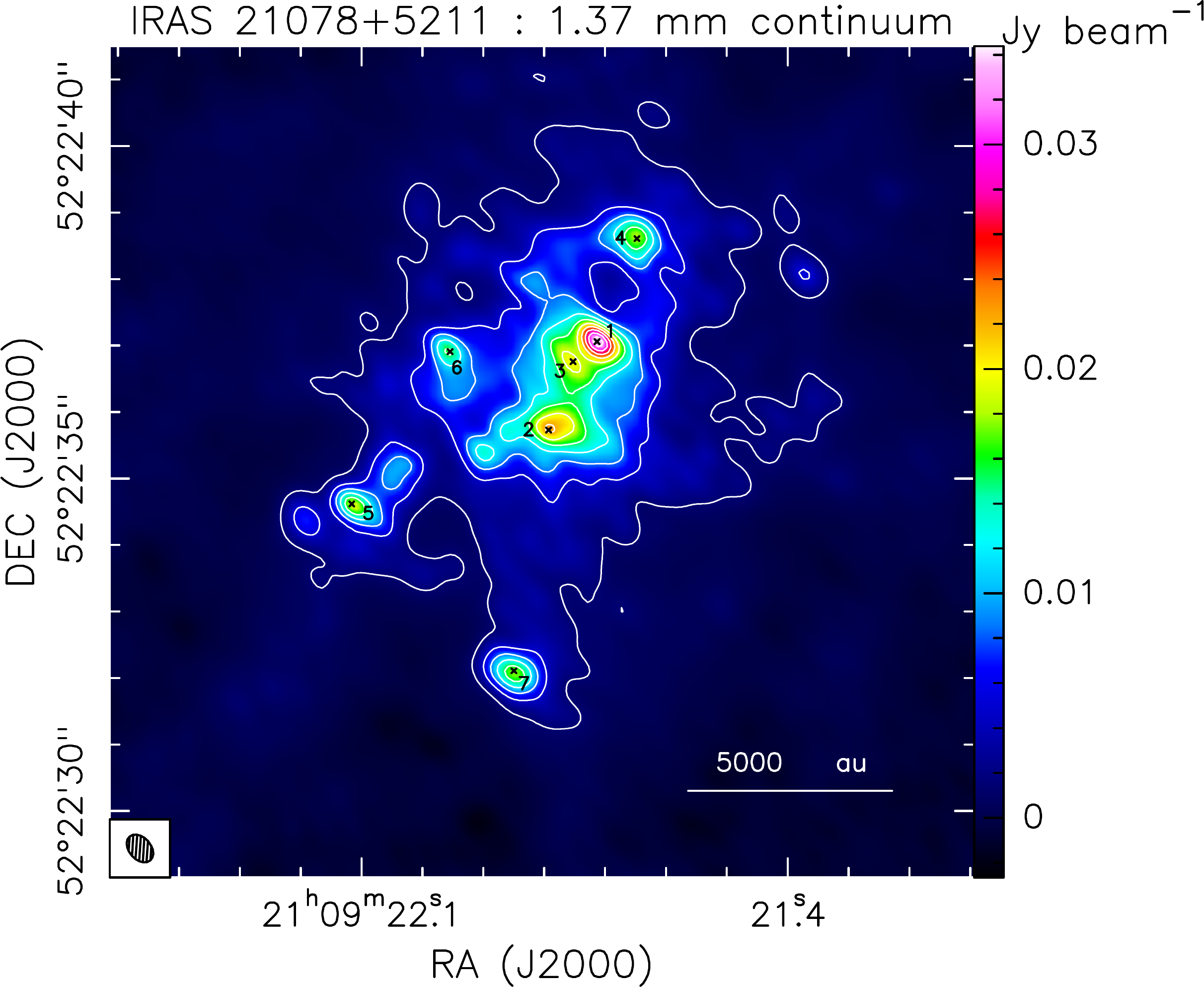}
\includegraphics[width=0.7\textwidth]{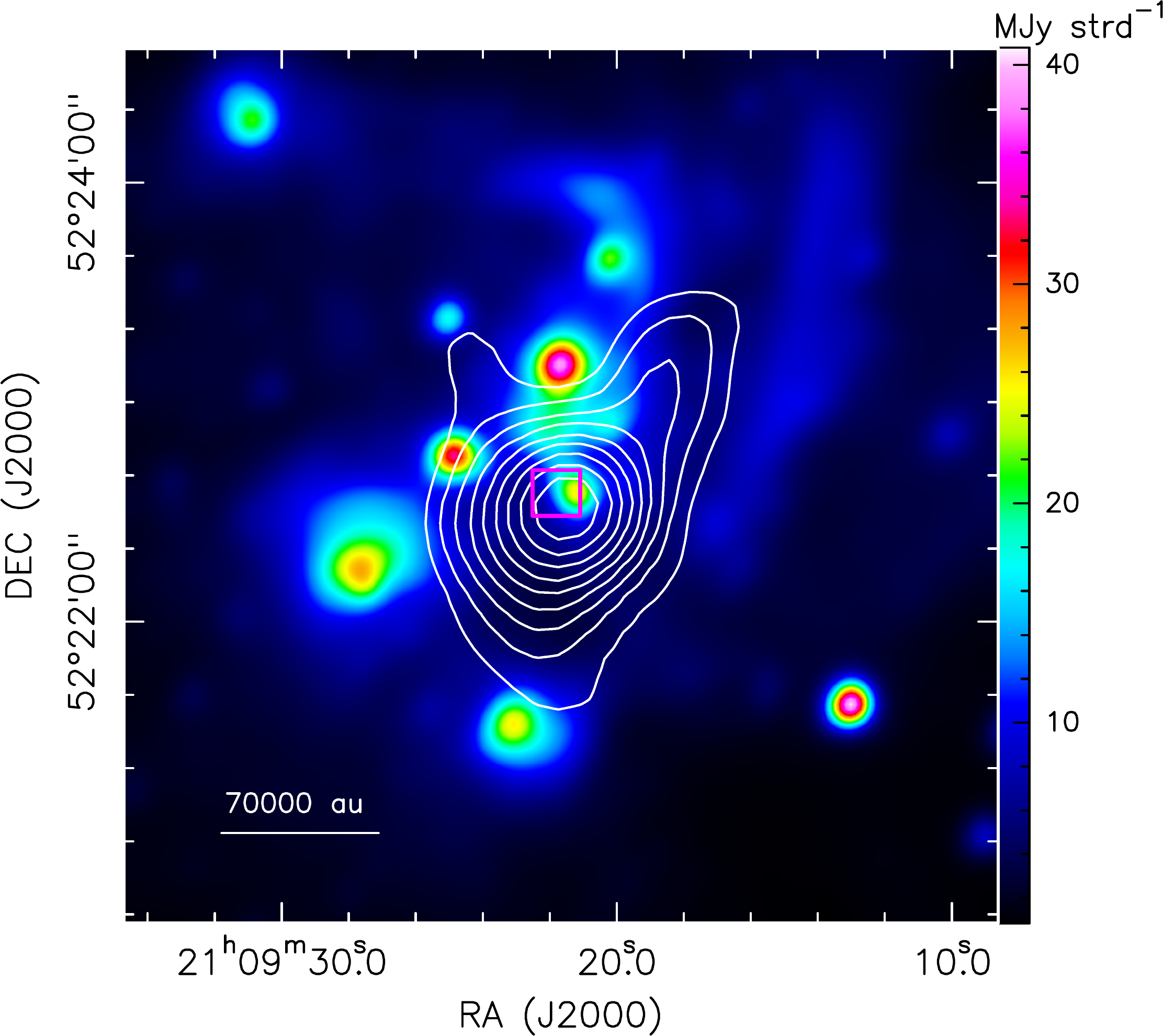}
\caption{Upper~panel:~NOEMA 1.37~mm continuum emission (color map) of \targ. The  white contours correspond to levels ranging from \ 3$\sigma$ ($\sigma$ = 0.5~m\Jyb) to the peak intensity of \ 35~m\Jyb, in steps of \ 7$\sigma$. The little black crosses indicate the positions of the seven strongest 1.37~mm peaks labeled as in \citet[][see Table~5]{Beu18}, with label numbers ordered in intensity. The restoring beam of the 1.37~mm continuum map is shown in the lower left corner. \ Lower~panel:~Wide-field Infrared Survey Explorer (WISE) 4.6~$\mu$m image \citep[color map,][]{Wri10} toward \targ. The white contours represent the  SCUBA 850~$\mu$m continuum \citep{DiFra08}, showing 10 levels increasing logarithmically from \ 7$\sigma$ \ ($\sigma$ = 0.1~\Jyb) to 11.4~\Jyb.
The magenta rectangle delimits the region plotted in the upper~panel.}
\label{1.3cont}
\end{figure*}

\subsection{POETS}
The POETS survey complements multi-epoch VLBI observations of water masers with high-angular resolution deep imaging of radio continuum emission with the JVLA in a relatively large sample (38) of massive protostellar outflows. \citet{Mos16} give a general description of the POETS program, sample selection, observations, and data analysis. Hereafter, we resume only the main observational parameters of \targ.

\subsubsection{VLBI of 22~GHz water masers}
The water maser emission (at 22.23508~GHz) in \targ \ was observed within the VLBA BeSSeL\footnote{The Bar and Spiral Structure Legacy (BeSSeL) survey is a VLBA key project, whose main goal is to derive the structure and kinematics of the Milky Way by measuring accurate positions, distances (via trigonometric parallaxes), and proper motions of methanol and water masers in hundreds of high-mass star-forming regions distributed over the Galactic disk \citep{Rei14}.} survey (project code: BR145E) at six epochs spanning about one~year (from May 2010 to May 2011) with a total observing time per epoch of approximately seven~hours.
A general description of the BeSSeL VLBA observational setup is given in \citet{Rei09}.
In the course of the observations of \targ, the FWHM size of the VLBA beam varied from epoch to epoch in the range \ 0.5--0.9~mas, and the instantaneous field of view, limited by time-smearing, was \ $\approx$~5\farcs4. The velocity resolution for the 22~GHz water maser data was \ 0.42~\kms. In \targ, maser absolute positions and proper motions are derived with an accuracy of  \ 2~mas \ and \ 1--3~\kms, respectively.

\subsubsection{JVLA continuum observations}
 \targ \ was observed using the JVLA of the National Radio Astronomy Observatory (NRAO\footnote{NRAO is a facility of the National Science Foundation operated under cooperative agreement by Associated Universities, Inc.}) between October 2012 and January 2013 in A configuration (project code: 12B-044), in C, Ku, and K bands  (centered at 6.2, 13.1, and 21.7~GHz, respectively). In the following, we also refer to C-~and~K-band observations as the JVLA 5~cm and 1.3~cm continuum, respectively. The total on-source time was 15~min at C~and~Ku bands, and 30~min at K band. We employed the capabilities of the new WIDAR correlator to record dual polarization across a total bandwidth per polarization of 2~GHz.
For the JVLA A-Array continuum images of \targ, the FWHM size of the round synthesized beam and the largest recoverable angular scale are \ 0\farcs44 \ and \ 4\farcs5, 0\farcs17 \ and \ 1\farcs8, 0\farcs08 \ and \ 1\farcs2 at \ C, Ku and K band, respectively.  
The absolute positions of the images are expected to be accurate to within \ 30--40, 20, and 10~mas at 6, 13, and 22~GHz, respectively. 
The rms noise level of the continuum images is \ 10.7, 9.2, and 13.9~$\mu$\Jyb, at \ C, Ku, and K band, respectively.

\begin{figure}
\centering
\includegraphics[width=0.5\textwidth]{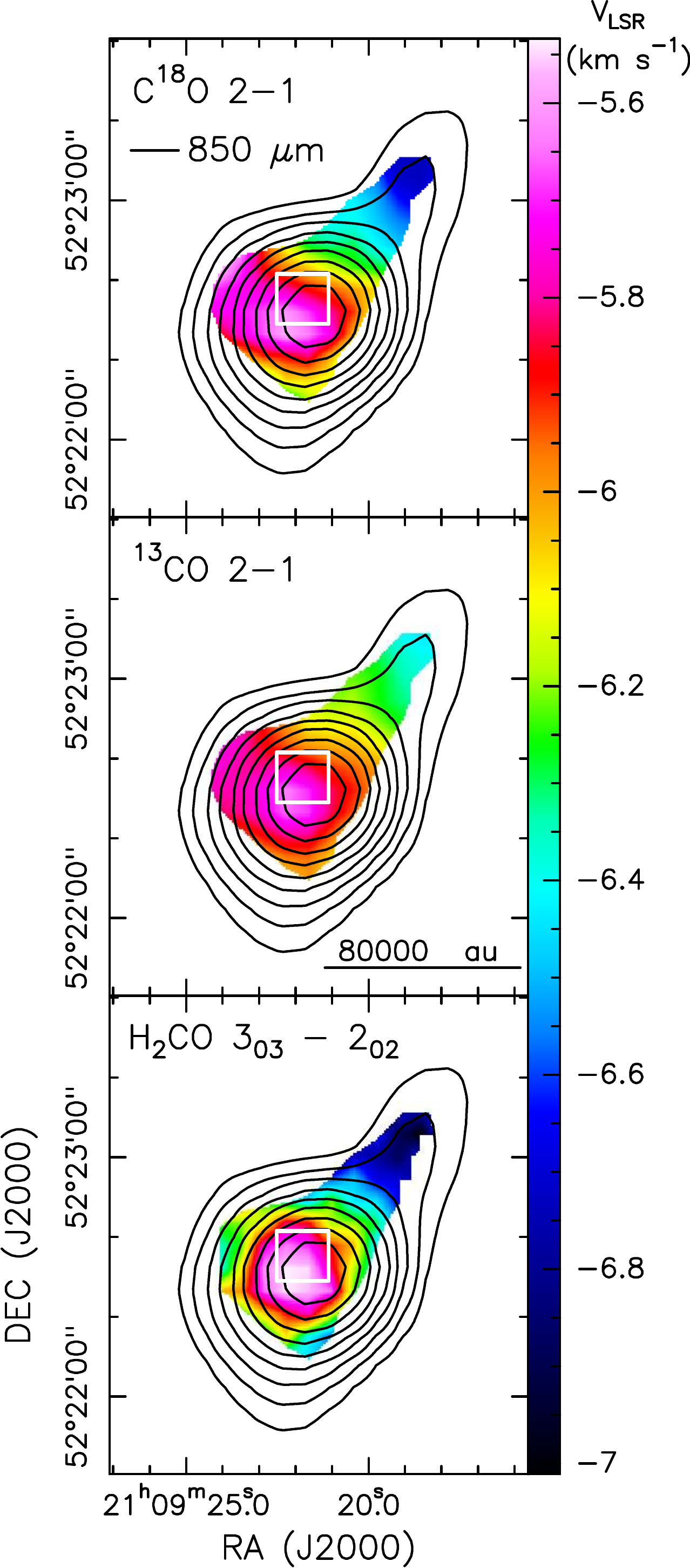}
\caption{IRAM~30~m data. Intensity-weighted velocity (color map) of the \ C$^{18}$O~$J=$ 2-1 (upper~panel), $^{13}$CO~$J=$ 2-1 
(middle~panel), and H$_2$CO~$J_{K_a,K_c}=$ 3$_{0,3}$-2$_{0,2}$ (lower~panel) lines determined over a small velocity range, [$-$7.3, $-$3.5]~\kms, around \ $V_{\rm sys}$ = $-$6.1~\kms. The plotted region corresponds to the area in which the velocity-integrated emission of the \ C$^{18}$O~$J=$ 2-1 \ line is higher than \ 44~\Jyb~\kms, $\approx$~10$\sigma$. In each panel, the black contours represent the  SCUBA 850~$\mu$m continuum \citep{DiFra08}, showing 10 levels increasing logarithmically from 1 to 11.4~\Jyb.
The white rectangle delimits the region plotted in the upper~panel of Fig.~\ref{1.3cont}.}
\label{cloud_vel}
\end{figure}

\subsection{LBT}


Near-infrared (NIR) observations toward \targ\
were carried out with the 
instrument LUCI~2 \citep{Age10,Seif10} at the 
$2 \times 8.4$~m LBT (Mount Graham, Arizona) on June 12, 2017,
as part of the ``Istituto Nazionale di Astrofisica'' (INAF) program 2016\_2017\_54 (PI: Moscadelli). A field
roughly centered on the position of \targ\ was imaged through
the broad-band $K_{\rm s}$ ($\lambda_{\rm c} = 2.16$~$\mu$m, 
FWHM = 0.27~$\mu$m) 
and narrow-band H$_{2}$ ($\lambda_{\rm c} = 2.127$~$\mu$m, 
FWHM = 0.023~$\mu$m) filters. We operated LUCI~2 in seeing-limited imaging, single
telescope mode, using camera N3.75, which provides a plate scale of
$\approx$~0.12\arcsec per pixel and a field of view of \ $ 4\arcmin \times 4\arcmin$.
The images were taken according to a dithering pattern alternating 
a pointing in which the nominal position of \targ \ is only
a few arcseconds from the frame center and a pointing in which this position is
$\approx$1\arcmin \ from the frame center, so that all the frames contain the
target position and skies can be obtained by selecting  subsets of
the more distant pointings. The H$_{2}$ observations consist of 31
dithered images with DIT$= 6$~s, NDIT$= 25$ (total integration time
4650~s) and the $K_{\rm s}$ observations consist of 13  dithered images
with DIT$= 2.74$~s, NDIT$= 18$ (total integration time 641~s). 

Data reduction was performed using standard {\small IRAF}\footnote{{\small
IRAF} is distributed
by the National Optical Astronomy Observatories, which are operated by the
Association of Universities for Research in Astronomy, Inc., under cooperative
agreement with the National Science Foundation.} routines. Each frame was flat-fielded
using dome flat images and corrected for bad pixels. As they are vignetted, we cut out
the outermost pixels before further steps. Skies were then constructed for each
frame by median-filtering a subset of four frames selected from the nearest in time 
that also have large pointing offsets. For each filter, all 
sky-subtracted frames were registered and averaged together. The final mosaicked
images exhibit point spread functions (PSFs) $\approx$1\arcsec\ wide. The continuum
emission in the H$_2$ filter was estimated by comparing photometry of stars in
the $K_{\rm s}$ and H$_2$ images. The two images were then scaled and subtracted from each other
to produce an image showing pure H$_{2}$ $2.12$~$\mu$m line emission. 

As the images showed a complex blob of line emission near the target 
(which in turn appears obscured in the NIR), we decided to perform further
adaptive optics (AO) assisted observations to try resolving possible bow-shock
features making up the blob. New NIR observations were then obtained with 
LUCI~1 and FLAO \citep{Esp12} at the LBT in diffraction-limited mode on October 1, 2017,
as part of the INAF program 
2017\_2018\_36 (PI: Massi). A field including both the target nominal position and the
line emission blob was imaged through the H$_2$ and $K_{\rm s}$ filters using camera
N30, which provides a plate scale of
$\approx$~0.015\arcsec per pixel and a field of view of \  $30\arcsec \times 30\arcsec$. A 
star with $R \approx 12$ at a distance of $\approx$1\arcmin\ from the imaged
field was selected as a guide star. A dithering pattern with random pointings 
a few arcseconds apart
was selected. The H$_2$ data consist of a set of 24 dithered frames with DIT$= 10$~s
and NDIT$=15$ (total integration time 1~hr) and the $K_{\rm s}$ data of a set of
6 dithered frames with DIT$= 5$~s
and NDIT$=20$ (total integration time 10~min). The frames were flat-fielded and
bad pixels were removed. For each frame, a sky was constructed by median filtering
the nearest (in time) four frames. The sky-subtracted frames were finally registered and
averaged together. Unfortunately the AO correction was not very efficient (owing to the lack of a suitable guide star) and a Strehl ratio of only $\approx$~1\%
was obtained. Nevertheless, these latest images (with a PSF FWHM $\approx$~0\farcs3)
exhibit a better spatial resolution 
than the previous seeing-limited images . 
An image showing pure H$_{2}$ $2.12$~$\mu$m line emission was obtained
following the same procedure as above. 

Comparing NIR images with radio and millimeter interferometric data, which usually have
high positional accuracy, requires an astrometric calibration as
precise as possible. First, we obtained an astrometric solution for the
seeing-limited, large field-of-view images by cross-matching them with 2Mass Point
Source Catalog entries. We were able to use about 100 relatively bright
stars all over the field.  Then, we cross-correlated the seeing-limited
images and the AO-assisted images, finding $\approx$20 suitable stars to
obtain an astrometric solution for the latter ones. We estimate that the
astrometric accuracy of the AO-assisted images is better than 0\farcs1.


\begin{figure*}
\centering
\includegraphics[width=0.76\textwidth]{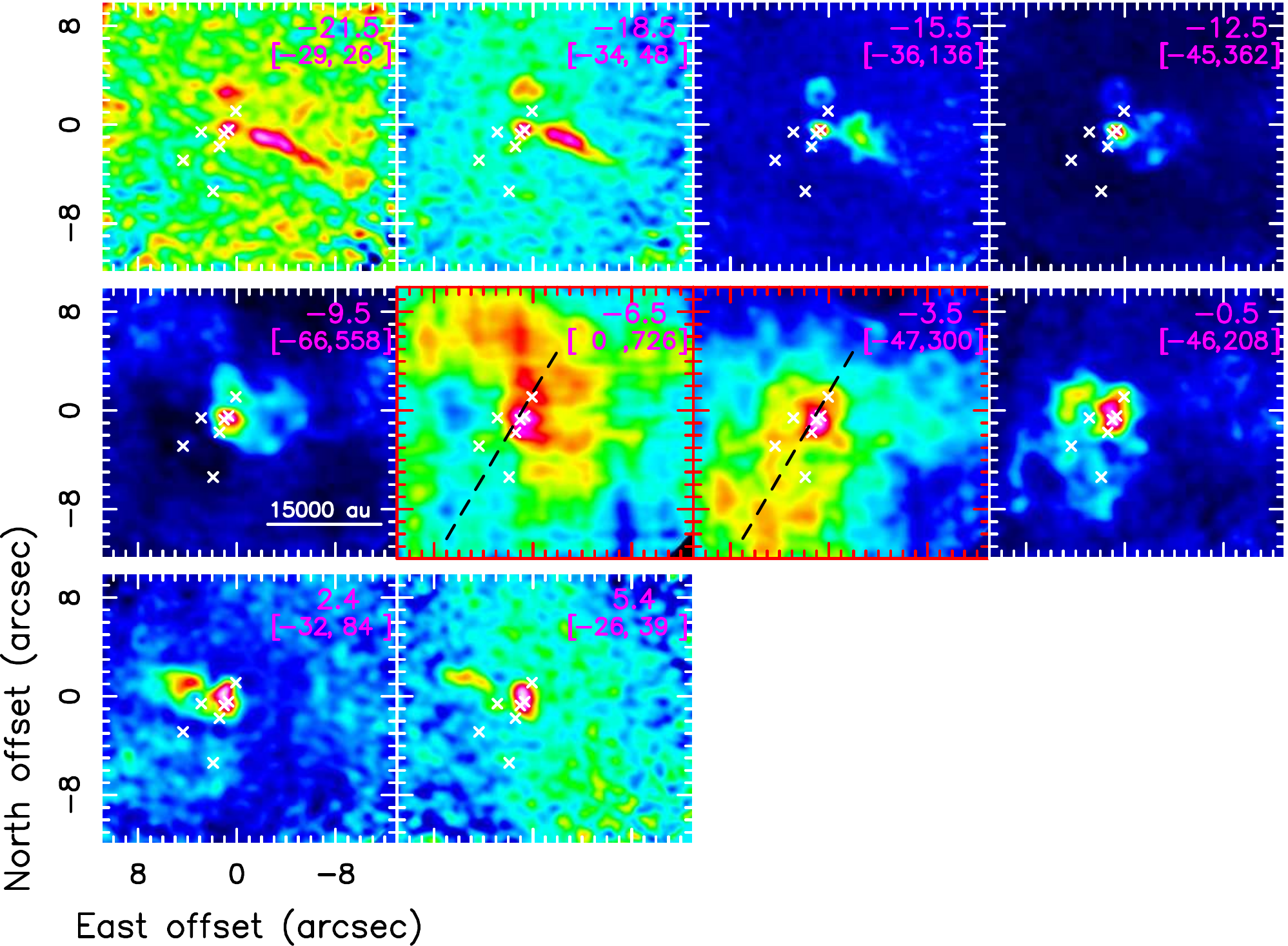}
\caption{Merged NOEMA--IRAM~30~m data. Each panel presents the emission of the \ $^{13}$CO~$J=$ 2-1 line (color map) in a different velocity channel. In the upper right corner of the panel, the channel \Vlsr\ (in kilometer per second) and the range of plotted intensity (in square parentheses in milliJansky per beam) are reported. The red boxes and ticks identify the panels corresponding to the central velocities, that is \ $ | V_{\rm LSR} - V_{\rm sys} | < 3$~\kms. The white crosses mark the positions of the seven strongest peaks of the 1.37~mm continuum emission. At the central velocities, the black dashed line gives the major axis of the 1.37~mm core cluster.}
\label{13COmap}
\end{figure*}

\begin{figure*}
\centering
\includegraphics[width=0.76\textwidth]{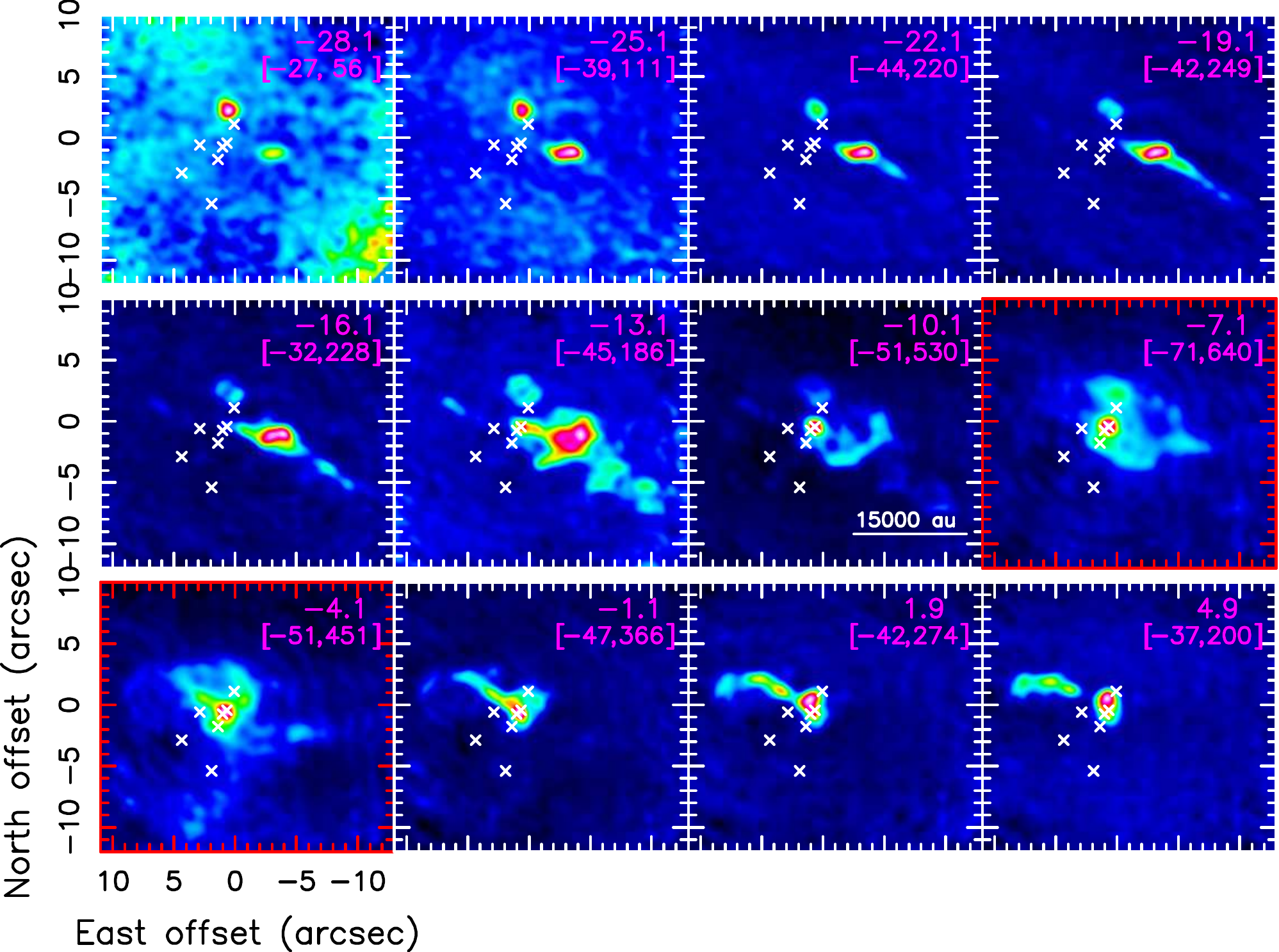}
\caption{Merged NOEMA--IRAM~30~m data. Each panel presents the emission of the \ SO~$J_N=$ 6$_5$-5$_4$ line (color map) in a different velocity channel. In the upper right corner of the panel,  the channel \Vlsr\ (in kilometer per second) and the range of plotted intensity (in square parentheses in milliJansky per beam) are reported. The red boxes and ticks identify the panels corresponding to the central velocities, that is \ $ | V_{\rm LSR} - V_{\rm sys} | < 3$~\kms. The white crosses mark the positions of the seven strongest peaks of the 1.37~mm continuum emission. }
\label{SOmap}
\end{figure*}

\section{Results}
\label{res}
\subsection{The cluster}
\label{obs_clu}



The upper panel of Figure~\ref{1.3cont} shows the 1.37~mm continuum emission in \targ, which is already presented in \citet[][see Fig.~B.3, upper panels]{Beu18}. Using the \textsc{CLUMPFIND} algorithm, \citet[][see Table~5]{Beu18} identify 20 compact cores with intensities in the range \ 3--35~m\Jyb, the seven strongest of which are indicated in Fig.~\ref{1.3cont} (upper~panel). The global pattern of the 1.37~mm continuum is elongated in the SE-NW direction: the PA of the major axis, determined through a linear fit to the positions of the seven strongest cores, is \ 149\degr $\pm$ 17\degr.
The three strongest cores are found inside a higher-emission plateau at the center of the cluster, and the lower-intensity cores draw an arc of weaker emission encircling the plateau clockwise from SE to NW, with no cores found to SW. 

The spatial distribution of the cores in the cluster revealed by the 1.37~mm continuum emission on linear scales of \ 10$^4$~au has several properties in common with the patterns of different SF indicators on larger scales.
The lower panel of Figure~\ref{1.3cont} shows that the core cluster is embedded within a molecular cloud traced by the 850~$\mu$m emission from cold dust, which presents a similar SE-NW elongation (PA $\approx$~140\degr). In turn, this molecular cloud is edged on the northeastern side with a necklace of compact 4.6~$\mu$m sources, associated with less embedded and, likely, more evolved YSOs in the region.
This distribution resembles that of the cores in the cluster on ten times smaller scales.


To study the kinematics of the gas in the parental molecular cloud enshrouding the core cluster, we use the IRAM~30~m maps of three low-excitation ($E_{\rm u}/k_{\rm B} \le 21$~K) molecular lines: \ $^{13}$CO~$J=$ 2-1, 
C$^{18}$O~$J=$ 2-1, \ and \ H$_2$CO~$J_{K_a,K_c}=$ 3$_{0,3}$-2$_{0,2}$.
Figure~\ref{cloud_vel} shows that, close to \ $V_{\rm sys}$ = $-$6.1~\kms, the velocities increase regularly from NW to SE along the major axis of the molecular cloud, reaching the highest values in correspondence of the core cluster. To investigate the gas kinematics inside the cluster, Fig.~\ref{13COmap} shows velocity-channel maps of the merged NOEMA--IRAM~30~m observations of the \ $^{13}$CO~$J=$ 2-1 \ transition. 
In the two channels closer to \ $V_{\rm sys}$, namely at
\Vlsr\ of \ $-$6.5 \ and \ $-$3.5~\kms, the \ $^{13}$CO~$J=$ 2-1 \ emission is prominent and extended on scales of \ a few 10$^4$~au along a SE-NW direction close to the similarly oriented, elongation axes of the 1.37~mm core cluster and 850~$\mu$m molecular cloud. Despite the limited velocity resolution of only \ 3~\kms, a \Vlsr\ gradient across the whole cluster, and its gaseous envelope, is clearly detected, the gas to N-NW emitting at lower \Vlsr\ than the gas to SE.  Thus, the merged NOEMA--IRAM~30~m observations of
the \ $^{13}$CO~$J=$ 2-1 \ line confirm the velocity gradient observed at larger scales using only the IRAM~30~m data.  


At high velocities, the spatial distribution of the more diffuse gas inside the cluster is revealed by the \ $^{13}$CO~$J=$ 2-1 (Fig.~\ref{13COmap}) \ and  \  SO~$J_{N}=$ 6$_5$-5$_4$ (Fig.~\ref{SOmap}) transitions.
 At relatively high absolute line of sight (LOS) velocities, that is \ $ | V_{\rm LSR} - V_{\rm sys} | \ge 3$~\kms, the channel emission in both lines is characterized by a SW-NE elongated feature pointing to core~1 (the strongest 1.37~mm continuum emitter), and placed either NE or SW of core~1 at high red-shifted ($\ge$$-$2~\kms) or blue-shifted ($\le$$-$10~\kms) \Vlsr. A straightforward interpretation for this emission feature in the \ $^{13}$CO \ and \ SO \ lines, as discussed in detail in Sect.~\ref{dis_jet}, is in terms of a collimated outflow emitted by a YSO inside core~1. In the \ SO~$J_N=$ 6$_5$-5$_4$ \ channel maps at \ \Vlsr $\approx$~$-$13~\kms, the emission is dominated by a strong bow shock of this outflow located \ $\approx$~4\arcsec\ SW of core~1. Aside from the prominent signature of the outflow from core~1, the only other notable emission feature at high absolute LOS velocities is a compact source visible (especially in the SO channel maps) at very blue-shifted velocities ($-$28~\kms\ $\le$ \Vlsr \ $\le$ $-$20~\kms) to N-NE of core~4 (the northernmost core).

\begin{figure*}
\centering
\includegraphics[width=\textwidth]{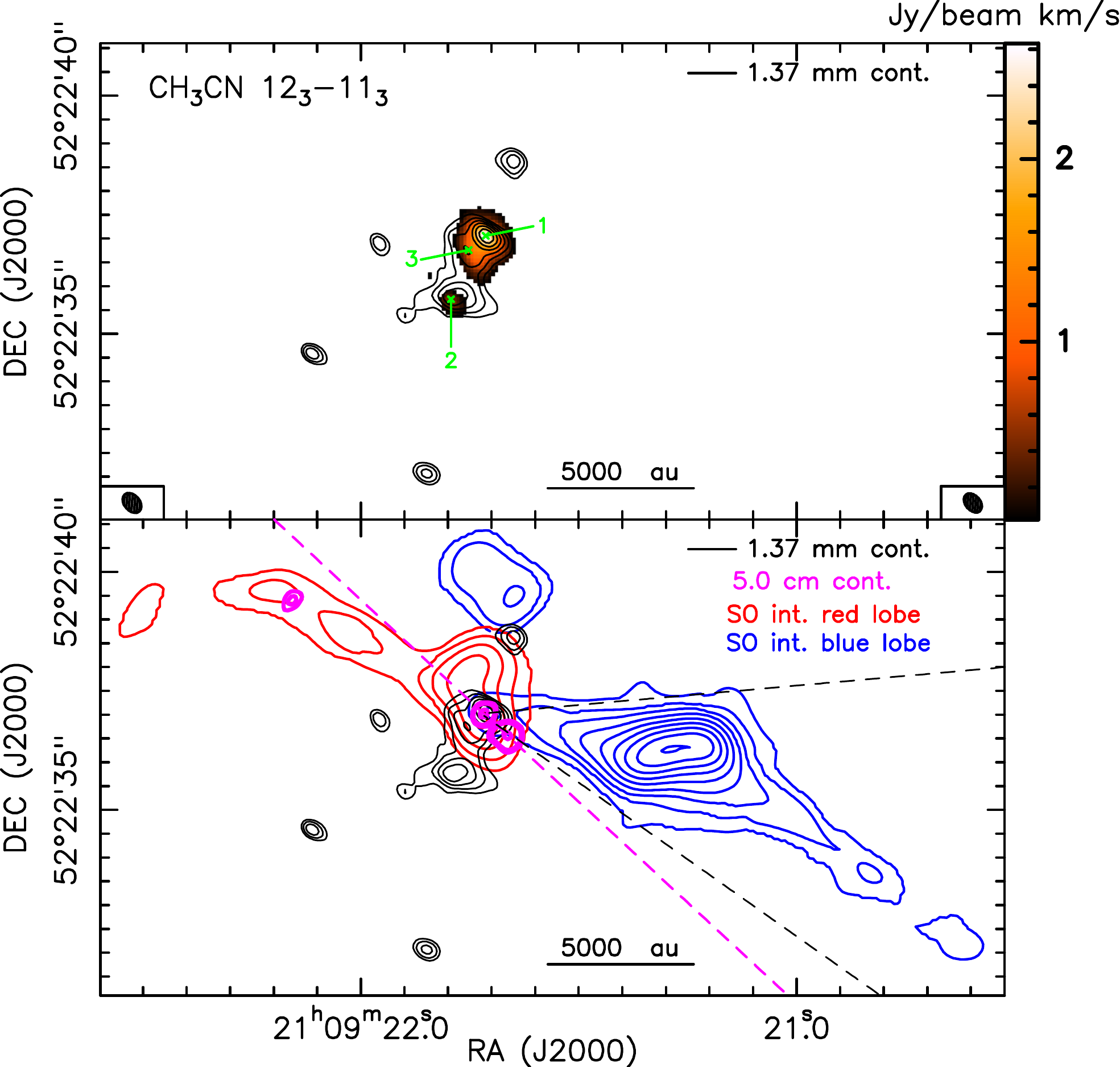}
\caption{Upper~panel:~NOEMA data. Velocity-integrated intensity (color map) of the \ CH$_3$CN~$J_K=$ 12$_3$-11$_3$ line. The 1.37~mm continuum emission is represented with  black contours, showing seven levels increasing logarithmically from 11 to 35~m\Jyb. The positions of the three strongest peaks of the 1.37~mm continuum emission are indicated, using the same labels as in \citet[][see Table~5]{Beu18}.
The restoring beams of the \ CH$_3$CN~$J_K=$ 12$_3$-11$_3$ line and 1.37~mm continuum maps are shown in the lower right~and~left corners, respectively. \  Lower~panel:~The  blue and red contours reproduce the emission of the \ SO~$J_N=$ 6$_5$-5$_4$ line integrated over the velocity ranges \ [$-$22,$-$13] \ and \ [$-$1, 5]~\kms, respectively. The plotted levels are from \ 0.03 to 2.5~\Jyb, in steps of \ 0.27~\Jyb, and from \ 0.6 to 2.2~\Jyb, in steps of \ 0.37~\Jyb, for the blue-~and~red-shifted SO emission, respectively. The magenta contours give the JVLA A-Array continuum at 5~cm \citep{Mos16}, showing levels at \ 40\%, 50\%, and 90\% of the peak emission of \ 95~$\mu$\Jyb: the magenta dashed line connects the two strongest (nearby, but resolved) 5~cm peaks. The two black dashed lines delimit the viewing angle of the blue-shifted emission of the \ SO~$J_N=$ 6$_5$-5$_4$ line from the NE 5~cm peak (aligned in position with core~1, see Sect.~\ref{obs_clu}). The black contours have the same meaning as in the upper panel.}
\label{CH3CN-SO}
\end{figure*}

%
\begin{figure*}
\centering
\includegraphics[width=\textwidth]{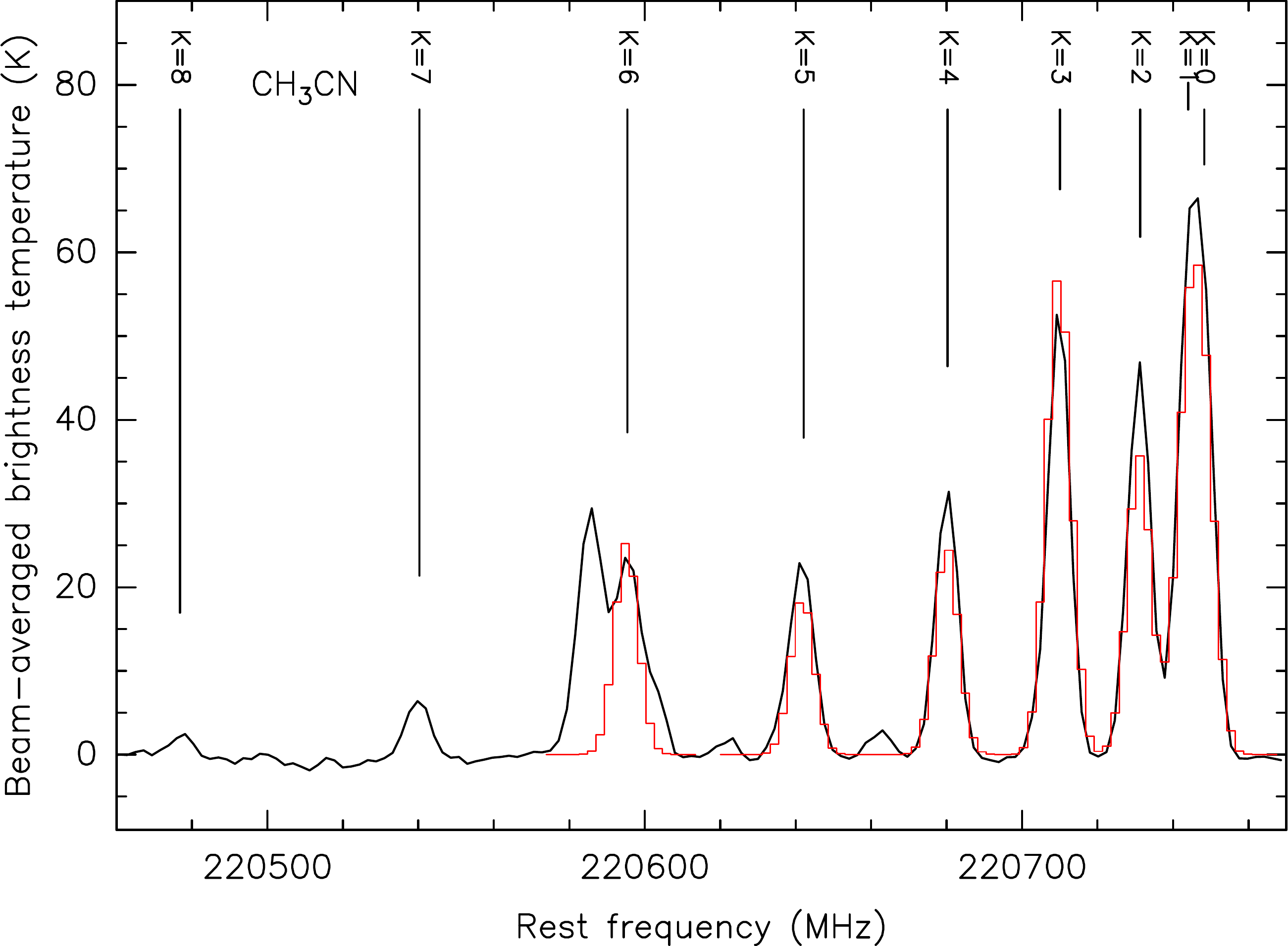}
\caption{Beam-averaged spectrum (black line) of the \ CH$_3$CN~$J_K=$ 12$_K$-11$_K$ ($K$ = 0--8) line emission toward the 1.37~mm continuum peak. 
The positions of the \ $K$ = 0--8 lines are indicated. The fits (red histograms) of the \ $K$ = 0--6 lines performed with \textsc{XCLASS} are also shown.}
\label{CH3CN_ratio}
\end{figure*}



In \targ, the higher-excitation ($E_{\rm u}/k_{\rm B} \ge 70$~K) molecular lines observed in the CORE program are only detected from inside the higher-emission plateau at the center of the 1.37~mm continuum distribution. The upper panel of Figure~\ref{CH3CN-SO} shows the velocity-integrated intensity of the CH$_3$CN~$J_K=$ 12$_3$-11$_3$ line, which is representative of the spatial distribution of all the observed dense-gas tracers. The CH$_3$CN emission peaks at the position of core~1 and it becomes progressively weaker moving from core~3 (near core~1) to core~2 (displaced further to S; see Fig.~\ref{1.3cont}, upper~panel). In the lower panel of  Fig.~\ref{CH3CN-SO}, the 1.37~mm continuum is overlaid with the tracers of the outflows discovered in this region, that is the SW-NE (PA = 43\degr $\pm$ 10\degr), double-component radio jet detected in the POETS survey and the red-~and~blue-shifted integrated emission of the CORE \ SO~$J_N=$ 6$_5$-5$_4$ \ line. It should be noted that the NE component of the radio jet with thermal emission from ionized gas near the YSO, coincides with core~1, and that the radio jet and the SO outflow are approximately parallel to each other. These two findings lead us to assume that the YSO embedded in core~1 is driving the radio jet, which, in turn, powers the larger-scale molecular outflow. In Sect.~\ref{dis_jet}, we check this assumption by comparing the properties of the radio jet and the SO outflow. Hereafter, we refer to the YSO inside core~1 as \ "YSO-1".



\begin{figure}
\centering
\includegraphics[width=0.5\textwidth]{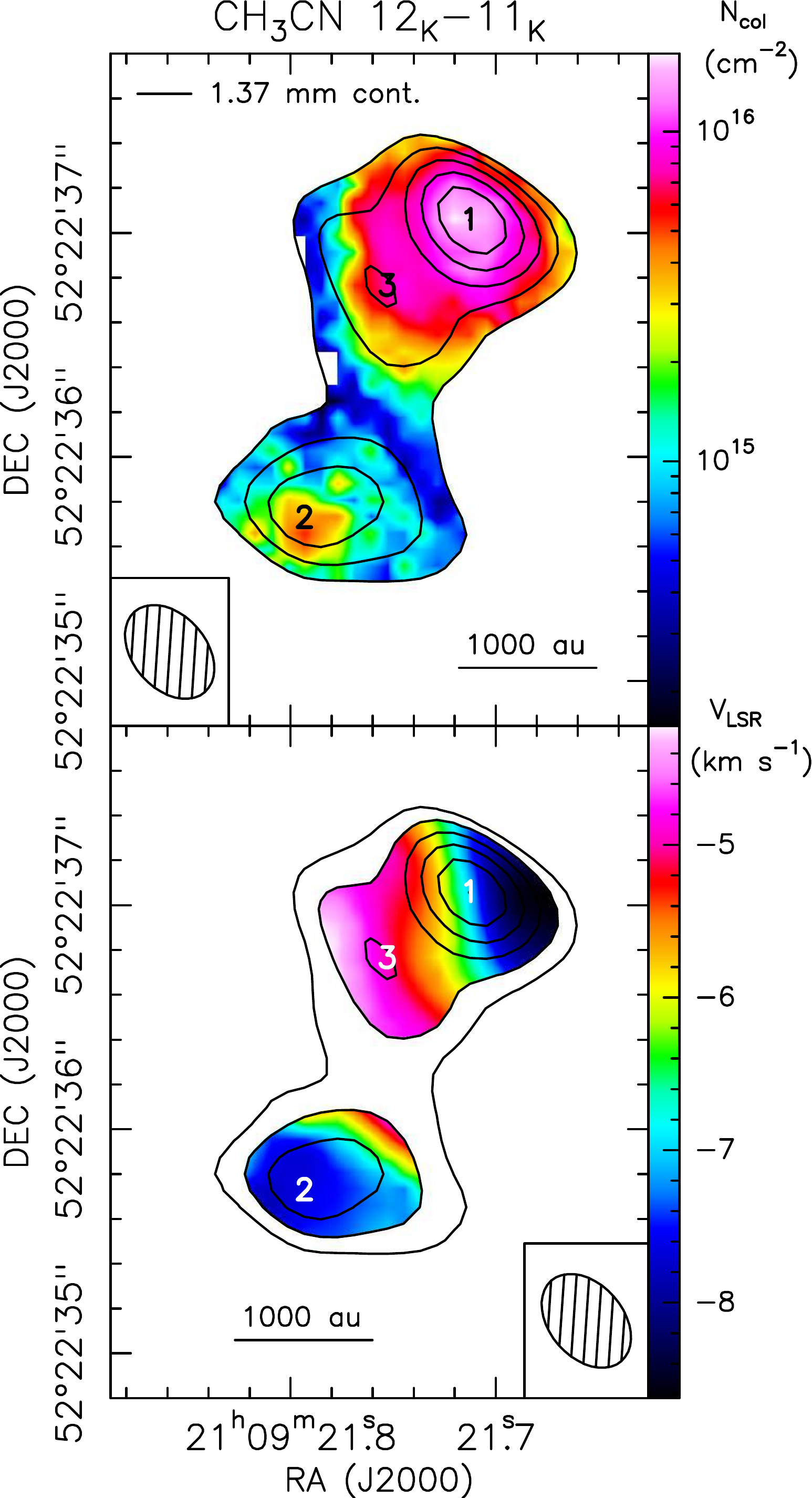}
\caption{NOEMA data. The color map reproduces the column density (upper~panel) and the velocity (lower~panel) of the \ CH$_3$CN \ emission determined by fitting with \textsc{XCLASS} the \ CH$_3$CN~$J_K=$ 12$_K$-11$_K$ (K= 0--6) lines (including the CH$_3$$^{13}$CN isotopologs), simultaneously. 
The plotted regions correspond to the areas in which the 1.37~mm continuum emission is higher than 13~and~16~m\Jyb \ for the column density and velocity plots, respectively. 
The 1.37~mm continuum emission is represented with black contours, showing six levels increasing logarithmically from 13 to 35~m\Jyb: the positions of the three strongest 1.37~mm peaks are denoted using the same labels as in \citet[][see Table~5]{Beu18}. The restoring beams of the 1.37~mm continuum and \ CH$_3$CN \ maps are shown in the lower left corner of the upper panel and lower right corner of the lower panel, respectively.}\label{CH3CN}
\end{figure}


\subsection{Cluster center: Cores~1,~2,~and~3}
\label{obs_bin}


Toward cores~1,~2,~and~3, the emission of the higher-excitation molecular lines is strong enough to allow us to study the kinematics and physical conditions of the gas.  For this purpose, we used the \textsc{XCLASS} (eXtended CASA Line Analysis Software Suite)
tool \citep{Moel17}. This tool models the data by solving the radiative transfer equation for an isothermal homogeneous object in local thermodynamic equilibrium (LTE) in one dimension. It can produce maps of column density, temperature, velocity, and line width by extracting spectra pixel-by-pixel from the image cubes and fitting all the unblended lines of a given molecular species simultaneously. 
The finite source size, dust attenuation, and line opacity are considered as well,
 which also permits us to properly fit the transitions of relatively lower-excitation energy and higher optical depths (as the \ CH$_3$CN~$J_K=$ 12$_K$-11$_K$, K= 0--2 lines). 

 To complement the \textsc{XCLASS} results, we also analyzed several, higher-excitation molecular lines, individually. We verified that  the selected  lines are sufficiently optically thin to reliably trace the more embedded gas kinematics. More specifically, in this analysis we made use of the following molecular lines: CH$_3$CN~$J_K=$ 12$_K$-11$_K$ ($K$ = 3--6), CH$_2$CO~$J_{K_a,K_c}=$ 11$_{1,11}$-10$_{1,10}$, and \ HC$_3$N~$J=$ 24-23.
We fitted the line profile of these transitions toward the 1.37~mm continuum peak in core~1 and found that the optical depth is always \ $<$~0.5 \ at the peak velocity and decreases to values \ $\lesssim$~0.05 \ across the line wings.
 The good quality of the fit for the CH$_3$CN~$J_K=$ 12$_K$-11$_K$ ($K$ = 0--6) lines is illustrated in Fig.~\ref{CH3CN_ratio}.

We used \textsc{XCLASS} to simultaneously fit the emission of the \ CH$_3$CN~$J_K=$ 12$_K$-11$_K$ (K= 0--6) lines (including the CH$_3$$^{13}$CN isotopologs). 
Figure~\ref{CH3CN} shows the maps of the \ CH$_3$CN \ column density and velocity toward the central plateau of the 1.37~mm continuum, where the three most intense cores reside. The \ CH$_3$CN \ column density reaches \ 10$^{16}$~cm$^{-2}$ \ toward core~1 and it is a factor of 2 lower toward core~2. Two \Vlsr\ gradients are clearly detected: the first across the adjacent cores~1~and~3, characterized by a change 
of \ $\approx$~4~\kms\ over \ $\approx$~1500~au; the second across core~2, with a change  of \ $\approx$~3~\kms\ over \ $\approx$~1000~au.
In the following, we determine the velocities of the cores~1~and~3 to assess the contribution of their relative motion to the observed velocity gradient.

\begin{figure*}
\centering
\includegraphics[width=0.85\textwidth]{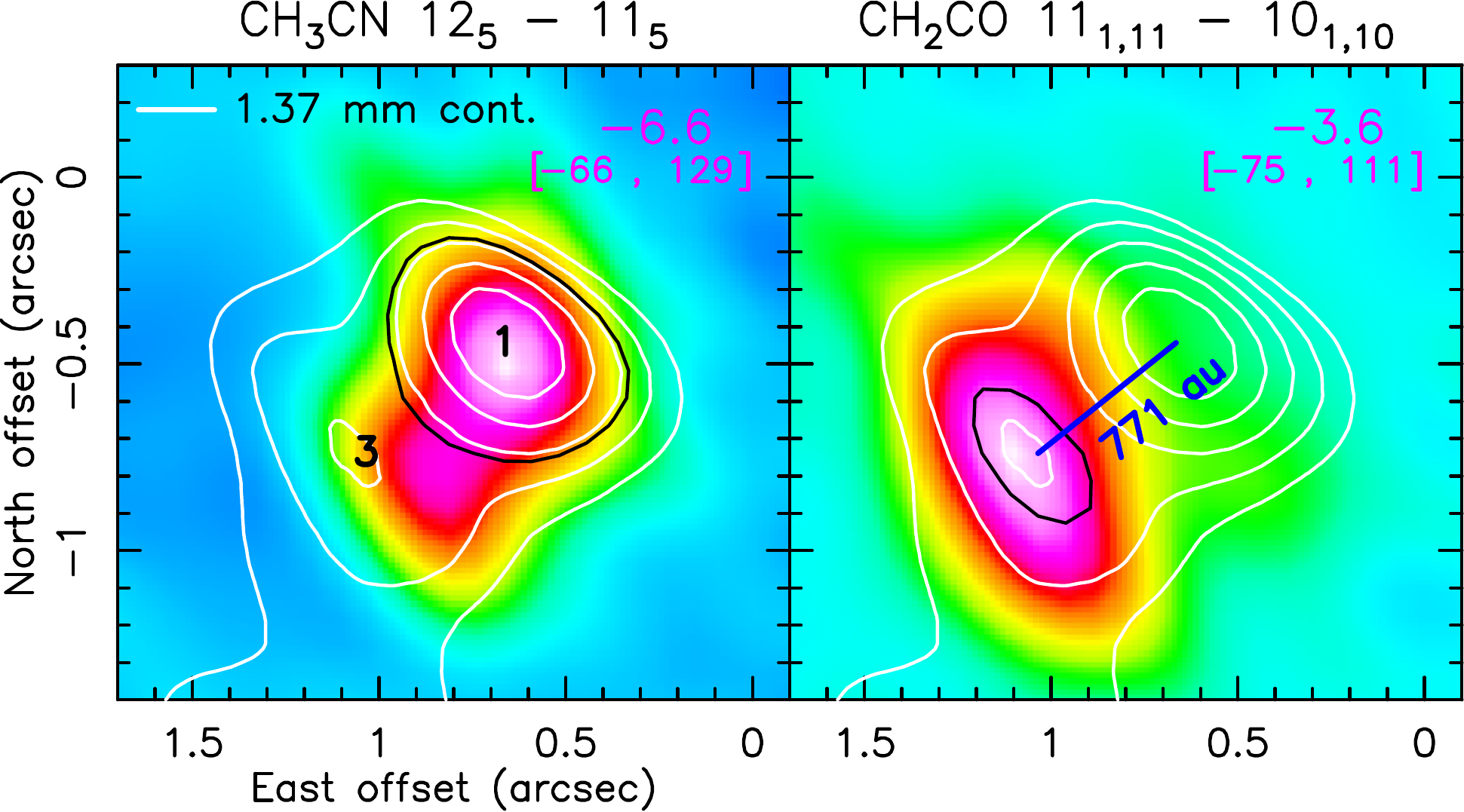}

\vspace*{0.5cm}
\includegraphics[width=0.8\textwidth]{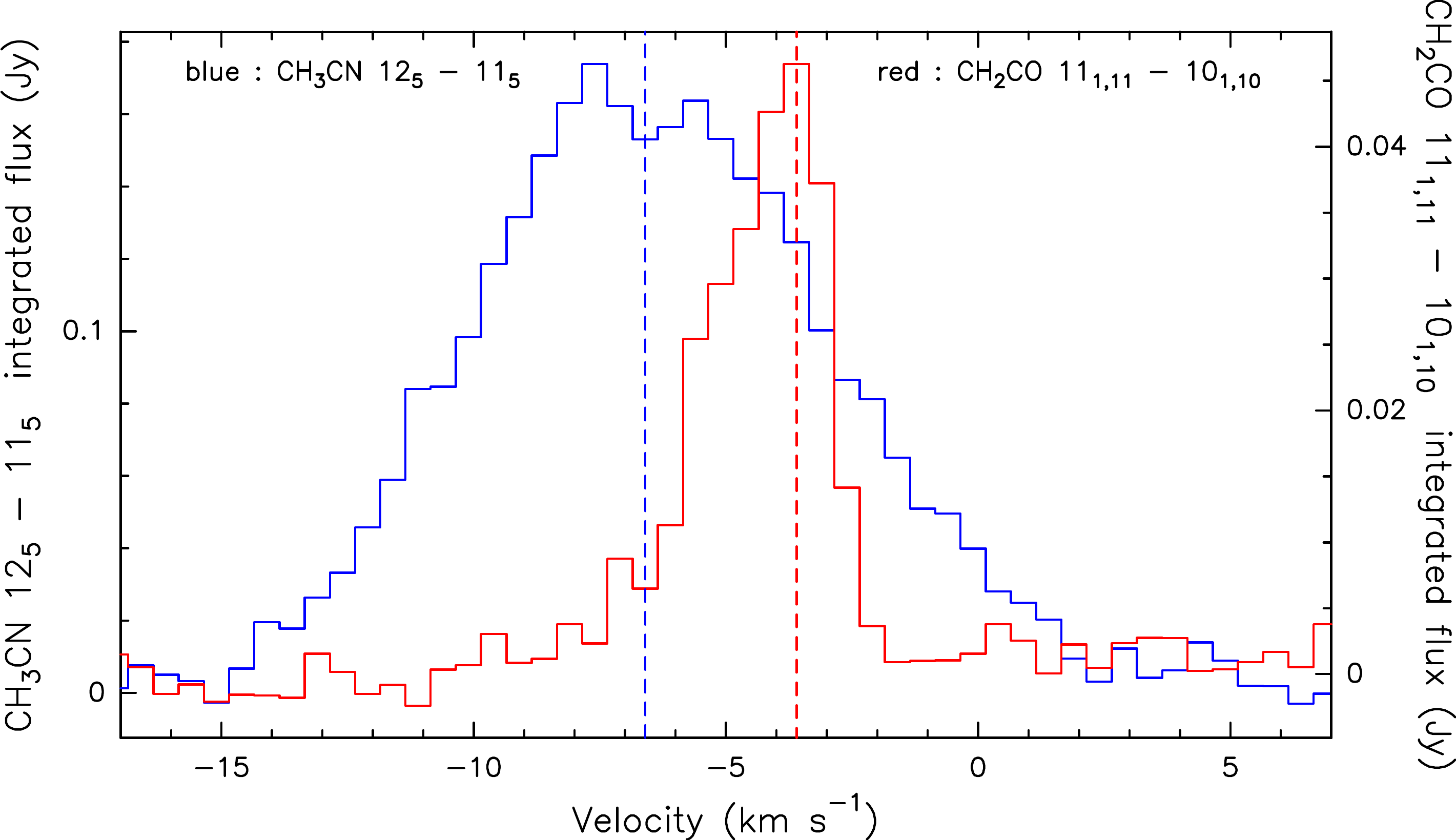}
\caption{NOEMA data. Upper~panels:~The color maps reproduce the emission of the CH$_3$CN~$J_K=$ 12$_5$-11$_5$ (left) and CH$_2$CO~$J_{K_a,K_c}=$ 11$_{1,11}$-10$_{1,10}$ (right) lines. In the upper right corner of the panel, the channel \Vlsr\ (in kilometer per second) and the range of plotted values (in  milliJansky per beam) are reported. The white contours reproduce the 1.37~mm continuum, with the same levels as in Fig.~\ref{CH3CN}: the positions of the two nearby 1.37~mm peaks are denoted using the same labels as in \citet[][see Table~5]{Beu18}, namely, 1 \ and \ 3 \ for the primary and secondary peak, respectively. In the right panel, the blue segment connects the two peaks. In the two panels, the black contours delimit the integration areas used to produce the spectra toward cores~1~and~3, presented in the lower panel.
\ Lower~panel:~Spectra of the \ CH$_3$CN~$J_K=$ 12$_5$-11$_5$ \ line toward core~1  (blue histogram) and 
the \ CH$_2$CO~$J_{K_a,K_c}=$ 11$_{1,11}$-10$_{1,10}$ \ line toward core~3 (red histogram). The flux scales for the CH$_3$CN and CH$_2$CO spectra are reported on the left and right, vertical axes, respectively. The blue and red dashed vertical lines indicate the approximate \Vlsr\ of the cores~1~and~3, respectively.}
\label{spec_bin}
\end{figure*}

While the emission of the \ CH$_3$CN~$J_K=$ 12$_K$-11$_K$ ($K$ = 2--6) lines (with \ $E_{\rm u}/k_{\rm B}$ \ varying in the range \ 97--326~K) is concentrated mainly toward core~1 (see Fig.~\ref{CH3CN}, upper~panel), we also identified a few molecular lines of slightly lower excitation,
such as the \  CH$_2$CO~$J_{K_a,K_c}=$ 11$_{1,11}$-10$_{1,10}$ \ transition ($E_{\rm u}/k_{\rm B} = 76$~K), which have comparable intensity toward cores~1~and~3 and can be employed to resolve the emission of the two cores in velocity. The upper~panels of Fig.~\ref{spec_bin} show the maps of the peak emission of the \ CH$_3$CN~$J_K=$ 12$_5$-11$_5$ \ line toward core~1 and the \ CH$_2$CO~$J_{K_a,K_c}=$ 11$_{1,11}$-10$_{1,10}$ \ line toward core~3, while the lower~panel of Fig.~\ref{spec_bin} reports the spectra of the two molecular lines integrated over the corresponding core. Among \ the CH$_3$CN transitions, we selected the  \ $K$ = 5 \ component  because it is intense and has a relatively high-excitation energy ($E_{\rm u}/k_{\rm B} = 247$~K), which allows us to better resolve the warm gas inside core~1. 
To determine the area of the two cores, we used the minimum level (18.6~m\Jyb, indicated by the black contours in Fig.~\ref{spec_bin}) of the 1.37~mm continuum in correspondence of which two disconnected emission islands are still obtained. 
 We stress that in this analysis we used the emissions of the  \ CH$_3$CN~$J_K=$ 12$_5$-11$_5$ \ and \   CH$_2$CO~$J_{K_a,K_c}=$ 11$_{1,11}$-10$_{1,10}$ \ lines only to derive the velocities of the cores~1~and~3, whose existence and geometrical properties are independently established from the 1.37~mm continuum emission. 
The two cores are separated by \ $\approx$~770~au \ and their LOS velocities differ by \ $\approx$~3~\kms. Therefore, it appears that the relative motion of the two cores can account for the observed \Vlsr\ gradient. This result is used in Sect.~\ref{TM_dis}, where we consider the possibility that the two cores are members of a binary system.

\begin{figure*}
\centering
\includegraphics[width=\textwidth]{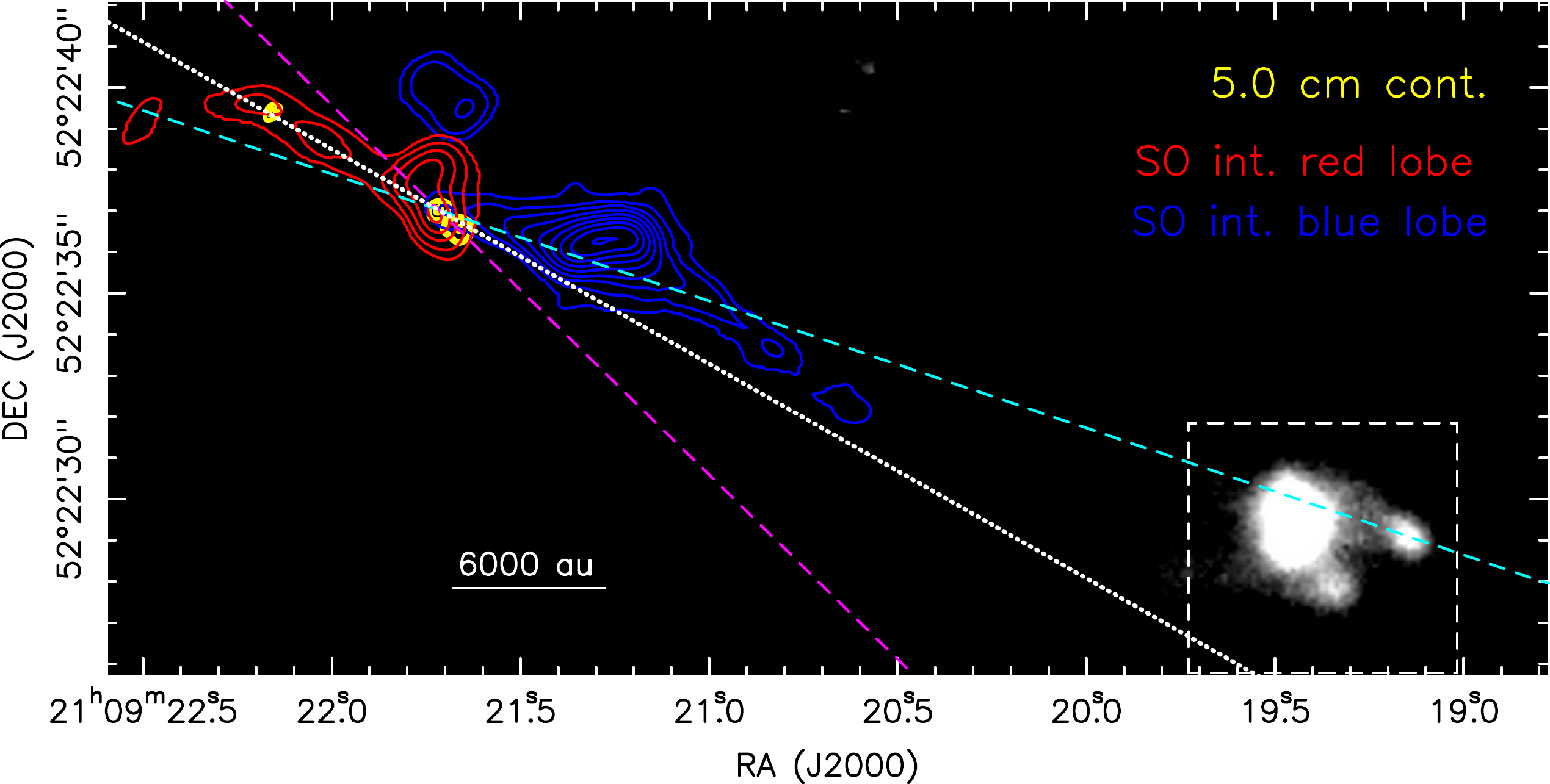}
\vspace*{0.5cm}
\includegraphics[width=0.7\textwidth]{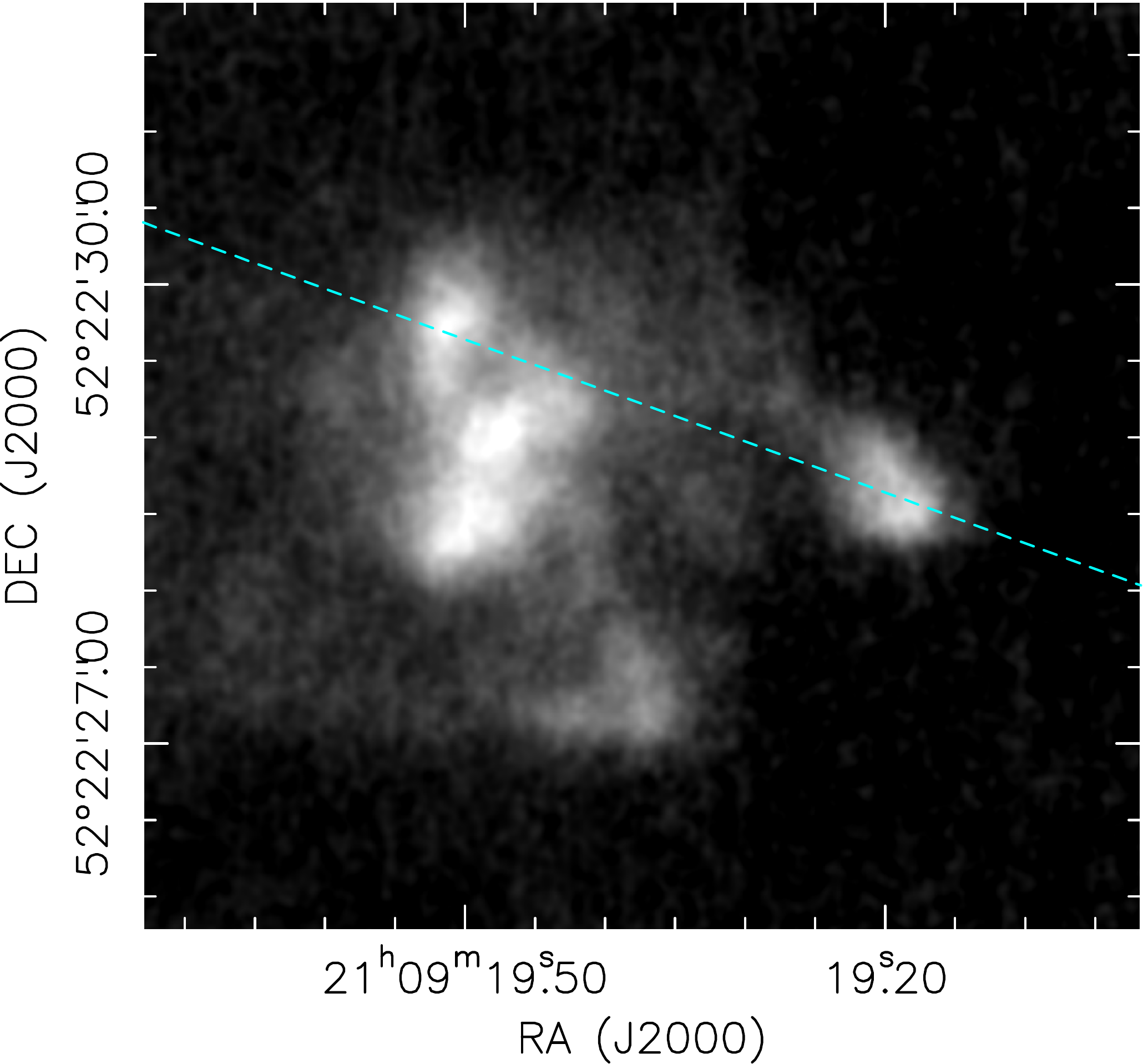}
\caption{Comparison of CORE, POETS and LBT observations. Upper~panel:~The gray-scale map reproduces the \ H$_2$~2.12~$\mu$m emission observed with LBT toward \targ. Referring to Fig.~\ref{CH3CN-SO}, lower~panel, the 
blue and red contours have the same meaning, and the yellow contours replace the 
magenta contours for the JVLA 5~cm continuum. The magenta dashed, white dotted, and  cyan dashed lines connect the main 5~cm peak with the secondary (nearby) 5~cm peak, the weak 5~cm spur to NE, and the bow shock traced by the LBT H$_2$~2.12~$\mu$m emission to SW, respectively. The white dashed box delimits the region plotted in the lower panel. \ Lower~panel:~Adaptive-optics assisted LBT observations of the H$_2$~2.12~$\mu$m emission (gray-scale map): the cyan dashed line has the same meaning as in the upper panel.}
\label{LBT-jet}
\end{figure*}

Our recent \ H$_2$~2.12~$\mu$m \ LBT observations toward \targ \ are shown in Fig.~\ref{LBT-jet} (upper~panel) \ together with the radio continuum and SO line emissions tracing the jet from YSO-1 and the molecular outflow from core~1, respectively. The H$_2$~2.12~$\mu$m emission arises from shock-excited, compressed, and hot molecular gas placed \ $\approx$~22\arcsec \ SW of core~1, pinpointed by the NE component of the radio jet at the center of the SO molecular outflow. Examining the lower panel of Fig.~\ref{LBT-jet}, which provides a higher-angular resolution view of the shock structure,  we note two well-shaped bow shocks at the SW tips of the H$_2$~2.12~$\mu$m emission. Since the major axes of these bow shocks approximately coincide with the directions to core~1, it is plausible that these shocks have been excited by an outflow emerging from core~1. Assuming that the outflow source resides in core~1, Fig.~\ref{LBT-jet} (upper~panel) shows that different outflow tracers, namely, the nearby SW lobe of synchrotron emission, the weak 5~cm spur \ $\approx$~7000~au to NE, and the \ H$_2$~2.12~$\mu$m bow shock \ $\approx$~40000~au to SW, are oriented at different PAs. In Sect.~\ref{dis_bin}, we discuss several scenarios to explain the spread in the directions of the various outflow tracers.


\begin{figure*}
\centering
\vspace*{-0.3cm}
\includegraphics[width=0.75\textwidth]{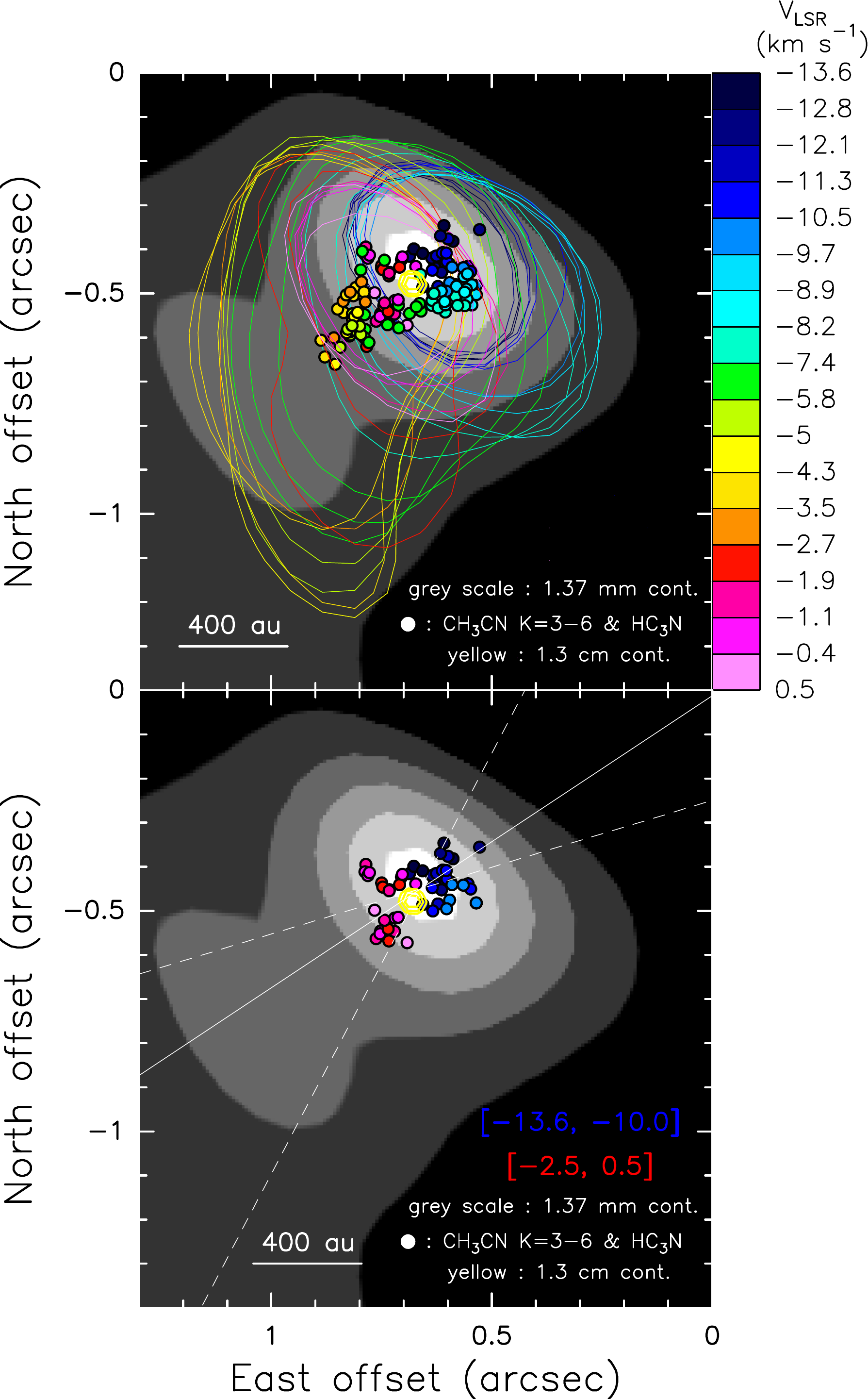}
\caption{NOEMA data. Upper~panel:~The gray-scale map reproduces the 1.37~mm continuum emission, plotting values in the range \ 10--35~m\Jyb. The colored dots give the (Gaussian-fitted) positions of the channel emission peaks for the CH$_3$CN~$J_K=$ 12$_K$-11$_K$ ($K$ = 3--6) and \ HC$_3$N~$J=$ 24-23 lines; the colors denote the channel \Vlsr. The colored contours indicate the half-peak level for the CH$_3$CN~$J_K=$ 12$_3$-11$_3$ emission in individual channels. The yellow contours show the JVLA A-Array continuum at 1.3~cm \citep{Mos16},
 showing levels at \ 70\%, 80\%, and 90\% of the peak emission of \ 0.50~m\Jyb. \ Lower~panel:~The gray-scale map, 
 yellow contours,  and colored dots have the same meaning as in the upper panel. Only the channel emission peaks at the extreme blue-~and~red-shifted velocities (reported in the lower right corner of the panel) are shown. The white continuous and dashed lines indicate the best-fit major axis of the distribution of the peaks and the corresponding fit uncertainty, respectively.}
\label{disk_vel}
\end{figure*}



\subsection{Most prominent core~1}
%



Toward \targ, the linear resolution achieved with the CORE data is \ $\approx$~625~au (for a FWHM beam size of \ $\approx$~0\farcs38). As seen in Sect.~\ref{obs_bin}, this is sufficient to resolve nearby molecular cores, on scales \ $\gtrsim$1000~au. Star formation models predict that the gravitational pull of high-mass YSOs influences the gas kinematics mainly at distances \ $\le$1000~au \citep[see, for instance,][]{Koe18}, where kinematic structures such as accretion disks and disk winds (DW) are expected. Toward core~1, the high 
signal-to-noise ratio of the \ CH$_3$CN~$J_K=$ 12$_K$-11$_K$ ($K$ = 2--6) \ and other high-excitation molecular lines (such as OCS~$J=$ 18-17, HC$_3$N~$J=$ 24-23,  and \ HNCO~$J_{K_a,K_c}=$ 10$_{1,10}$-9$_{1,9}$) permits us to study the gas kinematics on linear scales \ $\le$1000~au. For each of these strong molecular lines, we fit a 2D Gaussian profile to the emission in each velocity channel and determine the peak position as a function of velocity. If the emission is unresolved, the positional accuracy is equal to \ $ (\delta\theta \, / \,2) \times (\sigma \, / \, I) $ \citep[see, e.g.,][]{Rei88}, where \ $\delta\theta$ \ is the FWHM beam size and \ $I$ \ and \ $\sigma$ \ are the peak intensity and the rms noise, respectively, of a given channel. With an average value of the ratio \ $ (I \, / \, \sigma) \, = 27 $, the error  is \ $\lesssim$10~mas. For channels with resolved emission, the centroid positions convey more complex kinematic information because in these channels it is very likely that we are observing the combination of different types of motions and the shape of the emission can strongly deviate from a Gaussian. Consequently, for these channels the peak position obtained with the Gaussian fit is less reliable.

The upper panel of Figure~\ref{disk_vel} shows the distribution of channel centroids for the \ CH$_3$CN~$J_K=$ 12$_K$-11$_K$ ($K$ = 3--6) and \ HC$_3$N~$J=$ 24-23 lines, which are the most intense and compact molecular emissions in core~1. 
 In Fig.~\ref{disk_comp}, the distribution of channel centroids is plotted separately for the \ CH$_3$CN and \ HC$_3$N lines to show that the two molecules have a very similar \Vlsr \ pattern. This allows us to combine their emissions in our kinematical analysis.
The colored contours of Fig.~\ref{disk_vel}, reproducing the half-peak emission levels, clearly indicate that the structure is really compact only at the most extreme red-~and~blue-shifted velocities. At \ \Vlsr $\approx$~$-$4~\kms, the contamination from the nearby core~3 is evident. In the lower panel of Fig.~\ref{disk_vel} only the positions of the compact-emission channels at the extreme velocities are plotted. The distribution is bipolar and elongated along a direction, at PA = 123\degr $\pm$ 23\degr, approximately perpendicular to the radio jet. The compact 1.3~cm continuum emission, which best pinpoints the position of \  YSO-1, is located at the center of the distribution. In Sect.~\ref{dis_MY}, we propose that the derived velocity pattern traces the accretion disk around \ YSO-1.

\section{Discussion}
\label{discu}

\subsection{The cluster}
\label{dis_clu}


\subsubsection{Mass flow toward the core cluster}
\label{dis_flow}

The 1.37~mm continuum cluster is found at the density peak of a SE-NW elongated molecular cloud edged on the northeastern side with a necklace of less embedded YSOs and on the southwestern side with a filament of diffuse 4.6~$\mu$m emission (see Fig.~\ref{1.3cont}, lower~panel). These findings suggest that the molecular cloud is a density enhancement of a more extended, infrared dark filament. In Sect.~\ref{obs_clu} we showed that the velocities of the slow-moving  ($\le$~3~\kms) gas of the cloud and the cluster vary smoothly  along the major axis of the molecular cloud  over linear scales of \ 10$^4$--10$^{5}$~au (see Figs.~\ref{cloud_vel}~and~\ref{13COmap}). This regular change in \Vlsr\ with position makes us favor the interpretation in terms of a flow in the molecular gas (converging toward the density peak where the most massive cores reside) compared with cloud-cloud collision, which would rather appear as a sudden jump in velocity. 
In the assumption of a mass flow, the observed \Vlsr\ gradient can be easily explained if the blue-shifted NW side of the molecular cloud is farther away from us than the red-shifted SE side. The spatial and velocity extents of this \Vlsr\ gradient are comparable with those of the gradients, $\sim$1~\kms~per~0.1~pc, observed in infrared dark clouds (IRDC) at the sites of YSOs  \citep{Rag12}, which are often interpreted as infall. It is also consistent with simulations of large-scale accretion flows along filaments,  gravitationally accelerated toward local density peaks 
\citep[see, e.g.,][]{Tob12,Smi13}.

Assuming we are observing a mass infall toward the core cluster, we can use both the IRAM~30~m and merged NOEMA--IRAM~30~m observations of the \ $^{13}$CO~$J=$ 2-1 \ line to calculate the mass infall rate. The latter can be derived from the ratio of the momentum, $P_{\rm inf}$, to the length, $L_{\rm inf}$, of the flow. In Appendix~\ref{ap_mmo}, we describe the method to calculate the momentum of a molecular outflow, by employing the \ $^{13}$CO~$J=$ 2-1 \ emission.
The same equations hold to derive the momentum of a molecular infall as well. We employ the same parameters used for the \ $^{13}$CO~$J=$ 2-1 \ molecular outflow (see Table~\ref{out_mom})
with the following exceptions. First,~the velocities to be considered are those within a small interval about the systemic velocity, that is, [$-$7, $-$5.5]~\kms \ for the IRAM~30~m observations (see Fig.~\ref{cloud_vel}), and \ [$-$6.5, $-$3.5]~\kms \ for the merged NOEMA--IRAM~30~m data (see Fig.~\ref{13COmap}). Second,~the FWHM size of the map beam is \
$B_{\rm max} = B_{\rm min} =$~11\farcs8, for the IRAM~30~m observations. 
It is remarkable that, using either the \ IRAM~30~m \ or \ the merged NOEMA--IRAM~30~m \ observations, we obtain very consistent values for the infall momentum  \ $P_{\rm inf} \, \cos (i_{\rm inf}) \approx $~28~\ms~\kms, where \ $i_{\rm inf}$ \ is the inclination angle of the flow with respect to the LOS. 
Taking \ $L_{\rm inf} \, \sin (i_{\rm inf}) \approx$~6$\, \times \,$10$^4$~au, corresponding to the average value between the (sky-projected) flow lengths measured in the IRAM~30~m (8$\, \times \,$10$^4$~au, see Fig.~\ref{cloud_vel}) and NOEMA-IRAM~30~m (4$\, \times \,$10$^4$~au, see Fig.~\ref{13COmap}) maps, 
we finally derive a mass infall rate  \ $ \dot{M}_{\rm inf} \, \cot (i_{\rm inf}) = P_{\rm inf} \, \cos (i_{\rm inf}) \, / \, L_{\rm inf} \, \sin (i_{\rm inf})  \approx $~28~\ms~\kms / 6$\, \times \,$10$^4$~au $\approx$~10$^{-4}$~\ms~yr$^{-1}$. 
 Following the discussion in Appendix~\ref{ap_mmo}, owing to the uncertainties in the excitation temperature and abundance ratio of the \ $^{13}$CO~$J=$ 2-1 \ line, our estimate of the infall rate should be accurate to within a factor of 5.
The derived value of infall rate falls at the lower end of the range of  longitudinal flow rates measured in filaments associated with high-mass 
star-forming regions \citep{Chen19,Tre19}.

\subsubsection{Temperature and mass distribution over the cluster center}\label{TM_dis}

\begin{figure}
\centering
\includegraphics[width=0.5\textwidth]{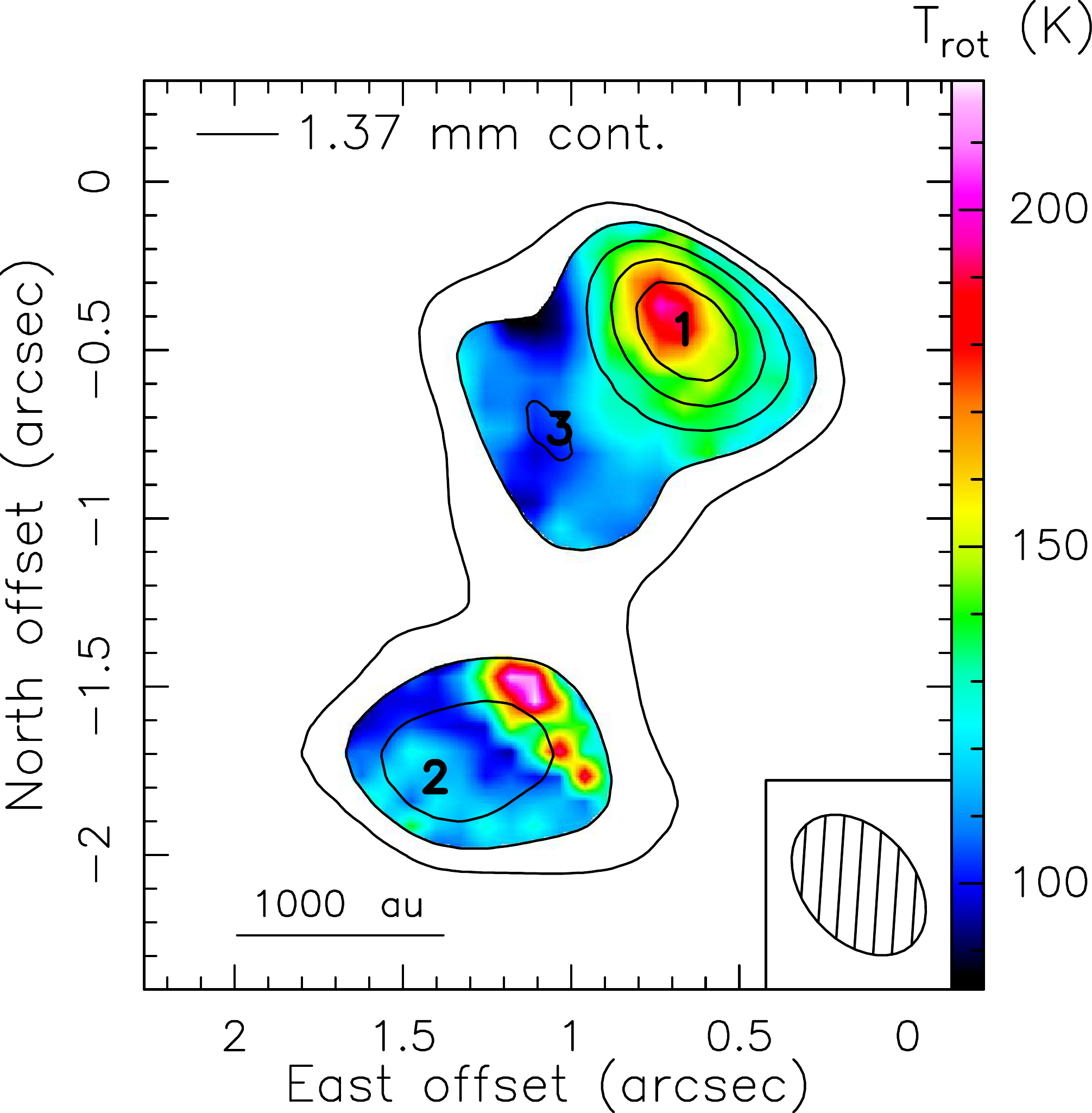}
\caption{NOEMA data. The color map shows the distribution of the rotational temperature derived by fitting with \textsc{XCLASS} the emission of the \ CH$_3$CN~$J_K=$ 12$_K$-11$_K$ (K= 0--6) lines (including the CH$_3$$^{13}$CN isotopologs). The plotted region corresponds to the area in which  the 1.37~mm continuum emission is higher than \ 16~m\Jyb. The black contours reproduce the 1.37~mm continuum, showing the same levels as in Fig.~\ref{CH3CN}: the positions of the three strongest 1.37~mm peaks are indicated using the same labels as in  \citet[][see Table~5]{Beu18}. The restoring beam of the \ CH$_3$CN \ maps is shown in the lower right corner of the panel.}
\label{T_rot}
\end{figure}

Figure~\ref{T_rot} shows the rotational temperature determined with \textsc{XCLASS} by simultaneously fitting the emission of the \ CH$_3$CN~$J_K=$ 12$_K$-11$_K$ (K= 0--6) lines (including the CH$_3$$^{13}$CN isotopologs).  While toward core~1 the gas temperature is always \ $\ge$115~K and may reach 200~K, toward core~3 it varies in the range \ 90--130~K. In correspondence of core~2 the temperature map is less homogeneous, probably owing to insufficient signal-to-noise ratio, and has an average temperature that is intermediate between that of core~1 and core~3.  By combining the maps of the 1.37~mm dust continuum and gas temperature, assuming that the dust emission is optically thin, we can make a reasonable estimate of the core masses. For consistency with \citet{Beu18}, we adopt a dust absorption coefficient of \ 0.9~cm$^2$~g$^{-1}$ \  \citep{Oss94} and a gas-to-dust mass ratio of 150 \citep{Dra11}. Over the areas of core~2 and the combined cores~1~and~3 (see Fig.~\ref{T_rot}), the mass is found to be  \ 0.46~and~0.92~\ms, respectively. Inside the individual cores~1~and~3, whose corresponding areas are delimited by the black contours in the upper panels of Fig.~\ref{spec_bin}, we obtain a gas mass of \ 0.42~\ms \ and \ 0.12~\ms, respectively. 
These values represent the mass inside the cores within radii \ $\le$500~au, and they are lower than the total core masses determined by \citet[][see Table~5]{Beu18} over radii $\ge$1000~au with a mean gas temperature of \ 66~K.

In Sect.~\ref{obs_bin}, using the \ CH$_3$CN~$J_K=$ 12$_K$-11$_K$ (K= 0--6) \ lines (see Fig.~\ref{CH3CN}), we traced the kinematics of the gas inside the three cores in the cluster center. Now, we use this result to derive the dynamical masses of the cores and estimate the masses of the embedded YSOs.
In Sect.~\ref{dis_YSO} we determine the mass 
of YSO-1 (the YSO inside core~1) to be \ $5.6~\pm~2.0$~\ms. In Sect.~\ref{obs_bin}, we showed that the \Vlsr\ gradient across the adjacent cores~1~and~3 can be well explained in terms of the relative motion of the cores. Assuming that we are observing rotation seen almost edge-on, in gravito-centrifugal equilibrium, from the spatial separation between the cores, $\approx$770~au, and their difference in velocity, $\approx$3~\kms\ (see Sect.~\ref{obs_bin}), we derive a total mass of \ $\approx$8~\ms. Subtracting both the mass of YSO-1 of  $\approx$5.6~\ms \ and that (0.92~\ms, see above) of the gas embedded inside cores~1~and~3, we obtain a rough estimate for the mass of the YSO inside core~3 of \  $\approx$~1--2~\ms.  Regarding core~2, interpreting the \Vlsr\ gradient 
($\approx$3~\kms \ across \ $\approx$1000~au, see Sect.~\ref{obs_bin}) observed toward this core as rotation, in gravito-centrifugal equilibrium, we derive a dynamical mass of \ $\approx$2.5~\ms. After subtracting the gas mass of \ 0.46~\ms \ (see above), the derived mass for the YSO inside core~2 is \ $\approx$2~\ms.   
Despite the large uncertainty in these determinations, the presence of relatively low-mass YSOs inside cores~2~and~3 agrees with the lower temperature (see Fig.~\ref{T_rot}) and mass of these cores with respect to core~1.

\subsubsection{Present state and future evolution}
\label{evo_dis}

Among the 20 cores identified in the 1.37~mm continuum emission of \targ\ by \citet[][see Table~5]{Beu18}, only cores~1,~2,~and~3, which are the most massive ($\ge$ 2~\ms), show signatures of an embedded protostar. As described in Sect.~\ref{obs_clu} (see also Fig.~\ref{CH3CN-SO}, upper~panel), the higher-excitation molecular lines are only detected toward these three cores, indicating that they are sufficiently warm to require local heating by a protostar.  Fitting the \ CH$_3$CN~$J_K=$ 12$_K$-11$_K$ \ lines, the gas temperature over the three cores is found everywhere \ $\gtrsim$100~K (see Sect.~\ref{TM_dis}), and, through the observed \Vlsr\  gradients (see Sect.~\ref{obs_bin}), the masses of the embedded YSOs are evaluated in the range \ 1--6~\ms \ (see Sect.~\ref{TM_dis}). The concentration of the most massive cores (and YSOs) toward the center of the 1.37~mm continuum cluster hints at mass segregation.

Judging the evolutionary state of the other, less massive cores is difficult with the present observations. Comparing with the most massive cores, the non-detection toward these cores of the higher-excitation molecular lines indicates an upper limit of  \ $\lesssim$~1~\ms \ for the mass of the embedded YSOs. The fact that no molecular outflows are detected from the less massive cores (see Sect.~\ref{obs_clu}) would be consistent with most of these cores still being in a prestellar phase. However, with our data, a molecular outflow is observed only from the most massive YSO, YSO-1 in core~1, and this hints at a sensitivity limit when using the  \ $^{13}$CO~$J=$ 2-1  \ and  \  SO~$J_N=$ 6$_5$-5$_4$ \ lines as outflow tracers. Future interferometric observations in the \ CO~$J=$ 2-1 \ line, more intense and with broader excitation conditions, could unveil less massive and energetic outflows. On a more speculative ground, the non-detection of outflows from the less massive cores could be also explained if the (putative) embedded low-mass YSOs were not actively accreting because most of the gas available across the cluster was gathered by YSO-1.

To discuss the future evolution of the \targ\ core cluster is useful to compare it with more evolved states as the well-studied, NIR stellar clusters around Ae or Be pre-main-sequence (PMS) intermediate-mass (1--6~\ms) stars \citep[see, for instance,][]{Tes99}. Ae and Be PMS stars are young enough, 0.5--5~Myr, that any population of lower-mass stars born in the same environment did not have enough time to move away from the birthplace. Therefore, the spatial distribution of the stars reflects that of the parental molecular cores. The typical size of these NIR stellar clusters 
is $\approx$0.2~pc, and the highest stellar densities, up to 10$^3$~pc$^{-3}$, are found in correspondence with the most massive ($\approx$6~\ms)  Ae or Be stars. In \targ, 20 cores are detected over an area of \ $\approx$~10$^{-2}$~pc$^{2}$, equivalent to a density of \ 2$\, \times \,$10$^4$~pc$^{-3}$, and the most massive YSO (YSO-1) has already attained a mass close to \ 6~\ms. In Sect.~\ref{dis_flow}, we calculated a mass infall rate of \ $\sim$10$^{-4}$~\ms~yr$^{-1}$ \ from the parental molecular cloud  \citep[total mass of \ 177~\ms,][see Table~1]{Beu18} to the core cluster, which, over the characteristic formation time of a few 10$^5$~yr \ of 
the  Ae and Be PMS stars \citep{Pal93}, would imply an infall toward the cluster of \ a few 10~\ms. If a relevant fraction of this mass is accreted by the most massive YSOs at the center of the 
cluster, it is very likely that at least one high-mass ($\ge$8~\ms) star will form. In conclusion, because of the relatively high stellar density and the likelihood of ultimately forming massive stars, we think that the \targ \ core cluster can be the progenitor of a stellar system whose most massive members are early B or late O-type high-mass stars, rather than Ae or Be intermediate-mass stars.

\begin{figure*}
\centering
\vspace*{-0.3cm}\includegraphics[width=0.65\textwidth]{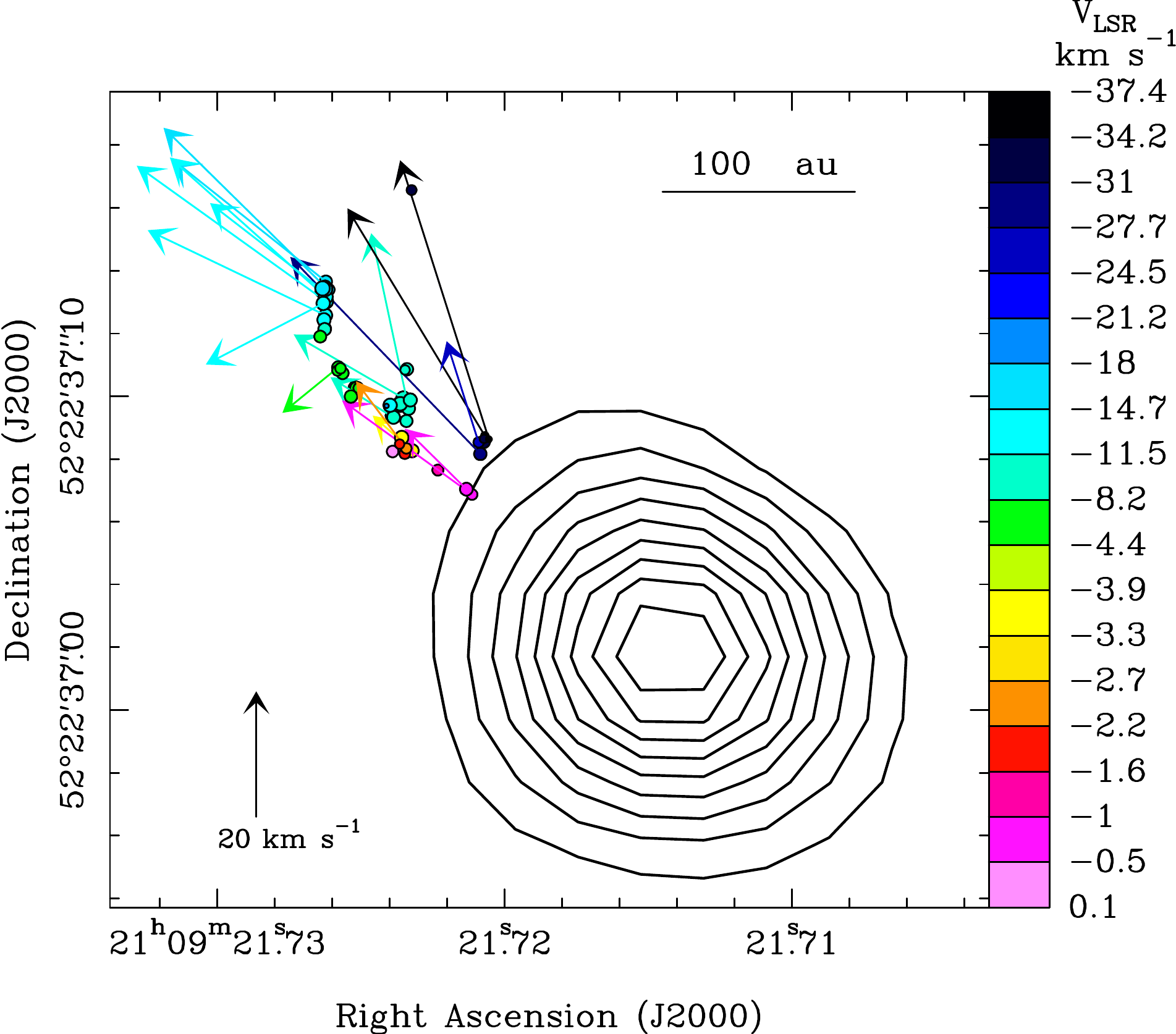}
\includegraphics[width=0.65\textwidth]{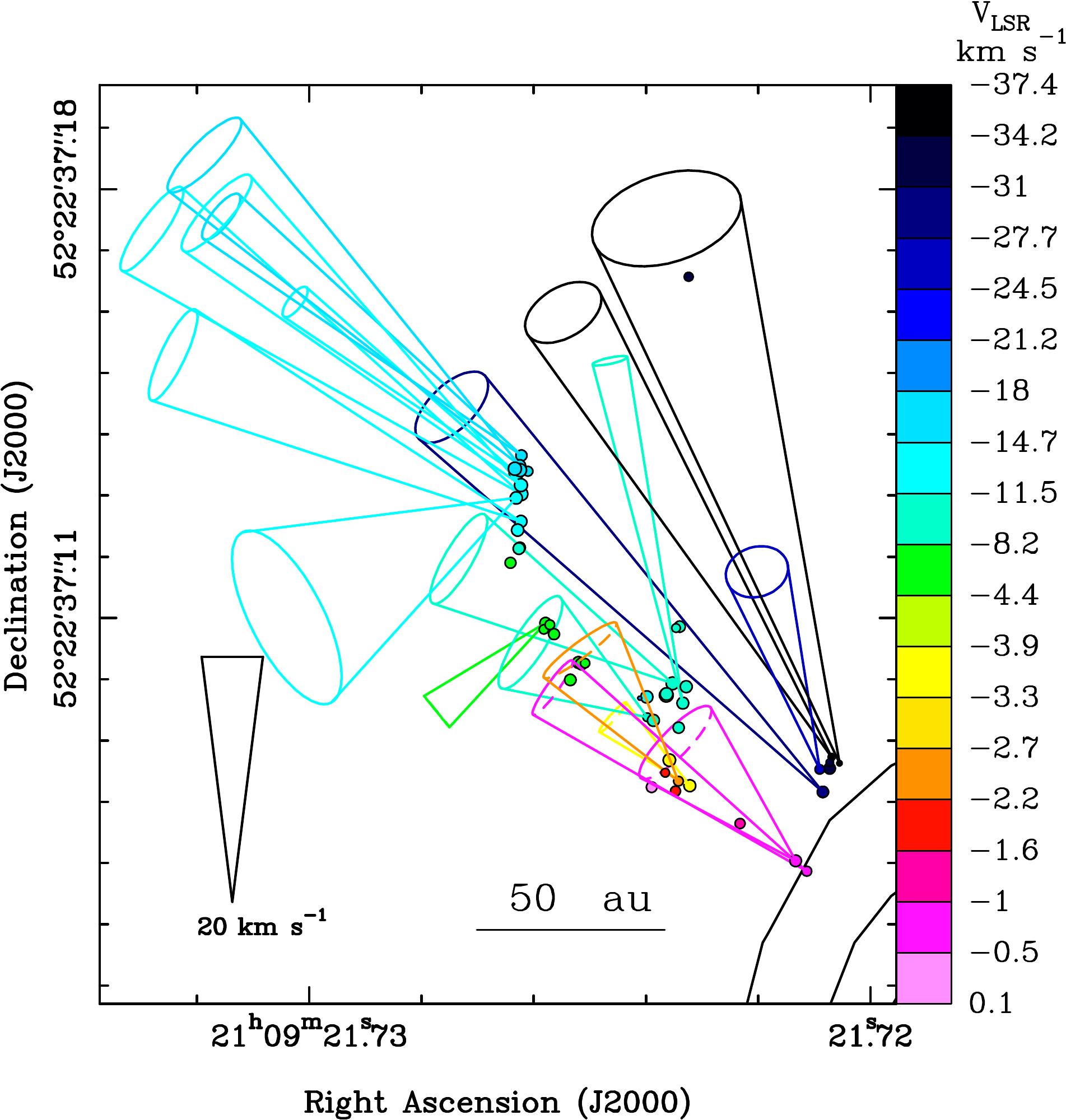}
\caption{POETS water maser observations. Upper~panel:~Colored dots and arrows give absolute positions and proper motions of the 22~GHz water masers determined with VLBI observations \citep{Mos16}; colors denote the maser \Vlsr. The dot area scales logarithmically with the maser intensity. The black contours indicate the JVLA A-Array continuum at 1.3~cm \citep{Mos16}, showing levels from 10\% \ to  90\% of the peak emission of \ 0.50~m\Jyb in steps of 10\%. \ Lower~panel:~Colored dots and black contours have the same meaning as in the upper~panel. Colored cones are employed to visualize the maser 3D velocities, representing the inclination with respect to the LOS through the ellipticity of the cone basis and the uncertainty in the direction by means of the cone aperture.}
\label{WM}
\end{figure*}

\subsection{Disk-jet system associated with YSO-1}
\label{dis_MY}

\subsubsection{Radio jet and molecular outflow}
\label{dis_jet}

Absolute positions and proper motions of the 22~GHz water masers observed in \targ\ with the POETS survey are presented in Fig.~\ref{WM}. The water masers show a SW-NE elongated distribution at radii between 100~au and~300~au from the peak of the compact 1.3~cm JVLA continuum, corresponding to the core of the radio jet and the YSO-1 position. The elongation, at PA $\approx$~38\degr, of the maser spatial distribution, and the average direction of the proper motions, PA $\approx$~49\degr\ \citep[][see Table~A1]{Mos19b}, are in good agreement with the orientation, PA $\approx$~43\degr, of the double-component radio jet (see Fig.~\ref{CH3CN-SO}). The faint, thermal, and nonthermal emissions from the NE and SW radio lobes, respectively, and the water masers are all manifestations of different types of shocks produced by the jet. In the jet core, close to YSO-1, we observe free-free emission, which likely originates from relatively weak (C-type), internal shocks of the jet corresponding to changes in the mass loss or ejection velocity \citep{Mos16,Ang18}; the nonthermal emission is likely synchrotron emission from electrons accelerated at relativistic velocities in strong jet shocks through first-order Fermi acceleration \citep{Bel78}; finally, the fast water masers emerge from relatively strong (J-type) shocks, whereas the jet impinges on very dense (molecular hydrogen number density, $n_{\rm H_2} \sim 10^8$~cm$^{-3}$) ambient material at high velocity \citep{Mos20}.

As reflected in Fig.~\ref{CH3CN-SO} (lower panel), the radio jet and the molecular outflow traced by the SO~$J_N=$ 6$_5$-5$_4$ line at larger scales are approximately parallel to each other and, as already anticipated in Sect.~\ref{obs_clu}, the molecular outflow could be powered by the jet. To assess the physical association between the jet and molecular outflow, in the following we determine and compare their physical properties. In Appendix~\ref{ap_mmo}, we describe the method to calculate the momentum of the molecular outflow, $P_{\rm out}$. We obtain \ $P_{\rm out} \, \cos (i_{\rm out}) \sim$50~\ms~\kms, where \ $i_{\rm out}$ \ is the inclination angle of the outflow with respect to the LOS. Employing the water maser positions and  
three-dimensional (3D) velocities, the momentum rate of the jet in \targ\ was estimated to be \ $\dot{P}_{\rm jet} \sim$ 1$\, \times \,$10$^{-3}$~\ms~\kms~yr$^{-1}$ \ by \citet[][see Sect.~6.2, Eq.~1, and Table~5]{Mos16}. In this calculation, a major source of uncertainty was the pre-shock, ambient density, which was assumed to be  \  $n_{\rm H_2} \sim$ 10$^8$~cm$^{-3}$. Now, thanks to the CORE data, we can better constrain this parameter. Averaging the derived mass distribution (see Sect.~\ref{TM_dis}) over a cylindric volume equal to the beam area times the beam size, in the direction of the continuum peak we find the value  \  $n_{\rm H_2} \approx$~3$\, \times \,$10$^8$~cm$^{-3}$, which should be accurate to within a factor of a few. Accordingly, our new, more precise, estimate of the momentum rate of the jet is \ $\dot{P}_{\rm jet} \sim$ 3$\, \times \,$10$^{-3}$~\ms~\kms~yr$^{-1}$.

We wish to derive the momentum rate of the molecular outflow, to be compared with that of the jet. Knowing the value of \  $P_{\rm out} \, \cos (i_{\rm out}) $ (see above), we need to estimate the inclination angle and timescale of the outflow. The clear spatial separation of the blue-~and~red-shifted lobes (see Fig.~\ref{CH3CN-SO}, lower~panel) indicates that the gas flows along directions significantly inclined with respect to both the LOS and the plane of the sky. Based on the opening angle of $\approx$~40\degr \ of the blue-shifted outflow lobe (see Fig.~\ref{CH3CN-SO}, lower~panel), we judge that the outflow axis has to be at least \ 20\degr \ away from both the LOS and the plane of the sky, that is \ 20\degr $ \le i_{\rm out} \le $ 70\degr. 
The dynamical timescale of the outflow, $t_{\rm out}$, can be estimated from the ratio of the size of the lobes to the corresponding maximum \Vlsr\ range. Referring to the SO~$J_N=$ 6$_5$-5$_4$  emission in Fig.~\ref{CH3CN-SO} (lower~panel), we have: $t_{\rm out} \; \tan (i_{\rm out}) \approx$~22\arcsec / 27~\kms\ = 6.3$\, \times \,$10$^3$~yr, at a distance of \ 1.63~kpc. We caution that previous studies have pointed out that the dynamical timescale of molecular outflows can largely underestimate (by a factor between 5 and 10) the true outflow age \citep[see, for instance,][]{Par91}. Accordingly, the upper limit for the momentum rate of the outflow is found to be \ $\dot{P}_{\rm out}$ $\cos^2 (i_{\rm out})$ / $\sin (i_{\rm out})$ $\lesssim$  8$\, \times \,$10$^{-3}$~\ms~\kms~yr$^{-1}$.
Considering that this value is only accurate to within an order of magnitude and that \ $0.4 \le \sin (i_{\rm out})$ / $\cos^2 (i_{\rm out}) \le 8$ \ for \ 20\degr $ \le i_{\rm out} \le $ 70\degr, we conclude that the momentum rate of the molecular outflow is consistent with that of the jet. It is then plausible that the jet powers the molecular outflow.


\begin{figure} 
\includegraphics[width=0.45\textwidth]{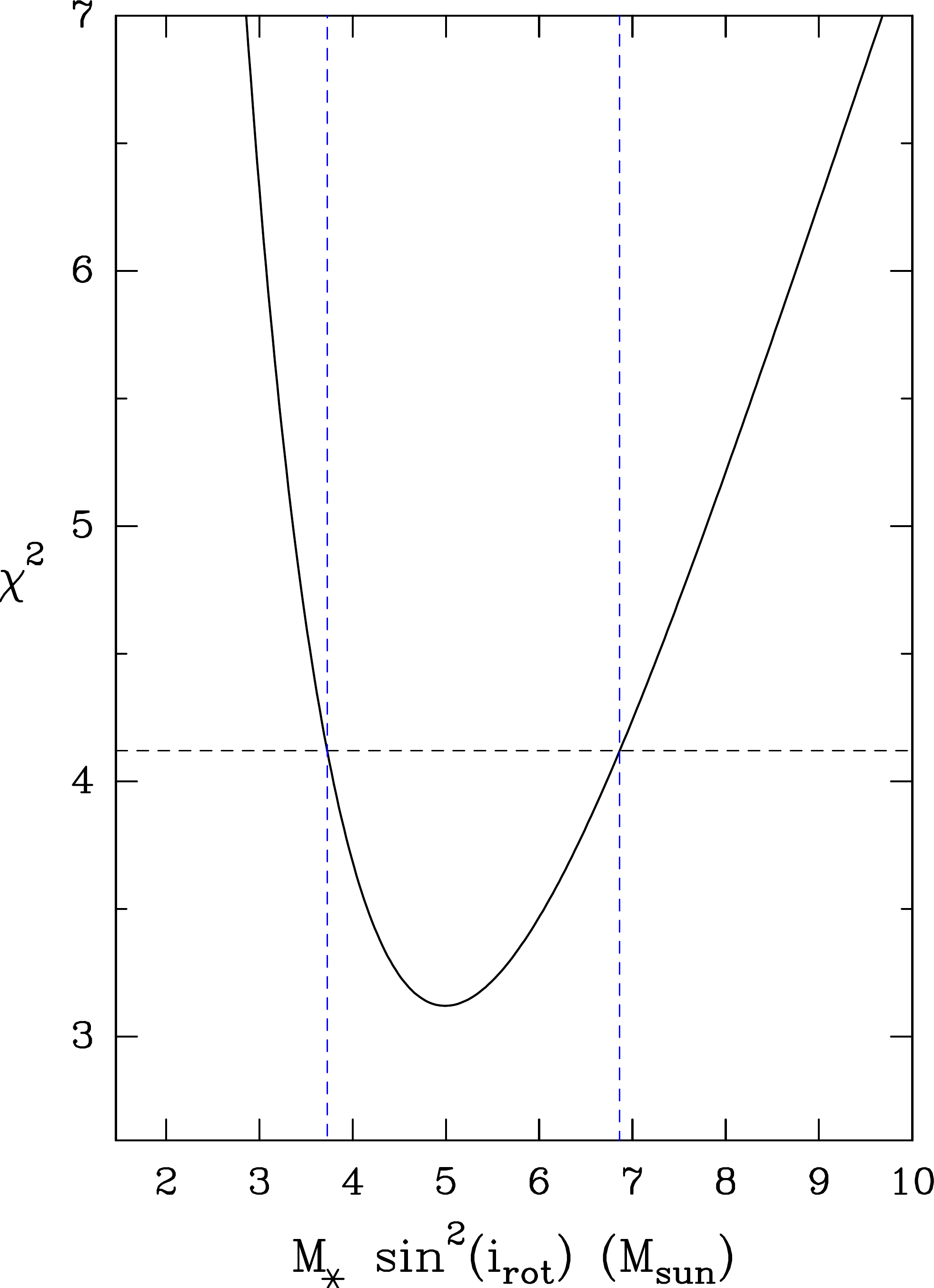}
\caption{Plot of the $\chi^2$ values from the Keplerian fit (see Eq.~\ref{chi2}) vs. the free parameter \ $M_{\star} \sin^2(i_{\rm rot})$. 
The horizontal black dashed line indicates the value \ $\chi^2_{\rm min}$+1.0 = 4.12, which corresponds to the 1$\sigma$ confidence level for the $\chi^2$-distribution with one free parameter \citep{Lam76}. The two vertical blue dashed lines indicate the 1$\sigma$ confidence interval for the best-fit value of the free parameter.}
\label{Chi2-plot}
\end{figure}


\subsubsection{Mass and luminosity of YSO-1}
\label{dis_YSO}

The velocity pattern in Fig.~\ref{disk_vel} (lower~panel) is consistent with that expected for a rotating disk around a YSO. It is elongated approximately perpendicular to the radio jet, the blue-~and~red-shifted velocities lie on the two sides of the YSO, and its size of \ $\approx$~400~au agrees with the model predictions for accretion disks around high-mass YSOs \citep[see, for instance,][]{Koe18}. The \Vlsr\ does not present a regular change with the position, but that likely results from the limited precision in tracing the velocity pattern using channel emission centroids and the insufficient angular resolution. 

Making the assumption (to be checked a posteriori, see below) of Keplerian rotation, we determine the central mass, $M_{\star}$, by minimizing the following \ $\chi^2$ expression:
\begin{equation}
\label{chi2}
\chi^2 = \sum_j  \frac{ [ V_j - ( V_{\star} \pm 0.74 \; (M_{\star} \sin^2(i_{\rm rot}) )^{0.5} \; \;  | S_j-S_{\star} |^{-0.5} \, ) ]^2 } {( \Delta V_{j} )^2} 
,\end{equation}

\noindent where \ $V_j$ \ and \ $S_j$ \ are the channel \Vlsr\ (in kilometer per second) and corresponding peak positions (in arcsecond), the index $j$ runs over all the fitted channels, and the $+$ and $-$ symbols hold for red-~and blue-shifted velocities, respectively. 
 $V_{\star}$ (in kilometer per second) \ and \ $S_{\star}$ (in arcsecond) are the \Vlsr\ and position of YSO-1, respectively.
 The central mass $M_{\star}$ \ is given in solar masses. 
 Indicating with \ $i_{\rm rot}$ the inclination of the disk rotation axis with respect to the LOS, the factor \ $ \sin^2(i_{\rm rot}) $ \ takes into account that the disk plane is seen at an angle \ $90-i_{\rm rot}$ \ from the LOS and the observed \Vlsr\ corresponds to the rotation velocity multiplied by the factor \ $\sin (i_{\rm rot})$. 
To take into account the uncertainty on the velocity and
that on the position, the global velocity error \ $ \Delta V_{j} $ \ is obtained by summing in quadrature two
errors: that on the velocity (taken equal to half of the channel width) and that obtained by converting the error on the offset into a velocity error through the function fitted to the data.

The \Vlsr, $V_{\star} = -6.4$~\kms,  and position $S_{\star} =$ 0\farcs81 of YSO-1 are assumed equal to the average of the \Vlsr\ and positions of all the channels. Thus, the term \ $M_{\star} \sin^2(i_{\rm rot})$ \ is the only free parameter in Eq.~\ref{chi2}. We allowed this term to vary over the range \ 2--14~\ms, which is consistent with the upper limit of \ $\approx$~12~\ms \ imposed by the bolometric luminosity of \ $\approx$10$^4$~\ls \  \citep{Dav11}. Figure~\ref{Chi2-plot} reports the plot of the $\chi^2$ \ vs. the free parameter; the blue dashed lines delimit the 1$\sigma$ confidence interval following \citet{Lam76}. This plot shows that we do find an absolute minimum of the $\chi^2$ and the determined 1$\sigma$ range for the values of the free parameter is \ $M_{\star} \, \sin^2(i_{\rm rot})~=~5.3~\pm~1.6$~\ms.

The fitted value of \ $M_{\star}$ \ is much larger than the gas mass, 0.42~\ms, inside core~1, derived from the dust emission in Sect.~\ref{TM_dis}. 
 As noted in Sect.~\ref{TM_dis}, this is the mass inside core~1 within a radius \ $\le$~500~au, which is comparable to the size of the disk structure traced by the \ CH$_3$CN \ and \ HC$_3$N \ lines around YSO-1 (see Fig.~\ref{disk_vel}, lower~panel). The YSO-1 envelope should extend at significantly larger radii and be much more massive.
The finding that the central mass is much greater than the mass of the disk supports our assumption of Keplerian rotation. The disk, traced by the bipolar \Vlsr\ pattern of molecular emissions (see Fig.~\ref{disk_vel}, lower~panel), and the radio-maser jet are approximately perpendicular to each other in the plane of the sky. This suggests that the jet is directed close to the rotation axis of the disk, or \ $i_{\rm jet} \approx i_{\rm rot}$ (where \ $i_{\rm jet}$ \ is the inclination of the jet with respect to the LOS). From the angular distribution of the maser proper motions, reported in Fig.~\ref{WM}, the semi-opening angle of the jet is evaluated to be \ $\approx$~18\degr\ by \citet[][see Table~5]{Mos16}. Figure~\ref{WM}, lower~panel, also shows that the water masers are both blue-~and~red-shifted, which indicates that the jet intersects the plane of the sky and the relation \ $| i_{\rm jet} - 90\degr | \lesssim$ 18\degr \ must hold. Following these considerations, we can write  \ $ \sin^2(i_{\rm rot}) \approx \sin^2(i_{\rm jet}) \gtrsim \sin^2(72\degr) = $ 0.9. 
Allowing for the variation of the factor \ $ \sin^2(i_{\rm rot})  $ \ in the range \ 0.9--1, the 1$\sigma$ interval
 for the fitted YSO-1 mass is \ $M_{\star}~=~5.6~\pm~2.0$~\ms.

By employing the relationship between bolometric luminosity, $L_{\rm bol}$, and outflow momentum rate, $\dot{P}$, for massive YSOs  by \citet{Mau15} (Log$_{10}\, [\dot{P} \, / \, \textrm{M$_{\odot}$~km~s$^{-1}$~yr$^{-1}$}] = -4.8 + 0.61 $ Log$_{10} \, [L_{\rm bol} \, / \, L_{\odot}]$), and our best estimate of the momentum rate \ $\dot{P}_{\rm jet} \sim$ 3$\, \times \,$10$^{-3}$~\ms~\kms~yr$^{-1}$ (see Sect.~\ref{dis_jet}), we infer a luminosity of \ 5$\, \times \,$10$^{3}$~\ls \ for YSO-1. This value is consistent with the luminosity  of \ 1.3$\, \times \,$10$^4$~\ls \ of the \targ\ region, which was derived by \citet{Mol96} using low-angular resolution IRAS data \citep[$\sim$1\arcmin,][]{Neu84} only\footnote{Neither Herschel infrared Galactic Plane Survey \citep[Hi-GAL;][]{Mol10b, Mol10a} nor Red MSX Source \citep[RMS;][]{Hoa05,Lum13} observations are available for \targ.}. Indeed, combining the far-infrared IRAS fluxes with the mid-infrared, Wide-Field Infrared Survey Explorer \citep[WISE;][]{Wri10} and Midcourse Space Experiment \citep[MSX;][]{Ega03} data at higher angular resolution ($\sim$10\arcsec), \citet[][see Table~1]{Mos16} determine a lower luminosity of \ 5$\, \times \,$10$^{3}$~\ls, which agrees well with that inferred for YSO-1. 

\subsection{Magnetized and rotating jet}
\label{dis_mag}
\begin{figure} 
\includegraphics[width=0.45\textwidth]{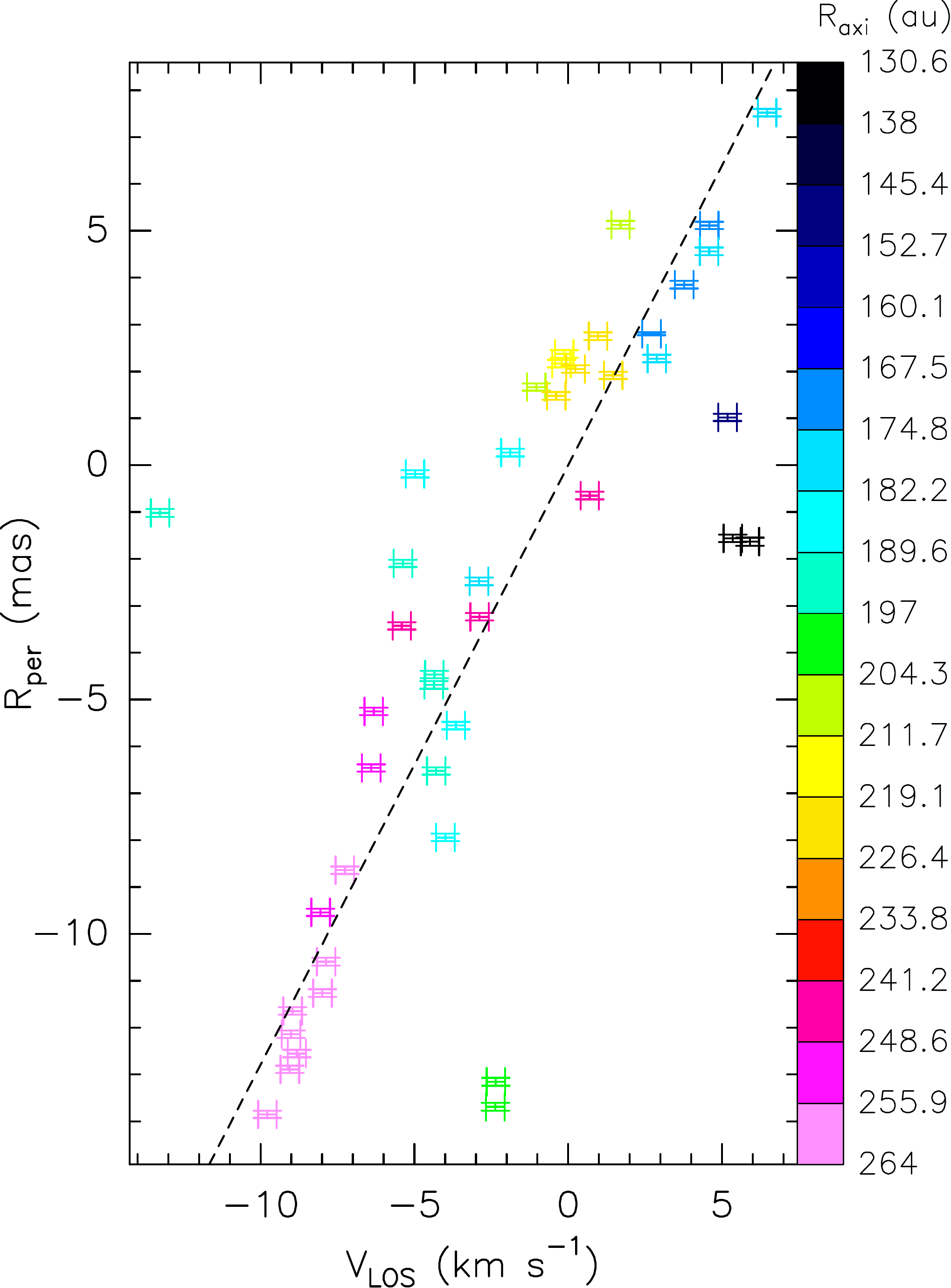}
\caption{Plot of the maser positions projected perpendicular to the jet axis, $R_{\rm per}$, vs. corresponding LOS velocities, $V_{\rm LOS}$, for the masers with \ $V_{\rm LOS} \ge$~$-$20~\kms. Colored error bars are used to indicate values and associated errors; the colors represent the distance from YSO-1 projected along the jet axis, $R_{\rm axi}$, as coded in the wedge on the right-hand side of the panel. The 
black dashed line shows the linear fit to the plotted values.}
\label{rp-vL}
\end{figure}

The radio-maser jet emerging from YSO-1 is unique in combining two properties that, to our knowledge, for the first time are observed associated with the same object: \ a lobe with nonthermal emission and jet rotation. We noted the first point several times through this article and this refers to the SW lobe of the double-component radio jet (see Fig.~\ref{CH3CN-SO}, lower~panel), whose spectral index over the frequency range \ 6--22~GHz is \ $\le$$-$0.7  \citep[][see Table~4 and Fig.~19]{Mos16}. Observations of nonthermal components in jets from high-mass YSOs are still rare, although good examples were recently discovered thanks to high-angular resolution, sensitive radio survey \citep[for instance, the POETS source G035.02$+$0.35,][]{San19b}. The observational evidence for the second point, as already discussed in \citet[][see Sect.~3.2]{Mos19b}, is the \Vlsr\ gradient of the water masers transversal to the jet direction. Figure~\ref{WM} (lower~panel) shows the monotonic change of the maser \Vlsr\ along a SE-NW direction perpendicular to the SW-NE collimation axis of the proper motions, with red-~and~blue-shifted \Vlsr\ toward SE and NW, respectively. The linear correlation between the maser LOS velocities, $V_{\rm LOS} \, = \, V_{\rm LSR} - V_{\star}$, and positions projected perpendicular to the jet direction, $R_{\rm per}$, is clearly illustrated in Fig.~\ref{rp-vL}.
A straightforward interpretation for the maser 3D velocity pattern is in terms of a composition of two motions: a flow along, and a rotation around, the jet axis. 


 \begin{figure} 
\includegraphics[width=0.45\textwidth]{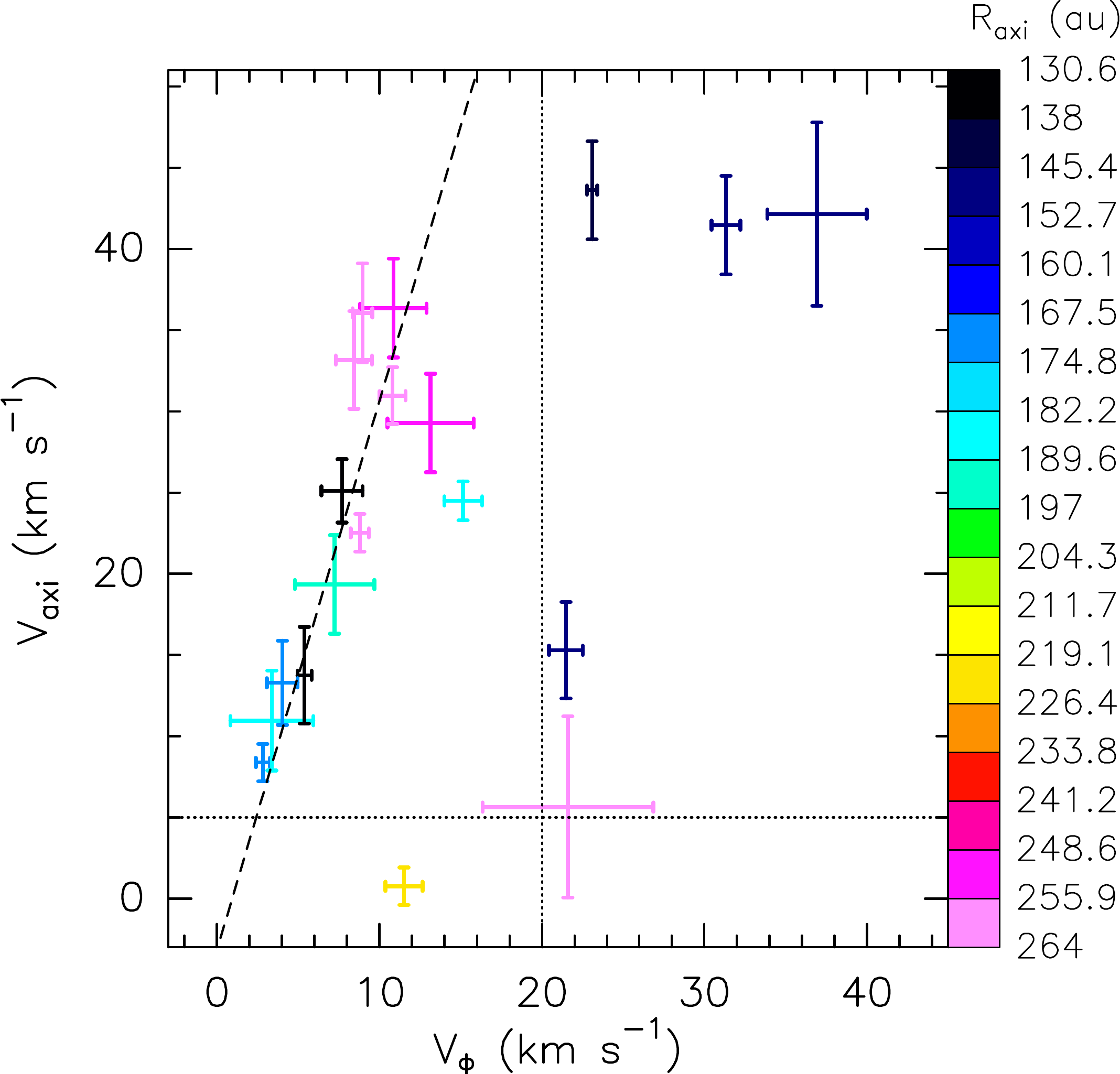}
\caption{Plot of the components \ $V_{\rm axi}$ \ vs.  \ $V_{\Phi}$ (see definitions in Sect.~\ref{dis_mag})  for the 3D maser velocities shown in Fig.~\ref{WM}. 
Colored error bars are used to indicate values and associated errors; the colors represent the distance from YSO-1 projected along the jet axis, $R_{\rm axi}$, as coded in the wedge on the right-hand side of the panel. The black dashed line shows the linear fit to the subset of points within the region delimited by the horizontal and vertical black dotted lines, corresponding to \ $V_{\Phi}$ $\le$ 20~\kms \ and \ $V_{\rm axi}$ $\ge$ 5~\kms.}
\label{va-vf}
\end{figure}


The findings of nonthermal emission from a jet lobe and jet rotation are strong indications that magnetic fields play\ an important role in both launching and accelerating the jet. The interpretation of the nonthermal continuum in terms of synchrotron emission requires the presence of a magnetic field sufficiently ordered and strong to trap electrons in jet shocks, confine them even at relativistic velocities, and warrant the detection of their synchrotron signal. Jet rotation is considered a critical test for magneto-centrifugal (MC) DW \citep[][]{Bla82,Pud07}, where the magneto-centrifugally launched jet extracts the excess angular momentum from the disk gas, which, then, can be accreted by the protostar. Comparing Figs.~\ref{disk_vel}~and~\ref{WM}, we evince that the jet and the disk of YSO-1 rotate about axes approximately parallel to each other and have the same sense of rotation, that is, toward and away from us to NW and SE, respectively. This result supports our idea that the jet rotation stems from the disk through the magnetic leverage of a MC~DW.

The measurement of the 3D velocities of the jet near the YSO through water maser VLBI observations permits a more quantitative comparison with the predictions of the MC~DW theory. According to the MC~DW models \citep[][]{Pud07}, the YSO wind is centrifugally launched along the magnetic field lines threading the accretion disk, it is accelerated to the Alfv\'{e}n speed sliding along the rotating field lines, and then is magnetically collimated. A large body of different MC~DW simulations have shown that, depending on the assumed  magnetic field configuration, the final jet structure can be either collimated toward a cylinder or at wide angle \citep[][]{Pud07}. In the following, we compare our water maser observations with wind models that recollimate within a cylindrical flow tube (see, for instance, the simulations in \citet{Sta15} and \citet{Koe18}). 
 We also note that a model of a cylindrical rotating jet has been recently proposed by \citet{Burn15} to interpret the 3D motion of the water masers observed with VERA \citep[VLBI exploration of radio astrometry;][]{Kob03} toward the massive YSO S235AB-MIR.

Let us indicate with \ $r_{0}$ \ the radius of the footpoint of the magnetic field line anchored to the disk and with \ $r_{\rm A}$ \ the distance from the rotation axis at which the wind velocity equals the  Alfv\'{e}n speed. At radii \ $r_{\rm c}$ $\ge$ $r_{\rm A}$, the wind recollimates, and its velocity has two (dominant) components: the main component oriented along the jet axis, $\Upsilon_{\rm axi}$, and the azimuthal component, $\Upsilon_{\Phi}$, representing the rotation around the axis. Combining Eqs.~12~and~13 of \citet{Pud07}, the ratio of these two components can be related to the radius \ $r_{\rm c}$ \ of the cylindrical flow through the expression
\begin{equation}
 \Upsilon_{\rm axi}$ / $\Upsilon_{\Phi} \approx  \sqrt{2} \; r_{\rm c} / r_{\rm A} 
 \label{eq_v_r}
.\end{equation}

The water masers originate in shocks that arise where the jet impinges on high-density material and propagate away along the jet direction at a speed, assuming momentum conservation, \ $V$~=~$\sqrt{\rho_{\rm jet}/\rho_{\rm amb}}$~$\Upsilon$ \ \citep[see, for instance,][]{Mas93}. In this equation, \ $\rho_{\rm jet}$ \ and \ $\Upsilon$ \ are the jet density and speed, respectively, and \ $\rho_{\rm amb}$ \ is the ambient density. Since the water masers move parallel to the wind, their motion can be described in terms of the composition of the same velocity components, that is, that directed along the jet axis and the azimuthal component. Let us indicate with \ $V_{\rm axi}$ \ and \ $V_{\rm per}$ \ the maser velocity components in the plane of the sky parallel and perpendicular to  the sky-projected jet axis (at PA = 43\degr). Since the jet is directed close ($\le$18\degr, see sect.~\ref{dis_YSO}) to the plane of the sky,   $V_{\rm axi}$ \ approximates the component along the jet axis with a maximum error of \ $\cos^{-1}(18\degr) \approx$~5\%. The  azimuthal component, $V_{\Phi}$, is equally well approximated with \ $V_{\Phi} \approx \sqrt{V^2_{\rm per}+V^2_{\rm LOS}}$. From the considerations above, it is clear that the ratio of the maser velocity components \ $V_{\rm axi}$ \ and \ $V_{\Phi}$ \ is an accurate measurement of the corresponding ratio of the wind velocity components, that is, \  $\Upsilon_{\rm axi} \, / \, \Upsilon_{\Phi}  \approx  V_{\rm axi} \, / \, V_{\Phi} $. 

Figure~\ref{va-vf} shows that the ratio \ $V_{\rm axi} \, / \, V_{\Phi} $ \ is approximately constant for the vast majority of the water masers (with measured proper motions). We obtain \ $ V_{\rm axi} \, / \, V_{\Phi} \, \approx$~3.4$\pm$1. The only notable exception is a small maser cluster at the most blue-shifted velocities ($V_{\rm LOS} \le$~$-$20~\kms) closer to YSO-1 (axis-projected separation \ $R_{\rm axi} \le$~150~au), for which \ 1~$\lesssim V_{\rm axi} \, / \, V_{\Phi} \lesssim$~2. On the basis of Eq.~\ref{eq_v_r}, a simple explanation for the small spread in the ratio of the maser velocity components is that most of the masers originate within a thin cylindrical shell, tracing shocked wind material MC-launched within a thin annulus of radius \ $r_0$ \ of the rotating disk.  This interpretation fits with the models predicting a shock origin for the water masers, since it is plausible that the most external layer of the jet mainly interacts with the surrounding ambient material and the water masers are mainly found on the jet wall. Thus, a reasonable estimate for the radius of the jet, $r_{\rm jet}$, can be obtained from half the spread in \ $R_{\rm per}$, the maser positions projected perpendicular to the jet direction, considering only the water masers with \ $V_{\rm LOS} \ge$~$-$20~\kms\ (see Fig.~\ref{rp-vL}). We derive \ $r_{\rm jet} \approx$~10~mas, or \ $\approx$~16~au.

The terminal poloidal velocity of a MC~DW, which for a cylindrical flow corresponds to \ $\Upsilon_{\rm axi}$, can be approximated as \citep[][see Eq.~12]{Pud07}
\begin{equation}
\label{eq_vk}
 \Upsilon_{\rm axi} \approx \sqrt{2} \; \; \frac{r_{\rm A}}{r_0} \; \; \upsilon_{\rm K,0}
,\end{equation}
 
\noindent where \ $ \upsilon_{\rm K,0}$ \ is the disk Keplerian velocity at the radius \ $r_0$. 
From Eq.~\ref{eq_v_r} and knowing both the ratio of \ $\Upsilon_{\rm axi}$ / $\Upsilon_{\Phi} \approx $~3.4 \ and the radius \ $ r_{\rm c} = r_{\rm jet} \approx$~16~au, we get \ $r_{\rm A} \approx$~6.7~au. Numerical and theoretical models \citep[for a review see][]{Pud07} supported by observations \citep{Bac02} constrain the ratio of \ $r_{\rm A} / r_0 \approx$~3. Then, using the above value of \ $r_{\rm A}$ \ we infer \ $r_0 \approx$~2.2~au. From the latter and the estimated mass of YSO-1 \ $M_{\star} \approx$~5.6~\ms, we find \ $ \upsilon_{\rm K,0} \approx$~50~\kms, which, through Eq.~\ref{eq_vk}, yields \ $ \Upsilon_{\rm axi} \approx$~200~\kms and \ $\Upsilon_{\Phi} \approx$~200~\kms~/~3.4~$\approx$~60~\kms.

The derived values of jet radius and terminal velocity are in reasonable agreement with the few measurements of spatially resolved radio jets from intermediate-mass YSOs \citep[][see Table~1]{Ang18} and the results from high-resolution, 3D numerical simulations of MC~DWs to 100~au scale \citep[][see Fig.~3]{Sta14}. It is also plausible that our estimate for the jet rotational velocity is larger than the values of a few 10~\kms\  typical for jets from low-mass YSOs \citep{Bac02,Cof04,Ray07}. The ratio of the maser to the jet speeds depends on the density contrast between the jet and the ambient medium through \ $V$ / $\Upsilon$~=~$\sqrt{\rho_{\rm jet}/\rho_{\rm amb}}$. High-angular resolution observations toward low-mass \citep{Pod15,Gom15} and high-mass \citep{Ang18} YSOs, which are also supported by recent numerical simulations of MC-DWs \citep[][see Fig.~5]{Sta19}, indicate values of \ $n_{\rm H_2} \sim 10^6$~cm$^{-3}$ \ for the density of the high-velocity gas in jets. From this value and our measurement of the average ambient density in core~1 \ $n_{\rm H_2} \approx 10^8$~cm$^{-3}$, we derive \ $V$~/~$\Upsilon$~$\sim$~0.1. This agrees with the ratio of the maser-averaged value of \ $V_{\rm axi}$~=~23~\kms\ to the estimate for \ $\Upsilon_{\rm axi} \approx$~200~\kms.

A key prediction of the MC~DW theory is that the wind mass flux is about 10\% of the accretion flux \citep{Pud19}.
Assuming that YSO-1 gathers most of the mass flowing toward the core cluster from the parental molecular cloud, its accretion  rate can be estimated as \ $ \dot{M}_{\rm acc} \approx  \dot{M}_{\rm inf}   \approx $~ $\tan (i_{\rm inf})$ 10$^{-4}$ ~\ms~yr$^{-1}$ (see Sect.~\ref{dis_flow}). Combining the jet momentum rate  \ $\dot{P}_{\rm jet} \sim$ 3$\, \times \,$10$^{-3}$~\ms~\kms~yr$^{-1}$ (see Sect.~\ref{dis_jet}) with the jet terminal velocity \ $ \Upsilon_{\rm axi} \approx$~200~\kms \ derived above, the mass ejection rate of YSO-1 is found to be \ $\dot{M}_{\rm eje} \approx$~1.5$\, \times \,$10$^{-5}$~\ms~yr$^{-1}$. 
Thus, the ratio of the ejection and the accretion rates is \ $\dot{M}_{\rm eje} \, / \, \dot{M}_{\rm acc} \approx$ 0.15  $\, \cot (i_{\rm inf})$. For an intermediate value of (the inclination of the infall) \ $i_{\rm inf} \approx $~55\degr, this ratio is consistent with the value of \ $\approx$~0.1 \ predicted by the MC~DW theory.

\subsection{Change in the jet direction}
\label{dis_bin}


Since there is no hint of multiple YSOs inside core~1 in our data, the different direction of the outflows from core~1 shown in
Fig.~\ref{LBT-jet} (upper~panel) is most likely due to a change in the orientation of the jet ejected from YSO-1. The angle between the direction, at PA = 71\degr, to the far \ H$_2$ bow shock and the axis of the compact radio jet, is \ $\approx$28\degr. From previous studies of jet precession \citep[see, for instance,][]{Fen98,She20} and recent 3D magnetohydrodynamics (MHD) simulations of protostar formation 
\citep[see, for instance,][]{Hir19}, it is known that several processes can produce a large ($\ge$ a few 10\degr) change in the jet orientation: \ 1)~radiation-induced warping of the protostellar accretion disk;\ 2)~tidal interactions between the disk around the primary and the secondary, within a close binary system; 3)~anisotropic accretion; and\ 4)~magnetic braking. In the following, we discuss each of these points. We can readily exclude the first point, since the mass and luminosity of YSO-1 is not sufficient to warp the disk. By employing Eq.~1 of \citet{She20}, we find that the size of the observed disk is orders of magnitude smaller than the critical value beyond which the disk can be unstable to warping. 

Concerning now the second point, we consider the binary system of YSOs inside the adjacent cores~1~and~3 (see Sect.~\ref{TM_dis}). Assuming that the disk surface density is uniform, for Keplerian rotation, the precession frequency is given by the expression \citep[][see Eq.~1]{Ter99}
\begin{equation}
w_{\rm p} = -\frac{15}{32} \; \frac{M_{\rm s}}{M_{\rm p}} \; \left( \frac{R_{\rm d}}{D} \right)^3 \; \cos(\delta) \; \sqrt{\frac{G M_{\rm p}}{R^3_{\rm d}}}
\label{pre_fre}
,\end{equation}

\noindent where \ $M_{\rm p}$ \ and \ $M_{\rm s}$ \ are the masses of the primary and secondary stars, respectively, $R_{\rm d}$ \ and \ $D$ \ are the disk radius and binary separation, respectively, and \ $\delta$ \ is the inclination angle of the binary orbit with respect to the plane of the disk. From Sects.~\ref{TM_dis}~and~\ref{dis_YSO}, we have \ $M_{\rm p} = M_{\star} \approx$~5.6~\ms, $M_{\rm s} / M_{\rm p} \approx$~0.25, $D  \approx$~770~au. Referring to Fig.~\ref{disk_vel} (lower~panel),  $R_{\rm d} \approx$~250~au \ is estimated by taking half the spread in position (projected along the major axis of the distribution at PA = 123\degr) for the high-velocity molecular emissions tracing the YSO-1 disk. The line, at PA = 129\degr, connecting the dust emission peaks of cores~1~and~3 and the disk major axis form a small angle of \ $\approx$~6\degr, which corresponds to the projection of \ $\delta$  \ on the plane of the sky: assuming \ $\cos(\delta) = 1 $, we calculate an upper limit for \ $w_{\rm p}$. Using the listed values for the parameters of Eq.~\ref{pre_fre}, we infer a lower limit for the precession period \ $T_{\rm p} = 2 \pi / w_{\rm p} \gtrsim$~4$\, \times \,$10$^5$~yr. Comparing this with the dynamical timescale of the outflow \ $t_{\rm out} \sim$~6.3$\, \times \,$10$^3$~yr (see Sect.~\ref{dis_jet}), we could argue that is very unlikely that the jet has precessed across an angle \ $\ge$~30\degr \ over a tiny fraction of the precession period.
The precession axis should be directed close to the LOS and the binary orbit should be seen almost face-on, with the consequence that the mass of the binary system inferred from the observed \Vlsr\ gradient would be exceedingly too high with respect to the mass of YSO-1.
 However, if the dynamical time underestimated the true outflow age by a factor as large as 10, the outflow lifetime could still be a significant fraction (i.e., up to one-sixth) of the precession period. Therefore, we cannot completely exclude that the change in the YSO-1 jet direction at different length scales is due to precession induced by the tidal interaction between the YSOs in cores~1~and~3.

Recent 3D nonideal MHD simulations of the collapse of a rotating, magnetized molecular clump \citep{Tsu18,Hir19,Mac20}, which aim to study the effects of a misalignment between the rotation axis of the clump and the (initially) uniform magnetic field, indicate that the evolution of the angular momentum of the central region is controlled by both anisotropic accretion and magnetic braking. While the former occurs during the earlier phases at lower density when the prestellar clump preferentially contracts along the magnetic field line due to flux freezing, the latter operates at higher density close to the protostar where an outer magnetically supported pseudo-disk and an inner centrifugally supported disk form.
The above simulations show that the elongation of the disk-like structure orbiting the protostar is approximately perpendicular to the magnetic field at large scales (10$^3$--10$^4$~au) and to the clump rotation axis at small scales ($\sim$10$^2$~au). The direction of the protostellar outflow changes with time by a few 10\degr\ following the evolution of the accretion disk, and converges to the rotation axis of the clump once the centrifugally supported disk has grown in size. 

The formation of the core cluster in \targ\ could have undergone processes similar to those predicted in these MHD simulations. This possibility is suggested by the finding that the most massive and evolved cores in the cluster have comparable sizes of a few 100~au and are all elongated approximately in the same direction. Indeed, Fig.~\ref{CH3CN} (upper~panel) shows that core~2 is extended along an axis about parallel to the direction of both the binary system in cores~1~and~3 (at PA = 129\degr) and the disk around YSO-1 in core~1 (at PA = 123\degr). According to the aforementioned simulations, the direction, at PA $\approx$~43\degr, of the jet from YSO-1 (the most evolved YSO in the cluster) should be close to the rotation axis of the parental clump. The large-scale magnetic field should be oriented approximately perpendicular to the SE-NW elongated, core cluster (see Fig.~\ref{1.3cont}, upper~panel), whose PA is \ 149\degr $\pm$ 17\degr \ (see Sect.~\ref{obs_clu}). Therefore, in this scenario, the direction from YSO-1 to the H$_2$ bow shock, at PA = 71\degr, would be about parallel to the magnetic field. The H$_2$ shocks trace the direction of the jet axis at an earlier time than the radio jet because they
are at a larger distance from YSO-1. The observed change in the jet orientation could then correspond to the model-predicted evolution of the protostellar outflow, from being collimated close to the magnetic field in the earliest phases to ultimately align with the rotation axis of the clump. This interpretation, relying only on marginal evidence, is speculative, but it can be tested with future measurements of the magnetic field configuration in \targ.



\section{Conclusions}
\label{conclu}

This work combines the data of the CORE NOEMA millimeter interferometer and POETS VLBA-JVLA radio surveys to study the SF in  \targ\ on scales from \ $\sim$10$^4$~au to \ $\sim$10~au. The CORE dust continuum and molecular line emissions allow us to map the cluster of molecular cores, identify the positions of the YSOs, and determine the kinematics of the associated accretion-ejection structures down to a few 100~au. The POETS radio continuum and water maser data complement the CORE information by unveiling the properties of the ionized gas and the 3D gas kinematics near the YSOs at radii as small as a few astronomical units. Our main results can be resumed as follows:

\begin{itemize}

\item The SE-NW elongated, core cluster (size $\sim$0.1~pc) of \targ\  is found inside a larger (size $\sim$1~pc) molecular cloud of similar SE-NW elongation. A \Vlsr\ gradient of amplitude of \ $\approx$~1~\kms\ per 0.1~pc \ is detected across the major axis of the molecular cloud and the cluster. Assuming we are observing a mass flow from the harboring cloud to the cluster, we derive a mass infall rate of
\ $\approx$~10$^{-4}$~\ms~yr$^{-1}$. 

\item A signature of protostellar activity, in terms of emission from high-excitation molecular lines or a molecular outflow, is found only for the most massive cores (labeled 1, 2~and~3) at the center of the cluster. The masses of the YSOs inside these three cores are estimated in the range \ 1--6~\ms. If a relevant fraction of the mass infalling onto the cluster is accreted by these more massive YSOs, it is likely that the \targ \ core cluster can be the progenitor of a stellar system whose most massive members are early B or late O-type high-mass stars.

\item We detect a SW-NE collimated, bipolar molecular outflow emerging from the most massive core~1. It is about parallel to the double-lobe (separation $\approx$~0\farcs5) radio jet detected in the POETS survey, ejected from a YSO embedded in core~1. We refer to this YSO as YSO-1. We show that the momentum rates of the radio jet and the molecular outflow are comparable, which, also because of the similar orientation, leads us to think that the molecular outflow is driven by the radio jet. 

\item A \Vlsr\ gradient of amplitude of \ $\approx$~14~\kms\ over 500~au, directed perpendicular to the radio jet, is revealed at the position of YSO-1. We propose an interpretation in terms of an almost edge-on, rotating disk and, by fitting a Keplerian rotation to the \Vlsr\ pattern, we obtain a mass for YSO-1 of \ $5.6~\pm~2.0$~\ms.

\item The water masers observed in \targ\ with the POETS survey are clearly tracing the YSO-1 jet. They emerge at radii of \ 100--300~au NE of YSO-1 and their spatial distribution and proper motions are directed within a few degrees from the axis of the radio jet.

\item The radio-maser jet from YSO-1 is unique in presenting a lobe with nonthermal emission and \ a signature of jet rotation. The latter is the monotonic change of the maser \Vlsr\ along a (SE-NW) direction perpendicular to the (SW-NE) collimation axis of the maser proper motions. 

\item We show that the maser 3D velocity pattern is consistent with the MC~DW models predicting the recollimation of the wind within a rotating cylindrical flow. The masers could trace shocked gas at the wall of the cylindrical jet. We determine the jet radius to be \ $\approx$~16~au and the corresponding launching radius and terminal (poloidal) velocity \ $\approx$~2.2~au \ and \ $\approx$~200~\kms, respectively. Assuming that YSO-1 gathers most of the mass flowing toward the core cluster from the parental molecular cloud, the ratio of the mass ejection and accretion rates of YSO-1 can be consistent with the value ($\approx$~0.1) predicted by the MC~DW theory.
 


\end{itemize}


\begin{acknowledgements}\\
A.P. acknowledges financial support from CONACyT and UNAM-PAPIIT IN113119 grant, M\'exico. \\
A.S.M. acknowledges support from the Collaborative Research Centre 956 (subproject A6), funded by the Deutsche Forschungsgemeinschaft (DFG) - project 184018867. \\
D.S. acknowledges support by the Deutsche Forschungsgemeinschaft 
through SPP 1833: ``Building a Habitable Earth'' (SE 1962/6-1). \\
F.B. acknowledges support from the project PRIN-INAF-MAIN-STREAM 2017 “Protoplanetary disks seen through the eyes of new-generation instruments”. \\
H.B. acknowledges support from the Deutsche Forschungsgemeinschaft (DFG) via Sonderforschungsbereich (SFB) 881 “The Milky Way System” (sub-project B1).\\
H.B., A.A. and S.S. acknowledge support from the European Research Council under the European Community's Horizon 2020 framework program (2014-2020) via the ERC Consolidator grant ‘From Cloud to Star Formation (CSF)' (project number 648505).\\
R.G.M. acknowledges support from UNAM-PAPIIT project IN104319.\\
R.K. acknowledges financial support via the Emmy Noether Research Group on Accretion Flows and Feedback in Realistic Models of Massive Star Formation funded by the German Research Foundation (DFG) under grant no. KU 2849/3-1 and KU 2849/3-2.\\
This work is based on observations carried out under project number L14AB with the IRAM NOEMA Interferometer and the IRAM~30~m telescope. IRAM is supported by INSU/CNRS (France), MPG (Germany) and IGN (Spain). \\
The LBT is an international collaboration among institutions in the United
 States, Italy and Germany. LBT Corporation partners are: The University of 
Arizona on behalf of the Arizona Board of Regents; 
Istituto Nazionale di Astrofisica, Italy; 
LBT Beteiligungsgesellschaft, Germany, representing the Max-Planck Society, 
The Leibniz Institute for Astrophysics Potsdam, and Heidelberg University; 
The Ohio State University, and The Research Corporation, on behalf of The 
University of Notre Dame, University of Minnesota and University of Virginia.
\end{acknowledgements}

%
   \bibliographystyle{aa} 
   \bibliography{biblio} 

\begin{thebibliography}{89}
\expandafter\ifx\csname natexlab\endcsname\relax\def\natexlab#1{#1}\fi

\bibitem[{{Ageorges} {et~al.}(2010){Ageorges}, {Seifert}, {J{\"u}tte},
  {Knierim}, {Lehmitz}, {Germeroth}, {Buschkamp}, {Polsterer}, {Pasquali},
  {Naranjo}, {Gemperlein}, {Hill}, {Feiz}, {Hofmann}, {Laun}, {Lederer},
  {Lenzen}, {Mall}, {Mand el}, {M{\"u}ller}, {Quirrenbach}, {Sch{\"a}ffner},
  {Storz}, \& {Weiser}}]{Age10}
{Ageorges}, N., {Seifert}, W., {J{\"u}tte}, M., {et~al.} 2010, in Society of
  Photo-Optical Instrumentation Engineers (SPIE) Conference Series, Vol. 7735,
  Ground-based and Airborne Instrumentation for Astronomy III, 77351L

\bibitem[{{Ahmadi} {et~al.}(2018){Ahmadi}, {Beuther}, {Mottram}, {Bosco},
  {Linz}, {Henning}, {Winters}, {Kuiper}, {Pudritz}, {S{\'a}nchez-Monge},
  {Keto}, {Beltran}, {Bontemps}, {Cesaroni}, {Csengeri}, {Feng},
  {Galvan-Madrid}, {Johnston}, {Klaassen}, {Leurini}, {Longmore}, {Lumsden},
  {Maud}, {Menten}, {Moscadelli}, {Motte}, {Palau}, {Peters}, {Ragan},
  {Schilke}, {Urquhart}, {Wyrowski}, \& {Zinnecker}}]{Ahm18}
{Ahmadi}, A., {Beuther}, H., {Mottram}, J.~C., {et~al.} 2018, \aap, 618, A46

\bibitem[{{Anglada} {et~al.}(2018){Anglada}, {Rodr{\'\i}guez}, \&
  {Carrasco-Gonz{\'a}lez}}]{Ang18}
{Anglada}, G., {Rodr{\'\i}guez}, L.~F., \& {Carrasco-Gonz{\'a}lez}, C. 2018,
  \aapr, 26, 3

\bibitem[{{Bacciotti} {et~al.}(2002){Bacciotti}, {Ray}, {Mundt},
  {Eisl{\"o}ffel}, \& {Solf}}]{Bac02}
{Bacciotti}, F., {Ray}, T.~P., {Mundt}, R., {Eisl{\"o}ffel}, J., \& {Solf}, J.
  2002, \apj, 576, 222

\bibitem[{{Barklem} \& {Collet}(2016)}]{Bar16}
{Barklem}, P.~S. \& {Collet}, R. 2016, \aap, 588, A96

\bibitem[{{Bell}(1978)}]{Bel78}
{Bell}, A.~R. 1978, \mnras, 182, 147

\bibitem[{{Beuther} {et~al.}(2018){Beuther}, {Mottram}, {Ahmadi}, {Bosco},
  {Linz}, {Henning}, {Klaassen}, {Winters}, {Maud}, {Kuiper}, {Semenov},
  {Gieser}, {Peters}, {Urquhart}, {Pudritz}, {Ragan}, {Feng}, {Keto},
  {Leurini}, {Cesaroni}, {Beltran}, {Palau}, {S{\'a}nchez-Monge},
  {Galvan-Madrid}, {Zhang}, {Schilke}, {Wyrowski}, {Johnston}, {Longmore},
  {Lumsden}, {Hoare}, {Menten}, \& {Csengeri}}]{Beu18}
{Beuther}, H., {Mottram}, J.~C., {Ahmadi}, A., {et~al.} 2018, \aap, 617, A100

\bibitem[{{Blandford} \& {Payne}(1982)}]{Bla82}
{Blandford}, R.~D. \& {Payne}, D.~G. 1982, \mnras, 199, 883

\bibitem[{{Bonnell} {et~al.}(2004){Bonnell}, {Vine}, \& {Bate}}]{Bon04}
{Bonnell}, I.~A., {Vine}, S.~G., \& {Bate}, M.~R. 2004, \mnras, 349, 735

\bibitem[{{Bosco} {et~al.}(2019){Bosco}, {Beuther}, {Ahmadi}, {Mottram},
  {Kuiper}, {Linz}, {Maud}, {Winters}, {Henning}, {Feng}, {Peters}, {Semenov},
  {Klaassen}, {Schilke}, {Urquhart}, {Beltr{\'a}n}, {Lumsden}, {Leurini},
  {Moscadelli}, {Cesaroni}, {S{\'a}nchez-Monge}, {Palau}, {Pudritz},
  {Wyrowski}, \& {Longmore}}]{Bos19}
{Bosco}, F., {Beuther}, H., {Ahmadi}, A., {et~al.} 2019, \aap, 629, A10

\bibitem[{{Burns} {et~al.}(2015){Burns}, {Imai}, {Handa}, {Omodaka},
  {Nakagawa}, {Nagayama}, \& {Ueno}}]{Burn15}
{Burns}, R.~A., {Imai}, H., {Handa}, T., {et~al.} 2015, \mnras, 453, 3163

\bibitem[{{Cesaroni} {et~al.}(2019){Cesaroni}, {Beuther}, {Ahmadi},
  {Beltr{\'a}n}, {Csengeri}, {Galv{\'a}n-Madrid}, {Gieser}, {Henning},
  {Johnston}, {Klaassen}, {Kuiper}, {Leurini}, {Linz}, {Longmore}, {Lumsden},
  {Maud}, {Moscadelli}, {Mottram}, {Palau}, {Peters}, {Pudritz},
  {S{\'a}nchez-Monge}, {Schilke}, {Semenov}, {Suri}, {Urquhart}, {Winters},
  {Zhang}, \& {Zinnecker}}]{Ces19}
{Cesaroni}, R., {Beuther}, H., {Ahmadi}, A., {et~al.} 2019, \aap, 627, A68

\bibitem[{{Cesaroni} {et~al.}(2005){Cesaroni}, {Neri}, {Olmi}, {Testi},
  {Walmsley}, \& {Hofner}}]{Ces05}
{Cesaroni}, R., {Neri}, R., {Olmi}, L., {et~al.} 2005, \aap, 434, 1039

\bibitem[{{Chen} {et~al.}(2019){Chen}, {Zhang}, {Wright}, {Busquet}, {Lin},
  {Liu}, {Olguin}, {Sanhueza}, {Nakamura}, {Palau}, {Ohashi}, {Tatematsu}, \&
  {Liao}}]{Chen19}
{Chen}, H.-R.~V., {Zhang}, Q., {Wright}, M.~C.~H., {et~al.} 2019, \apj, 875, 24

\bibitem[{{Clark}(1980)}]{Cla80}
{Clark}, B.~G. 1980, \aap, 89, 377

\bibitem[{{Coffey} {et~al.}(2004){Coffey}, {Bacciotti}, {Woitas}, {Ray}, \&
  {Eisl{\"o}ffel}}]{Cof04}
{Coffey}, D., {Bacciotti}, F., {Woitas}, J., {Ray}, T.~P., \& {Eisl{\"o}ffel},
  J. 2004, \apj, 604, 758

\bibitem[{{Csengeri} {et~al.}(2017){Csengeri}, {Bontemps}, {Wyrowski}, {Motte},
  {Menten}, {Beuther}, {Bronfman}, {Commer{\c{c}}on}, {Chapillon},
  {Duarte-Cabral}, {Fuller}, {Henning}, {Leurini}, {Longmore}, {Palau},
  {Peretto}, {Schuller}, {Tan}, {Testi}, {Traficante}, \& {Urquhart}}]{Cse17}
{Csengeri}, T., {Bontemps}, S., {Wyrowski}, F., {et~al.} 2017, \aap, 600, L10

\bibitem[{{Davies} {et~al.}(2011){Davies}, {Hoare}, {Lumsden}, {Hosokawa},
  {Oudmaijer}, {Urquhart}, {Mottram}, \& {Stead}}]{Dav11}
{Davies}, B., {Hoare}, M.~G., {Lumsden}, S.~L., {et~al.} 2011, \mnras, 416, 972

\bibitem[{{Di Francesco} {et~al.}(2008){Di Francesco}, {Johnstone}, {Kirk},
  {MacKenzie}, \& {Ledwosinska}}]{DiFra08}
{Di Francesco}, J., {Johnstone}, D., {Kirk}, H., {MacKenzie}, T., \&
  {Ledwosinska}, E. 2008, \apjs, 175, 277

\bibitem[{{Draine}(2011)}]{Dra11}
{Draine}, B.~T. 2011, {Physics of the Interstellar and Intergalactic Medium}

\bibitem[{{Egan} {et~al.}(2003){Egan}, {Price}, \& {Kraemer}}]{Ega03}
{Egan}, M.~P., {Price}, S.~D., \& {Kraemer}, K.~E. 2003, in Bulletin of the
  American Astronomical Society, Vol.~35, American Astronomical Society Meeting
  Abstracts, 1301

\bibitem[{{Esposito} {et~al.}(2012){Esposito}, {Riccardi}, {Pinna}, {Puglisi},
  {Quir{\'o}s-Pacheco}, {Arcidiacono}, {Xompero}, {Briguglio}, {Busoni},
  {Fini}, {Argomedo}, {Gherardi}, {Agapito}, {Brusa}, {Miller}, {Guerra Ramon},
  {Boutsia}, \& {Stefanini}}]{Esp12}
{Esposito}, S., {Riccardi}, A., {Pinna}, E., {et~al.} 2012, in Society of
  Photo-Optical Instrumentation Engineers (SPIE) Conference Series, Vol. 8447,
  Adaptive Optics Systems III, 84470U

\bibitem[{{Federrath}(2015)}]{Fed15}
{Federrath}, C. 2015, \mnras, 450, 4035

\bibitem[{{Fendt} \& {Zinnecker}(1998)}]{Fen98}
{Fendt}, C. \& {Zinnecker}, H. 1998, \aap, 334, 750

\bibitem[{{Gieser} {et~al.}(2019){Gieser}, {Semenov}, {Beuther}, {Ahmadi},
  {Mottram}, {Henning}, {Beltran}, {Maud}, {Bosco}, {Leurini}, {Peters},
  {Klaassen}, {Kuiper}, {Feng}, {Urquhart}, {Moscadelli}, {Csengeri},
  {Lumsden}, {Winters}, {Suri}, {Zhang}, {Pudritz}, {Palau}, {Menten},
  {Galvan-Madrid}, {Wyrowski}, {Schilke}, {S{\'a}nchez-Monge}, {Linz},
  {Johnston}, {Jim{\'e}nez-Serra}, {Longmore}, \& {M{\"o}ller}}]{Gie19}
{Gieser}, C., {Semenov}, D., {Beuther}, H., {et~al.} 2019, \aap, 631, A142

\bibitem[{{Goddi} {et~al.}(2011){Goddi}, {Humphreys}, {Greenhill}, {Chand ler},
  \& {Matthews}}]{God11c}
{Goddi}, C., {Humphreys}, E.~M.~L., {Greenhill}, L.~J., {Chand ler}, C.~J., \&
  {Matthews}, L.~D. 2011, \apj, 728, 15

\bibitem[{{Goldsmith} \& {Langer}(1999)}]{Gol99}
{Goldsmith}, P.~F. \& {Langer}, W.~D. 1999, \apj, 517, 209

\bibitem[{{G{\'o}mez-Ruiz} {et~al.}(2015){G{\'o}mez-Ruiz}, {Codella},
  {Lefloch}, {Benedettini}, {Busquet}, {Ceccarelli}, {Nisini}, {Podio}, \&
  {Viti}}]{Gom15}
{G{\'o}mez-Ruiz}, A.~I., {Codella}, C., {Lefloch}, B., {et~al.} 2015, \mnras,
  446, 3346

\bibitem[{{Hennebelle} {et~al.}(2019){Hennebelle}, {Lee}, \&
  {Chabrier}}]{Hen19}
{Hennebelle}, P., {Lee}, Y.-N., \& {Chabrier}, G. 2019, \apj, 883, 140

\bibitem[{{Hirano} \& {Machida}(2019)}]{Hir19}
{Hirano}, S. \& {Machida}, M.~N. 2019, \mnras, 485, 4667

\bibitem[{{Hoare} {et~al.}(2005){Hoare}, {Lumsden}, {Oudmaijer}, {Urquhart},
  {Busfield}, {Sheret}, {Clarke}, {Moore}, {Allsopp}, {Burton}, {Purcell},
  {Jiang}, \& {Wang}}]{Hoa05}
{Hoare}, M.~G., {Lumsden}, S.~L., {Oudmaijer}, R.~D., {et~al.} 2005, in IAU
  Symposium, Vol. 227, Massive Star Birth: A Crossroads of Astrophysics, ed.
  R.~{Cesaroni}, M.~{Felli}, E.~{Churchwell}, \& M.~{Walmsley}, 370--375

\bibitem[{{Klessen} \& {Glover}(2016)}]{Kle16}
{Klessen}, R.~S. \& {Glover}, S. C.~O. 2016, Saas-Fee Advanced Course, 43, 85

\bibitem[{{Kobayashi} {et~al.}(2003){Kobayashi}, {Sasao}, {Kawaguchi},
  {Manabe}, {Omodaka}, {Kameya}, {Shibata}, {Miyaji}, {Honma}, {Tamura},
  {Hirota}, {Kuji}, {Horiai}, {Sakai}, {Sato}, {Iwadate}, {Kanya}, {Ujihara},
  {Jike}, {Fujii}, {Oyama}, {Kurayama}, {Suda}, {Sakakibara}, {Kamohara}, \&
  {Kasuga}}]{Kob03}
{Kobayashi}, H., {Sasao}, T., {Kawaguchi}, N., {et~al.} 2003, in Astronomical
  Society of the Pacific Conference Series, Vol. 306, New technologies in VLBI,
  ed. Y.~C. {Minh}, P48

\bibitem[{{K{\"o}lligan} \& {Kuiper}(2018)}]{Koe18}
{K{\"o}lligan}, A. \& {Kuiper}, R. 2018, \aap, 620, A182

\bibitem[{{Lampton} {et~al.}(1976){Lampton}, {Margon}, \& {Bowyer}}]{Lam76}
{Lampton}, M., {Margon}, B., \& {Bowyer}, S. 1976, \apj, 208, 177

\bibitem[{{Longmore} {et~al.}(2014){Longmore}, {Kruijssen}, {Bastian}, {Bally},
  {Rathborne}, {Testi}, {Stolte}, {Dale}, {Bressert}, \& {Alves}}]{Long14}
{Longmore}, S.~N., {Kruijssen}, J.~M.~D., {Bastian}, N., {et~al.} 2014, in
  Protostars and Planets VI, ed. H.~{Beuther}, R.~S. {Klessen}, C.~P.
  {Dullemond}, \& T.~{Henning}, 291

\bibitem[{{Lu} {et~al.}(2018){Lu}, {Zhang}, {Liu}, {Sanhueza}, {Tatematsu},
  {Feng}, {Smith}, {Myers}, {Sridharan}, \& {Gu}}]{Lu18}
{Lu}, X., {Zhang}, Q., {Liu}, H.~B., {et~al.} 2018, \apj, 855, 9

\bibitem[{{Lumsden} {et~al.}(2013){Lumsden}, {Hoare}, {Urquhart}, {Oudmaijer},
  {Davies}, {Mottram}, {Cooper}, \& {Moore}}]{Lum13}
{Lumsden}, S.~L., {Hoare}, M.~G., {Urquhart}, J.~S., {et~al.} 2013, \apjs, 208,
  11

\bibitem[{{Machida} {et~al.}(2020){Machida}, {Hirano}, \& {Kitta}}]{Mac20}
{Machida}, M.~N., {Hirano}, S., \& {Kitta}, H. 2020, \mnras, 491, 2180

\bibitem[{{Masson} \& {Chernin}(1993)}]{Mas93}
{Masson}, C.~R. \& {Chernin}, L.~M. 1993, \apj, 414, 230

\bibitem[{{Maud} {et~al.}(2015){Maud}, {Moore}, {Lumsden}, {Mottram},
  {Urquhart}, \& {Hoare}}]{Mau15}
{Maud}, L.~T., {Moore}, T.~J.~T., {Lumsden}, S.~L., {et~al.} 2015, \mnras, 453,
  645

\bibitem[{{McKee} \& {Tan}(2003)}]{McK03}
{McKee}, C.~F. \& {Tan}, J.~C. 2003, \apj, 585, 850

\bibitem[{{Molinari} {et~al.}(1996){Molinari}, {Brand}, {Cesaroni}, \&
  {Palla}}]{Mol96}
{Molinari}, S., {Brand}, J., {Cesaroni}, R., \& {Palla}, F. 1996, \aap, 308,
  573

\bibitem[{{Molinari} {et~al.}(2010{\natexlab{a}}){Molinari}, {Swinyard},
  {Bally}, {Barlow}, {Bernard}, {Martin}, {Moore}, {Noriega-Crespo}, {Plume},
  {Testi}, {Zavagno}, {Abergel}, {Ali}, {Anderson}, {Andr{\'e}}, {Baluteau},
  {Battersby}, {Beltr{\'a}n}, {Benedettini}, {Billot}, {Blommaert}, {Bontemps},
  {Boulanger}, {Brand}, {Brunt}, {Burton}, {Calzoletti}, {Carey}, {Caselli},
  {Cesaroni}, {Cernicharo}, {Chakrabarti}, {Chrysostomou}, {Cohen},
  {Compiegne}, {de Bernardis}, {de Gasperis}, {di Giorgio}, {Elia}, {Faustini},
  {Flagey}, {Fukui}, {Fuller}, {Ganga}, {Garcia-Lario}, {Glenn}, {Goldsmith},
  {Griffin}, {Hoare}, {Huang}, {Ikhenaode}, {Joblin}, {Joncas}, {Juvela},
  {Kirk}, {Lagache}, {Li}, {Lim}, {Lord}, {Marengo}, {Marshall}, {Masi},
  {Massi}, {Matsuura}, {Minier}, {Miville-Desch{\^e}nes}, {Montier}, {Morgan},
  {Motte}, {Mottram}, {M{\"u}ller}, {Natoli}, {Neves}, {Olmi}, {Paladini},
  {Paradis}, {Parsons}, {Peretto}, {Pestalozzi}, {Pezzuto}, {Piacentini},
  {Piazzo}, {Polychroni}, {Pomar{\`e}s}, {Popescu}, {Reach}, {Ristorcelli},
  {Robitaille}, {Robitaille}, {Rod{\'o}n}, {Roy}, {Royer}, {Russeil},
  {Saraceno}, {Sauvage}, {Schilke}, {Schisano}, {Schneider}, {Schuller},
  {Schulz}, {Sibthorpe}, {Smith}, {Smith}, {Spinoglio}, {Stamatellos},
  {Strafella}, {Stringfellow}, {Sturm}, {Taylor}, {Thompson}, {Traficante},
  {Tuffs}, {Umana}, {Valenziano}, {Vavrek}, {Veneziani}, {Viti}, {Waelkens},
  {Ward-Thompson}, {White}, {Wilcock}, {Wyrowski}, {Yorke}, \&
  {Zhang}}]{Mol10b}
{Molinari}, S., {Swinyard}, B., {Bally}, J., {et~al.} 2010{\natexlab{a}}, \aap,
  518, L100

\bibitem[{{Molinari} {et~al.}(2010{\natexlab{b}}){Molinari}, {Swinyard},
  {Bally}, {Barlow}, {Bernard}, {Martin}, {Moore}, {Noriega-Crespo}, {Plume},
  {Testi}, {Zavagno}, {Abergel}, {Ali}, {Andr{\'e}}, {Baluteau}, {Benedettini},
  {Bern{\'e}}, {Billot}, {Blommaert}, {Bontemps}, {Boulanger}, {Brand},
  {Brunt}, {Burton}, {Campeggio}, {Carey}, {Caselli}, {Cesaroni}, {Cernicharo},
  {Chakrabarti}, {Chrysostomou}, {Codella}, {Cohen}, {Compiegne}, {Davis}, {de
  Bernardis}, {de Gasperis}, {Di Francesco}, {di Giorgio}, {Elia}, {Faustini},
  {Fischera}, {Fukui}, {Fuller}, {Ganga}, {Garcia-Lario}, {Giard}, {Giardino},
  {Glenn}, {Goldsmith}, {Griffin}, {Hoare}, {Huang}, {Jiang}, {Joblin},
  {Joncas}, {Juvela}, {Kirk}, {Lagache}, {Li}, {Lim}, {Lord}, {Lucas},
  {Maiolo}, {Marengo}, {Marshall}, {Masi}, {Massi}, {Matsuura}, {Meny},
  {Minier}, {Miville-Desch{\^e}nes}, {Montier}, {Motte}, {M{\"u}ller},
  {Natoli}, {Neves}, {Olmi}, {Paladini}, {Paradis}, {Pestalozzi}, {Pezzuto},
  {Piacentini}, {Pomar{\`e}s}, {Popescu}, {Reach}, {Richer}, {Ristorcelli},
  {Roy}, {Royer}, {Russeil}, {Saraceno}, {Sauvage}, {Schilke},
  {Schneider-Bontemps}, {Schuller}, {Schultz}, {Shepherd}, {Sibthorpe},
  {Smith}, {Smith}, {Spinoglio}, {Stamatellos}, {Strafella}, {Stringfellow},
  {Sturm}, {Taylor}, {Thompson}, {Tuffs}, {Umana}, {Valenziano}, {Vavrek},
  {Viti}, {Waelkens}, {Ward-Thompson}, {White}, {Wyrowski}, {Yorke}, \&
  {Zhang}}]{Mol10a}
{Molinari}, S., {Swinyard}, B., {Bally}, J., {et~al.} 2010{\natexlab{b}},
  \pasp, 122, 314

\bibitem[{{M{\"o}ller} {et~al.}(2017){M{\"o}ller}, {Endres}, \&
  {Schilke}}]{Moel17}
{M{\"o}ller}, T., {Endres}, C., \& {Schilke}, P. 2017, \aap, 598, A7

\bibitem[{{Moscadelli} {et~al.}(2016){Moscadelli}, {S{\'a}nchez-Monge},
  {Goddi}, {Li}, {Sanna}, {Cesaroni}, {Pestalozzi}, {Molinari}, \&
  {Reid}}]{Mos16}
{Moscadelli}, L., {S{\'a}nchez-Monge}, {\'A}., {Goddi}, C., {et~al.} 2016,
  \aap, 585, A71

\bibitem[{{Moscadelli} {et~al.}(2019){Moscadelli}, {Sanna}, {Goddi},
  {Krishnan}, {Massi}, \& {Bacciotti}}]{Mos19b}
{Moscadelli}, L., {Sanna}, A., {Goddi}, C., {et~al.} 2019, \aap, 631, A74

\bibitem[{{Moscadelli} {et~al.}(2020){Moscadelli}, {Sanna}, {Goddi},
  {Krishnan}, {Massi}, \& {Bacciotti}}]{Mos20}
{Moscadelli}, L., {Sanna}, A., {Goddi}, C., {et~al.} 2020, \aap, 635, A118

\bibitem[{{Mottram} {et~al.}(2020){Mottram}, {Beuther}, {Ahmadi}, {Klaassen},
  {Beltr{\'a}n}, {Csengeri}, {Feng}, {Gieser}, {Henning}, {Johnston}, {Kuiper},
  {Leurini}, {Linz}, {Longmore}, {Lumsden}, {Maud}, {Moscadelli}, {Palau},
  {Peters}, {Pudritz}, {Ragan}, {S{\'a}nchez-Monge}, {Semenov}, {Urquhart},
  {Winters}, \& {Zinnecker}}]{Mot20}
{Mottram}, J.~C., {Beuther}, H., {Ahmadi}, A., {et~al.} 2020, \aap, 636, A118

\bibitem[{{M{\"u}ller} {et~al.}(2005){M{\"u}ller}, {Schl{\"o}der}, {Stutzki},
  \& {Winnewisser}}]{Muel05}
{M{\"u}ller}, H. S.~P., {Schl{\"o}der}, F., {Stutzki}, J., \& {Winnewisser}, G.
  2005, Journal of Molecular Structure, 742, 215

\bibitem[{{M{\"u}ller} {et~al.}(2001){M{\"u}ller}, {Thorwirth}, {Roth}, \&
  {Winnewisser}}]{Muel01}
{M{\"u}ller}, H.~S.~P., {Thorwirth}, S., {Roth}, D.~A., \& {Winnewisser}, G.
  2001, \aap, 370, L49

\bibitem[{{Myers}(2009)}]{Mye09}
{Myers}, P.~C. 2009, \apj, 700, 1609

\bibitem[{{Neugebauer} {et~al.}(1984){Neugebauer}, {Habing}, {van Duinen},
  {Aumann}, {Baud}, {Beichman}, {Beintema}, {Boggess}, {Clegg}, {de Jong},
  {Emerson}, {Gautier}, {Gillett}, {Harris}, {Hauser}, {Houck}, {Jennings},
  {Low}, {Marsden}, {Miley}, {Olnon}, {Pottasch}, {Raimond}, {Rowan-Robinson},
  {Soifer}, {Walker}, {Wesselius}, \& {Young}}]{Neu84}
{Neugebauer}, G., {Habing}, H.~J., {van Duinen}, R., {et~al.} 1984, \apjl, 278,
  L1

\bibitem[{{Neupane} {et~al.}(2020){Neupane}, {Garay}, {Contreras},
  {Guzm{\'a}n}, \& {Rodr{\'\i}guez}}]{Neu20}
{Neupane}, S., {Garay}, G., {Contreras}, Y., {Guzm{\'a}n}, A.~E., \&
  {Rodr{\'\i}guez}, L.~F. 2020, \apj, 890, 76

\bibitem[{{Ossenkopf} \& {Henning}(1994)}]{Oss94}
{Ossenkopf}, V. \& {Henning}, T. 1994, \aap, 291, 943

\bibitem[{{Padoan} {et~al.}(2020){Padoan}, {Pan}, {Juvela}, {Haugb{\o}lle}, \&
  {Nordlund}}]{Pad20}
{Padoan}, P., {Pan}, L., {Juvela}, M., {Haugb{\o}lle}, T., \& {Nordlund},
  {\r{A}}. 2020, \apj, 900, 82

\bibitem[{{Palla} \& {Stahler}(1993)}]{Pal93}
{Palla}, F. \& {Stahler}, S.~W. 1993, \apj, 418, 414

\bibitem[{{Parker} {et~al.}(1991){Parker}, {Padman}, \& {Scott}}]{Par91}
{Parker}, N.~D., {Padman}, R., \& {Scott}, P.~F. 1991, \mnras, 252, 442

\bibitem[{{Pickett} {et~al.}(1998){Pickett}, {Poynter}, {Cohen}, {Delitsky},
  {Pearson}, \& {M{\"u}ller}}]{Pick98}
{Pickett}, H.~M., {Poynter}, R.~L., {Cohen}, E.~A., {et~al.} 1998, \jqsrt, 60,
  883

\bibitem[{{Podio} {et~al.}(2015){Podio}, {Codella}, {Gueth}, {Cabrit},
  {Bachiller}, {Gusdorf}, {Lee}, {Lefloch}, {Leurini}, {Nisini}, \&
  {Tafalla}}]{Pod15}
{Podio}, L., {Codella}, C., {Gueth}, F., {et~al.} 2015, \aap, 581, A85

\bibitem[{{Pudritz} {et~al.}(2007){Pudritz}, {Ouyed}, {Fendt}, \&
  {Brandenburg}}]{Pud07}
{Pudritz}, R.~E., {Ouyed}, R., {Fendt}, C., \& {Brandenburg}, A. 2007, in
  Protostars and Planets V, ed. B.~{Reipurth}, D.~{Jewitt}, \& K.~{Keil}, 277

\bibitem[{{Pudritz} \& {Ray}(2019)}]{Pud19}
{Pudritz}, R.~E. \& {Ray}, T.~P. 2019, Frontiers in Astronomy and Space
  Sciences, 6, 54

\bibitem[{{Ragan} {et~al.}(2012){Ragan}, {Heitsch}, {Bergin}, \&
  {Wilner}}]{Rag12}
{Ragan}, S.~E., {Heitsch}, F., {Bergin}, E.~A., \& {Wilner}, D. 2012, \apj,
  746, 174

\bibitem[{{Ray} {et~al.}(2007){Ray}, {Dougados}, {Bacciotti}, {Eisl{\"o}ffel},
  \& {Chrysostomou}}]{Ray07}
{Ray}, T., {Dougados}, C., {Bacciotti}, F., {Eisl{\"o}ffel}, J., \&
  {Chrysostomou}, A. 2007, in Protostars and Planets V, ed. B.~{Reipurth},
  D.~{Jewitt}, \& K.~{Keil}, 231

\bibitem[{{Reid} {et~al.}(2014){Reid}, {Menten}, {Brunthaler}, {Zheng}, {Dame},
  {Xu}, {Wu}, {Zhang}, {Sanna}, {Sato}, {Hachisuka}, {Choi}, {Immer},
  {Moscadelli}, {Rygl}, \& {Bartkiewicz}}]{Rei14}
{Reid}, M.~J., {Menten}, K.~M., {Brunthaler}, A., {et~al.} 2014, \apj, 783, 130

\bibitem[{{Reid} {et~al.}(2009){Reid}, {Menten}, {Brunthaler}, {Zheng},
  {Moscadelli}, \& {Xu}}]{Rei09}
{Reid}, M.~J., {Menten}, K.~M., {Brunthaler}, A., {et~al.} 2009, \apj, 693, 397

\bibitem[{Reid {et~al.}(1988)Reid, Schneps, Moran, Gwinn, Genzel, Downes, \&
  R{\"o}nn{\"a}ng}]{Rei88}
Reid, M.~J., Schneps, M.~H., Moran, J.~M., {et~al.} 1988, ApJ, 330, 809

\bibitem[{{Sanhueza} {et~al.}(2019){Sanhueza}, {Contreras}, {Wu}, {Jackson},
  {Guzm{\'a}n}, {Zhang}, {Li}, {Lu}, {Silva}, {Izumi}, {Liu}, {Miura},
  {Tatematsu}, {Sakai}, {Beuther}, {Garay}, {Ohashi}, {Saito}, {Nakamura},
  {Saigo}, {Veena}, {Nguyen-Luong}, \& {Tafoya}}]{Sanh19}
{Sanhueza}, P., {Contreras}, Y., {Wu}, B., {et~al.} 2019, \apj, 886, 102

\bibitem[{{Sanna} {et~al.}(2019){Sanna}, {Moscadelli}, {Goddi}, {Beltr{\'a}n},
  {Brogan}, {Caratti o Garatti}, {Carrasco-Gonz{\'a}lez}, {Hunter}, {Massi}, \&
  {Padovani}}]{San19b}
{Sanna}, A., {Moscadelli}, L., {Goddi}, C., {et~al.} 2019, \aap, 623, L3

\bibitem[{{Sanna} {et~al.}(2018){Sanna}, {Moscadelli}, {Goddi}, {Krishnan}, \&
  {Massi}}]{San18}
{Sanna}, A., {Moscadelli}, L., {Goddi}, C., {Krishnan}, V., \& {Massi}, F.
  2018, \aap, 619, A107

\bibitem[{{Schw{\"o}rer} {et~al.}(2019){Schw{\"o}rer}, {S{\'a}nchez-Monge},
  {Schilke}, {M{\"o}ller}, {Ginsburg}, {Meng}, {Schmiedeke}, {M{\"u}ller},
  {Lis}, \& {Qin}}]{Schw19}
{Schw{\"o}rer}, A., {S{\'a}nchez-Monge}, {\'A}., {Schilke}, P., {et~al.} 2019,
  \aap, 628, A6

\bibitem[{{Seifert} {et~al.}(2010){Seifert}, {Ageorges}, {Lehmitz},
  {Buschkamp}, {Knierim}, {Polsterer}, {Germeroth}, {Pasquali}, {Naranjo},
  {J{\"u}tte}, {Feiz}, {Gemperlein}, {Hofmann}, {Laun}, {Lederer}, {Lenzen},
  {Mall}, {Mandel}, {M{\"u}ller}, {Quirrenbach}, {Sch{\"a}ffner}, {Storz}, \&
  {Weiser}}]{Seif10}
{Seifert}, W., {Ageorges}, N., {Lehmitz}, M., {et~al.} 2010, in Society of
  Photo-Optical Instrumentation Engineers (SPIE) Conference Series, Vol. 7735,
  Ground-based and Airborne Instrumentation for Astronomy III, 77357W

\bibitem[{{Shepherd} {et~al.}(2000){Shepherd}, {Yu}, {Bally}, \&
  {Testi}}]{She20}
{Shepherd}, D.~S., {Yu}, K.~C., {Bally}, J., \& {Testi}, L. 2000, \apj, 535,
  833

\bibitem[{{Smith} {et~al.}(2013){Smith}, {Shetty}, {Beuther}, {Klessen}, \&
  {Bonnell}}]{Smi13}
{Smith}, R.~J., {Shetty}, R., {Beuther}, H., {Klessen}, R.~S., \& {Bonnell},
  I.~A. 2013, \apj, 771, 24

\bibitem[{{Staff} {et~al.}(2014){Staff}, {Koning}, {Ouyed}, \&
  {Pudritz}}]{Sta14}
{Staff}, J., {Koning}, N., {Ouyed}, R., \& {Pudritz}, R. 2014, in European
  Physical Journal Web of Conferences, Vol.~64, European Physical Journal Web
  of Conferences, 05006

\bibitem[{{Staff} {et~al.}(2015){Staff}, {Koning}, {Ouyed}, {Thompson}, \&
  {Pudritz}}]{Sta15}
{Staff}, J.~E., {Koning}, N., {Ouyed}, R., {Thompson}, A., \& {Pudritz}, R.~E.
  2015, \mnras, 446, 3975

\bibitem[{{Staff} {et~al.}(2019){Staff}, {Tanaka}, \& {Tan}}]{Sta19}
{Staff}, J.~E., {Tanaka}, K. E.~I., \& {Tan}, J.~C. 2019, \apj, 882, 123

\bibitem[{{Steer} {et~al.}(1984){Steer}, {Dewdney}, \& {Ito}}]{Ste84}
{Steer}, D.~G., {Dewdney}, P.~E., \& {Ito}, M.~R. 1984, \aap, 137, 159

\bibitem[{{Terquem} {et~al.}(1999){Terquem}, {Eisl{\"o}ffel}, {Papaloizou}, \&
  {Nelson}}]{Ter99}
{Terquem}, C., {Eisl{\"o}ffel}, J., {Papaloizou}, J.~C.~B., \& {Nelson}, R.~P.
  1999, \apjl, 512, L131

\bibitem[{{Testi} {et~al.}(1999){Testi}, {Palla}, \& {Natta}}]{Tes99}
{Testi}, L., {Palla}, F., \& {Natta}, A. 1999, \aap, 342, 515

\bibitem[{{Tobin} {et~al.}(2012){Tobin}, {Hartmann}, {Bergin}, {Chiang},
  {Looney}, {Chandler}, {Maret}, \& {Heitsch}}]{Tob12}
{Tobin}, J.~J., {Hartmann}, L., {Bergin}, E., {et~al.} 2012, \apj, 748, 16

\bibitem[{{Trevi{\~n}o-Morales} {et~al.}(2019){Trevi{\~n}o-Morales}, {Fuente},
  {S{\'a}nchez-Monge}, {Kainulainen}, {Didelon}, {Suri}, {Schneider},
  {Ballesteros-Paredes}, {Lee}, {Hennebelle}, {Pilleri},
  {Gonz{\'a}lez-Garc{\'\i}a}, {Kramer}, {Garc{\'\i}a-Burillo}, {Luna},
  {Goicoechea}, {Tremblin}, \& {Geen}}]{Tre19}
{Trevi{\~n}o-Morales}, S.~P., {Fuente}, A., {S{\'a}nchez-Monge}, {\'A}.,
  {et~al.} 2019, \aap, 629, A81

\bibitem[{{Tsukamoto} {et~al.}(2018){Tsukamoto}, {Okuzumi}, {Iwasaki},
  {Machida}, \& {Inutsuka}}]{Tsu18}
{Tsukamoto}, Y., {Okuzumi}, S., {Iwasaki}, K., {Machida}, M.~N., \& {Inutsuka},
  S. 2018, \apj, 868, 22

\bibitem[{{V{\'a}zquez-Semadeni} {et~al.}(2019){V{\'a}zquez-Semadeni}, {Palau},
  {Ballesteros-Paredes}, {G{\'o}mez}, \& {Zamora-Avil{\'e}s}}]{Vaz19}
{V{\'a}zquez-Semadeni}, E., {Palau}, A., {Ballesteros-Paredes}, J.,
  {G{\'o}mez}, G.~C., \& {Zamora-Avil{\'e}s}, M. 2019, \mnras, 490, 3061

\bibitem[{{Wang} {et~al.}(2014){Wang}, {Zhang}, {Testi}, {van der Tak}, {Wu},
  {Zhang}, {Pillai}, {Wyrowski}, {Carey}, {Ragan}, \& {Henning}}]{Wan14}
{Wang}, K., {Zhang}, Q., {Testi}, L., {et~al.} 2014, \mnras, 439, 3275

\bibitem[{{Winter} {et~al.}(2018){Winter}, {Booth}, \& {Clarke}}]{Win18}
{Winter}, A.~J., {Booth}, R.~A., \& {Clarke}, C.~J. 2018, \mnras, 479, 5522

\bibitem[{{Wright} {et~al.}(2010){Wright}, {Eisenhardt}, {Mainzer}, {Ressler},
  {Cutri}, {Jarrett}, {Kirkpatrick}, {Padgett}, {McMillan}, {Skrutskie},
  {Stanford}, {Cohen}, {Walker}, {Mather}, {Leisawitz}, {Gautier}, {McLean},
  {Benford}, {Lonsdale}, {Blain}, {Mendez}, {Irace}, {Duval}, {Liu}, {Royer},
  {Heinrichsen}, {Howard}, {Shannon}, {Kendall}, {Walsh}, {Larsen}, {Cardon},
  {Schick}, {Schwalm}, {Abid}, {Fabinsky}, {Naes}, \& {Tsai}}]{Wri10}
{Wright}, E.~L., {Eisenhardt}, P.~R.~M., {Mainzer}, A.~K., {et~al.} 2010, \aj,
  140, 1868

\bibitem[{{Xu} {et~al.}(2013){Xu}, {Li}, {Reid}, {Menten}, {Zheng},
  {Brunthaler}, {Moscadelli}, {Dame}, \& {Zhang}}]{Xu13}
{Xu}, Y., {Li}, J.~J., {Reid}, M.~J., {et~al.} 2013, \apj, 769, 15

\end{thebibliography}
%

\begin{appendix}

\section{Velocity distributions of \ CH$_3$CN and HC$_3$N close to YSO-1}\label{CH3CN-HC3N}

\begin{figure*}
\centering
\vspace*{-0.3cm}
\includegraphics[width=0.75\textwidth]{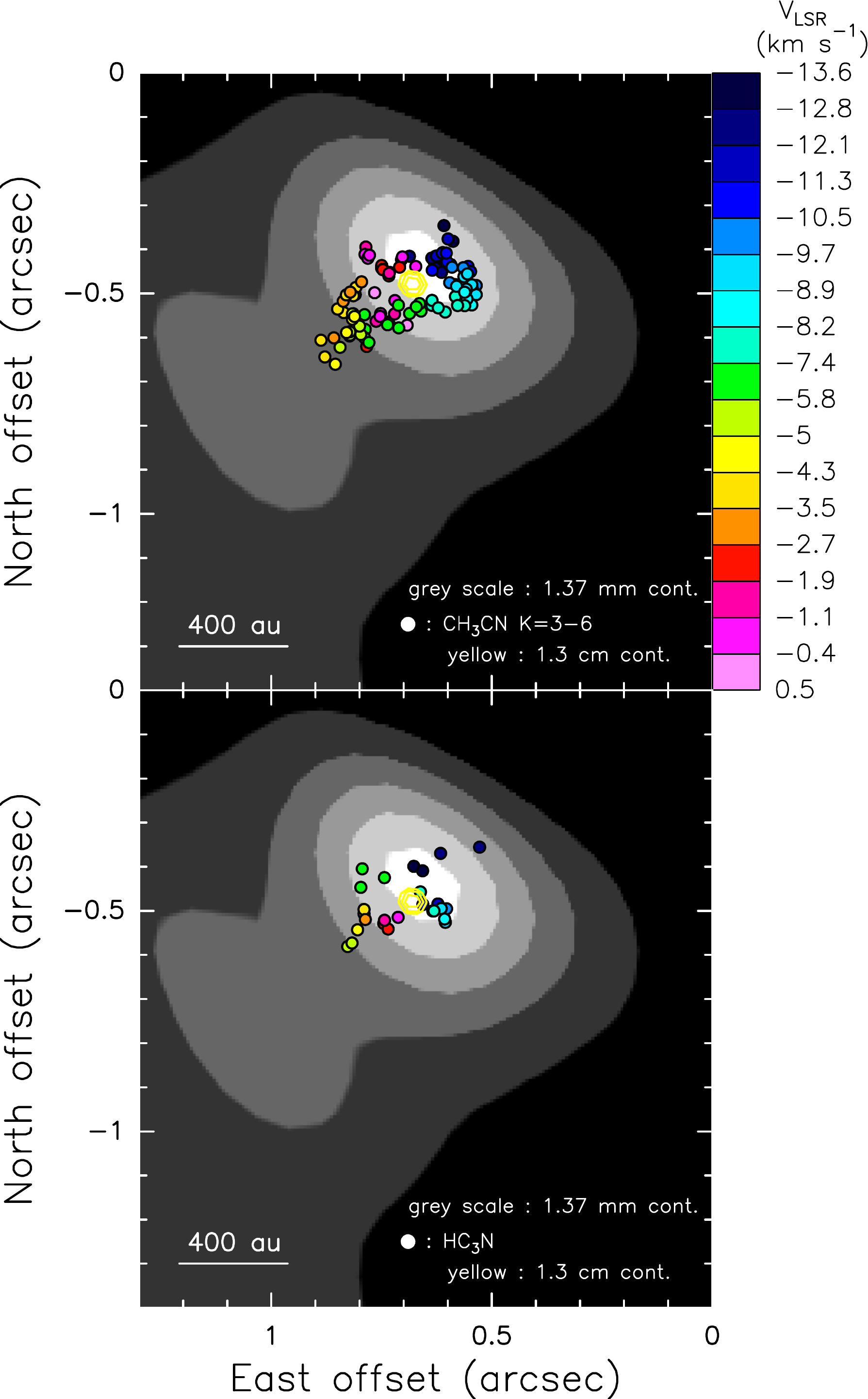}
\caption{NOEMA data. The gray-scale map reproduces the 1.37~mm continuum emission, plotting values in the range \ 10--35~m\Jyb. The colored dots give the (Gaussian-fitted) positions of the channel emission peaks for the CH$_3$CN~$J_K=$ 12$_K$-11$_K$ ($K$ = 3--6)  (upper~panel) and \ HC$_3$N~$J=$ 24-23 (lower~panel) lines; the 
color denotes the channel \Vlsr. The yellow contours show the JVLA A-Array continuum at 1.3~cm \citep{Mos16}, showing levels 
at \ 70\%, 80\%, and 90\% of the peak emission of \ 0.50~m\Jyb.}
\label{disk_comp}
\end{figure*}

Figure~\ref{disk_comp} shows that the \Vlsr\ distributions of the CH$_3$CN and  HC$_3$N emissions in proximity of YSO-1 are similar. For both molecules, the emission of the blue-shifted channels peak W-NW of YSO-1 (pinpointed by the compact 1.3~cm emission)  and that of the red-shifted channels E-SE of YSO-1. 

\section{Momentum of the molecular outflow}
\label{ap_mmo}

The momentum, $P_{\rm out}$, of the molecular outflow can be calculated using the \ $^{13}$CO~$J=$ 2-1 \ and \ SO~$J_N=$ 6$_5$-5$_4$ \ emission lines, which at sufficiently high absolute LOS velocities ($ | V_{\rm LSR} - V_{\rm sys} | \ge 3$~\kms; see Figs.~\ref{13COmap}~and~\ref{SOmap}) mainly trace the outflowing gas. In the limit of optically thin emission, we can use the following expressions \citep[see, for instance,][Eq.~10]{Gol99}:
\begin{align}
P_{\rm out} \, \cos (i_{\rm out}) &= F \; \sum_{k = 1}^{2} \; \sum_{n = n_{1,k}}^{n_{2,k}} \; \sum_{j = 1}^{J}  I_{n,j} \; \delta V \; \ |V_n-V_{\star}| \; d^2 \; \delta \Omega \label{Po_1} ,\\ 
F &= \frac{16 \, \ln(2)}{B_{\rm max} \; B_{\rm min}} \; \; \frac{Z_{\rm mol}(T_{\rm ex}) \; e^{E_{\rm u}/(k_{\rm B} T_{\rm ex})}}{h \; c \; g_{\rm u} \; A_{\rm ul}} \; \; \frac{\mu \; m_{\rm H}}{X_{\rm mol}},
\label{Po_2}
\end{align}

\noindent where \ $i_{\rm out}$ \ is the inclination angle of the outflow with respect to the LOS, the index \ $k$ \ considers separately the two, blue-~($k =$ 1)~and~red-shifted~($k =$ 2), outflow lobes, the sum in \ $n$ \ extends over the velocity channels ($n_{1,k} \le n \le n_{2,k}$) of each lobe, and \ $j$ \ runs over all the $J$ pixels of the maps shown in Figs.~\ref{13COmap}~and~\ref{SOmap}.  In Eq.~\ref{Po_1}, \ $ I_{n,j}$ \ and \ $V_n$ \ are, respectively, the map intensity and \Vlsr\ at pixel $j$ \ for channel $n$, $V_{\star}$ ($\approx$$-$6.4~\kms, see Sect.~\ref{dis_YSO}) is the \Vlsr\ of YSO-1, $\delta V$ (0.5~\kms) is the channel width, $d$  the distance (1.63~kpc) and \ $\delta \Omega$ the solid angle of the pixel. In Eq.~\ref{Po_2}, $B_{\rm max}$ \ and \ $B_{\rm min}$ \ are the FWHM sizes of the map beam along the major and minor axes, respectively, $T_{\rm ex}$ the excitation temperature of the line, $E_{\rm u}$, $g_{\rm u}$ \ and \ $A_{\rm ul}$ \ the energy and the statistical weight of the upper level and the spontaneous emission coefficient, respectively, of the transition, $ Z_{\rm mol}(T_{\rm ex}$) \ the molecular partition function, $X_{\rm mol}$ \ the molecular abundance with respect to the \ $H_2$ molecule, $\mu =$ 2.8 \ the mean molecular weight, $m_{\rm H}$ \ the mass of the hydrogen atom, $h$ \ the Planck constant, and \ $c$ \ the speed of light.

Table~\ref{out_mom} reports the parameters of the \ $^{13}$CO~$J=$ 2-1 \ and \ SO~$J_N=$ 6$_5$-5$_4$ \ transitions, recovered from databases of molecular spectroscopy. Assuming LTE conditions, the $T_{\rm ex}$ \ of the two lines can be approximated with the gas kinetic temperature. A lower limit is the value of \ $\approx$66~K, averaged over the full extent of the cluster, derived by \citet[][see Table~3]{Beu18} by fitting  the IRAM~30~m \ H$_2$CO data at an angular resolution of \ 11\farcs8  \ with  \textsc{XCLASS}; an upper limit is the maximum temperature of \ $\approx$200~K \ toward core~1 from the \textsc{XCLASS} fit of the \ CH$_3$CN~$J_K=$ 12$_K$-11$_K$ (K= 0--6) lines (see Fig.~\ref{T_rot}). We then use for both lines \ $T_{\rm ex} =$ 100~K, which should be accurate to within a factor of 2.
We verified that the error on the outflow momentum owing to this uncertainty on \ $T_{\rm ex}$ \ is less than a factor of 2.  
By fitting the emission of the two lines and the dust continuum with \textsc{XCLASS}, and assuming the same dust absorption coefficient and gas-to-dust mass ratio of Sect.~\ref{TM_dis}, we estimated the
 abundance ratios (with respect to \ H$_2$) of the \ $^{13}$CO \ and \ SO \ molecules to be \ 2$\, \times \,$10$^{-6}$ \ and \ 10$^{-8}$, respectively. Employing a single line for each molecular species does not allow us to correct for optical depth effects, which can be significant in particular for the more abundant \ $^{13}$CO. 
The \ H$_2$ \ column density is derived from NOEMA-only data, so due to potential missing flux 
 \citep[by about a factor of 2 for \targ,][see Table~3]{Beu18}, it is likely a lower limit as well. Based  on our experience with the CORE data, we judge that these estimates of the abundance ratios should be correct within a factor of \ 3.

The \Vlsr\ ranges \ [$V_{1,k}$, $V_{2,k}$] (where \ $V_{1,k}$ \ and \ $V_{2,k}$ \ are the velocities of the channels \ $n_{1,k}$ \ and \ $n_{2,k}$, respectively) for the blue-~($k =$ 1)~and~red-shifted~($k =$ 2), outflow lobes (see Table~\ref{out_mom}) are selected in such a way to include only the channels where the \ $^{13}$CO~$J=$ 2-1 \ and \ SO~$J_N=$ 6$_5$-5$_4$ \ emissions are detected with a high signal-to-noise ratio (see Figs.~\ref{13COmap}~and~\ref{SOmap}; the  channel at \Vlsr\ = $-$28.1~\kms\ of the \ SO~$J_N=$ 6$_5$-5$_4$ line is excluded because it is too noisy). 

Summing over the two outflow lobes, we derive a momentum for the molecular outflow of \ $\sim$17~\ms~\kms\ and \ $\sim$82~\ms~\kms\ using the \ $^{13}$CO~$J=$ 2-1 \ and \ SO~$J_N=$ 6$_5$-5$_4$ \ line, respectively. In the following, we adopt the intermediate value of \ $P_{\rm out} \, \cos (i_{\rm out}) \sim$50~\ms~\kms, which, based on the above considerations, should be accurate to within an order of magnitude.

%
\begin{table*}
\caption{Parameters employed to calculate the momentum of the molecular outflow}             
\label{out_mom}      
\centering          
\begin{tabular}{c c c c c c c c c c c}     
\hline\hline       
Transition & $E_{\rm u}$\tablefootmark{a}$ / k_{\rm B}$ & $g_{\rm u}$\tablefootmark{a} & $A_{\rm ul}$\tablefootmark{a} & $T_{\rm ex}$ & $Z_{\rm mol}$ & $X_{\rm mol}$ & $B_{\rm max}$ & $B_{\rm min}$ & $[V_{1,1}, V_{2,1}]$ & $[V_{1,2}, V_{2,2}]$ \\ 
     & K          &       & s$^{-1}$ &  K       &           &          & arcsec  & arcsec   &  \kms  &  \kms \\
\hline                     
SO~$J_N=$ 6$_5$-5$_4$  & 35.0 & 13 & 1.34$\, \times \,$10$^{-4}$ & 100 & 292\tablefootmark{b} & 10$^{-8}$ & 0.70 & 0.61 & [$-$25.1, $-$10.1]  & [$-$1.1, $+$4.9]  \\  
$^{13}$CO~$J=$~2-1   & 15.9 & 5  & 6.08$\, \times \,$10$^{-7}$ & 100 & 38\tablefootmark{c} & 2$\, \times \,$10$^{-6}$ & 0.70 & 0.61 & [$-$21.5, $-$9.5] & [$-$0.5, $+$5.4]  \\
\hline                  
\end{tabular}
\tablefoot{.\\
\tablefoottext{a}{Molecular parameters from the Cologne Database for Molecular
Spectroscopy \citep[http://www.astro.uni-koeln.de/cdms/,][]{Muel01,Muel05}.}
\tablefoottext{b}{Derived from the partition function tabulated by \citet{Bar16}.}
\tablefoottext{c}{Calculated using data from the Jet Propulsion Laboratory Catalog of Molecular Spectroscopy \citep[https://spec.jpl.nasa.gov/,][]{Pick98}.}
}
\end{table*}
%

\end{appendix} 

\end{document}